\documentclass[12pt]{JHEP}
\usepackage{epsfig}
\usepackage{amsmath}
\usepackage{amssymb}

\font\blackboard=msbm10 at 12pt
\font\blackboards=msbm7
\font\blackboardss=msbm5
\newfam\black
\textfont\black=\blackboard
\scriptfont\black=\blackboards
\scriptscriptfont\black=\blackboardss

\newcommand{\ba}{\begin{array}}
\newcommand{\ea}{\end{array}}
\newcommand{\be}{\begin{equation}}
\newcommand{\ee}{\end{equation}}
\newcommand{\bea}{\begin{eqnarray}}
\newcommand{\eea}{\end{eqnarray}}
\newcommand{\beas}{\begin{eqnarray*}}
\newcommand{\eeas}{\end{eqnarray*}}
\newcommand{\0}{\ket{0}}
\newcommand{\bra}[1]{\langle #1 |}
\newcommand{\onehalf}{\frac{1}{2}}
\newcommand{\er}[1]{(\ref{eq:#1})}
\newcommand{\ket}[1]{| #1 \rangle}
\newcommand{\tX}{\widetilde{X}}

\newcommand{\tV}{\widetilde{V}}
\newcommand{\adag}{a^{\dag}}
\newcommand{\bdag}{b^{\dag}}
\newcommand{\cdag}{c^{\dag}}
\newcommand{\hV}{\hat{V}}
\newcommand{\hX}{\hat{X}}
\newcommand{\barray}{\begin{eqnarray}}
\newcommand{\earray}{\end{eqnarray}}

\def\laplace{{\kern1pt\vbox{\hrule height 1.2pt\hbox{\vrule width
1.2pt\hskip
  3pt\vbox{\vskip 6pt}\hskip 3pt\vrule width 0.6pt}\hrule height
  0.6pt}
  \kern1pt}}
\def\scriptlap{{\kern1pt\vbox{\hrule height 0.8pt\hbox{\vrule width
  0.8pt
  \hskip2pt\vbox{\vskip 4pt}\hskip 2pt\vrule width 0.4pt}\hrule height
  0.4pt}
  \kern1pt}}

\def\roughly#1{\raise.3ex\hbox{$#1$\kern-
.75em\lower1ex\hbox{$\sim$}}}

\newcommand{\NP}{{\em Nucl.\ Phys.\ }}

\newcommand{\PL}{{\em Phys.\ Lett.\ }}
\newcommand{\PR}{{\em Phys.\ Rev.\ }}

\newcommand{\PRL}{{\em Phys.\ Rev.\ Lett.\ }}

\newcommand{\gone}[1]{}

\title{Tadpoles and Closed String Backgrounds in Open String Field
Theory}

\author{Ian Ellwood, Jessie Shelton and Washington Taylor\\
{Center for Theoretical Physics} \\ {MIT, Bldg.  6-308} \\ {Cambridge,
MA 02139, U.S.A.} \\ {\tt iellwood@mit.edu, jshelton@mit.edu,
wati@mit.edu}}

\abstract{We investigate the quantum structure of Witten's cubic open
bosonic string field theory by computing the one-loop contribution to
the open string tadpole using both oscillator and conformal field
theory methods.  We find divergences and a breakdown of BRST
invariance in the tadpole diagram arising from tachyonic and massless
closed string states, and we discuss ways of treating these problems.
For a D$p$-brane with sufficiently many transverse dimensions, the
tadpole can be rendered finite by analytically continuing the closed
string tachyon by hand; this diagram then naturally incorporates the
(linearized) shift of the closed string background due to the presence
of the brane.  We observe that divergences at higher loops will doom
any straightforward attempt at analyzing general quantum effects in
bosonic open string field theory on a D$p$-brane of any dimension, but
our analysis does not uncover any potential obstacles to the existence
of a sensible quantum open string field theory in the supersymmetric
case.}

\keywords{String Field Theory} \preprint{MIT-CTP-3368,
hep-th/0304259}

\begin{document}

%%%%%%%%%%%%%%%%%%%%%%%%%%%%%%%%%%%%%%%%%%%%%%%%%%%%%%%%%%%%%%%%%%%%%%%%%%%%
%%%%%%%%%%%%%%%%%%%%%%%%%%% Introduction %%%%%%%%%%%%%%%%%%%%%%%%%%%%%%%%%%%
%%%%%%%%%%%%%%%%%%%%%%%%%%%%%%%%%%%%%%%%%%%%%%%%%%%%%%%%%%%%%%%%%%%%%%%%%%%%

\section{Introduction}

String field theory is a space-time formulation of string theory which
may have the capacity to describe all string backgrounds in terms of a
common set of degrees of freedom.  Much recent interest in Witten's
open string field theory (OSFT) \cite{Witten:1985cc} has been centered
around the discovery that this theory can describe D-branes as
classical solitons, so that distinct open string backgrounds not
related through marginal deformations can appear as solutions of a
single set of equations of motion (for reviews
see~\cite{Sen:1999xm,Ohmori:2001am,DeSmet:2001af,Zwiebach:nj,Taylor:2002uv}).
Most of the recent work in this area has focused on classical aspects
of OSFT (although, for some recent papers which address quantum
features of the theory, see \cite{quantum}).

In order for string field theory to have a real chance at addressing
any of the deep unsolved problems in string theory/quantum gravity, it
is clearly necessary that the theory should be well defined quantum
mechanically.  In an earlier phase of work, some progress was made in
understanding the quantum structure of OSFT.  This work is summarized
and described in the language of BV quantization by Thorn in his
review \cite{Thorn:1988hm}.  In this paper we extend this earlier work
by carrying out a systematic analysis of the one-loop open string
tadpole diagram in Witten's bosonic OSFT.  We analyze the divergence
structure of this diagram and the role which closed strings play in
the structure of the tadpole, and we describe the implications of this
analysis for the quantum theory.

An important aspect of quantum open string field theory is the role
which closed strings play in the theory.  As has been known since
their first discovery \cite{NGS,Lovelace}, closed strings appear as poles in
nonplanar one-loop amplitudes of open strings.  An analysis of these
poles in the one-loop nonplanar two-point function of OSFT was given
in \cite{Freedman:fr}.  Because of the existence of these intermediate
closed string states, any unitary quantum open string field theory
must include some class of composite asymptotic states which can be
identified with closed strings.  These asymptotic states have not yet
been explicitly identified in OSFT, although related open string
states which can be used to compute amplitudes including closed
strings in OSFT are described in
\cite{Shapiro:gq,Shapiro:ac,Zwiebach:1992bw,Hashimoto:2001sm,Gaiotto:2001ji,
a-Garousi};
other approaches to understanding how closed strings appear in OSFT
were pursued in
\cite{closed}.  In this
paper, we consider the 
appearance of closed strings in OSFT from a different point of view
than has been taken in previous work on the subject.  We show that an
important part of the structure of the open string tadpole comes from
the closed string tadpole, which in the presence of a D-brane
describes the linearized gravitational fields of that D-brane.  This
demonstrates that not only do the closed strings appear as poles in
the open string theory, but that they also take expectation values in
response to D-brane sources within the context of OSFT; this provides
a new perspective on the role of closed strings in OSFT.

The relationship between open and closed strings is central to the
concept of holography and the AdS/CFT correspondence
\cite{Maldacena:1997re,Aharony:1999ti,D'Hoker:2002aw}.  In the AdS/CFT
correspondence, a decoupling limit is taken where open strings on a
brane are described by a conformal field theory; this theory has a
dual description as a near-horizon limit of the closed string
(gravity) theory around the D-brane.  A complete quantum open string
field theory would generalize this picture; if OSFT can be shown to be
unitary without explicitly including the closed strings as additional
dynamical degrees of freedom, we would have a more general holographic
theory in which the open string field theory on a D-brane would encode
the gravitational physics in the full D-brane geometry in a precise
fashion.  While we do not directly address these ideas in this paper,
some further discussion in this direction is included at the end of
the paper.

Until recently, only a few diagrams had been explicitly evaluated in
OSFT: the Veneziano amplitude \cite{Giddings,Sloan,Samuel-off}, and
the non-planar two-point function~\cite{Freedman:fr}.  These diagrams
were computed by explicitly mapping the Witten parameterization of
string field theory to a parameterization more natural for conformal
field theory, and then computing the diagram explicitly in CFT.  There
has recently been some renewed interest in studying perturbative
aspects of OSFT by developing new techniques for calculating diagrams
in the theory \cite{Taylor:2002bq,Bars:2002qt}.  Using these methods
it is possible to compute any OSFT diagram to a high degree of
accuracy using the level truncation method on oscillators.  This
method provides an alternative to the CFT method, and gives some
information about a wider range of diagrams while lacking the analytic
control of the CFT method.  In this paper we use both methods, finding
that each gives useful information.

The one-loop tadpole diagram we consider in this paper is perhaps the
simplest of the one-loop diagrams in OSFT.  While a preliminary study
of this diagram was done in
\cite{Thorn:1988hm,Taylor:2002bq}, we expand on the
analysis presented in \cite{Taylor:2002bq} and augment it by using the
conformal field theory method to give an alternate expression for the
diagram.  We also generalize the discussion by computing the diagram
for OSFT defined on a D$p$-brane background for any $p$.  

The one-loop tadpole diagram has divergences of several kinds.  As the
modular parameter $T$ describing the
length of the internal open string propagator becomes large, there
is a divergence from the open string tachyon.  This
divergence is easy to understand, and can be removed by
analytic continuation in the oscillator approach to OSFT.  In addition
to the large $T$ divergence of the diagram, there are divergences as
$T \rightarrow 0$.  
In the conformal
frame natural to OSFT this limit corresponds to a pinching off of the
world-sheet.  In an alternate conformal frame, however, the small open
string loop gives rise to a long closed string tube. These two
conformal frames are displayed in figure~\ref{f:conformalframes}.
\FIGURE{ \epsfig{file=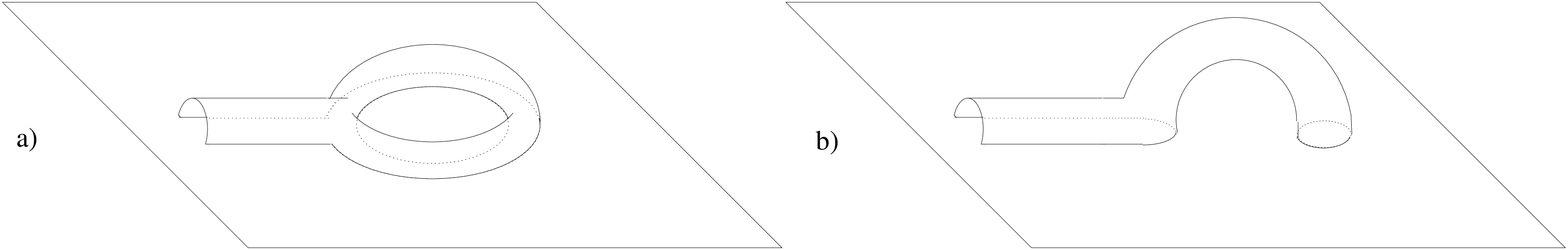,width=16cm}
\caption{\footnotesize
Two conformally equivalent pictures of the one-loop open string tadpole.
a) The open string tadpole is represented as a 
purely open string process in which a single open string splits into two
open strings which then collide.
b) The open string tadpole is represented as a transition between an open
string and a closed string. The closed string is absorbed into the brane.
}
\label{f:conformalframes}
} Since the propagation over long distances of massive fields is
suppressed, only the tachyon and the massless sector of the closed
strings contribute to the $T \rightarrow 0$ divergences of the tadpole.  
Using both the conformal field theory and oscillator approaches, we
isolate the  $T \rightarrow 0$ divergences in the one-loop
tadpole diagram.
By extracting
the leading divergences of the tadpole, we can separate the divergence
arising from the tachyon from any divergences associated with the
graviton/dilaton in the massless sector.

The divergence from the closed string tachyon arises because of the
usual problem that the Euclidean theory has a real exponent in the
Schwinger parameterization of the propagator, and diverges for
tachyonic modes.  This problem is usually dealt with by a simple
analytic continuation.  In this case, however, the analytic
continuation is rather subtle, as the closed string degrees of freedom
which are causing the divergence are not fundamental degrees of
freedom in the theory.  In the one-loop diagram we study in this
paper, the analytic continuation can be done by hand in the CFT
approach by explicitly using our understanding of the closed string
physics underlying the divergence.  Even for this relatively simple
diagram, however, there are a number of subtleties in this analytic
continuation, and to ensure that we completely remove all the
tachyonic divergences we are forced to consider lower-dimensional
brane backgrounds.  In a more general context, such as for higher-loop
diagrams, it would be difficult to systematically treat this type of
divergence using open string field theory.

Assuming that the divergence from the closed string tachyon is dealt
with by a form of analytic continuation, we are left with possible
divergences from the massless closed string states.  Such divergences
appear only when considering the open string theory on a D$p$-brane
with $p\geq 23$.  This is essentially because the open string tadpole
is generated directly from the closed string tadpole, and the closed
string tadpole arises from the solution of the linearized
gravitational equations with a D$p$-brane source
\cite{Polchinski:jq,DiVecchia}.  Since a brane of 
codimension 2 has a long-range potential which goes as $\ln r$, while
a brane of codimension 3 has a potential going as $1/r$, we need at
least three codimensions to remove the divergences from the massless
sector.  In general, whether the tadpole is finite or divergent, we
find that the structure of the linearized closed string fields in the
D-brane background is encoded in the open string tadpole.

After analyzing the divergence structure of the one-loop tadpole, we
also consider briefly what one might expect of OSFT at two or more
loops.  We consider in particular the two loop non-planar diagram
which represents a a torus with a hole in it.  This diagram contains
as a subdiagram the closed string one point function which suffers
from a BRST anomaly \cite{Fischler:ci,Das:dy,Fischler:1987gz,Polchinski:jq}
in world-sheet perturbation theory.  We conjecture that in OSFT this
diagram will lead to a divergence and BRST anomaly.  Since this
divergence is purely closed-string in nature, it should occur for the
theory on a D$p$-brane for any $p$, and poses a serious problem for
any attempt to make sense of the bosonic open string field theory as a
complete quantum theory.

In Section 3 we compute the one-loop tadpole using conformal field
theory methods.  In section 4 we compute the same diagram using
oscillator methods.  Section 5 contains a discussion of the one-loop
open string tadpole in Zwiebach's open-closed string field theory,
where the structure of the diagram is somewhat more transparent.
Section 6 synthesizes the results of the preceding sections, and
contains a general discussion of the divergences of the tadpole and
the role of closed strings in the tadpole.  Section 7 contains some
concluding remarks.  In two appendices we include some technical points.
Appendix A contains a discussion of the BRST anomaly in the D25-brane
theory.  Appendix B contains some comments on the infinite-level limit
of the level truncation method used in Section 4.

%%%%%%%%%%%%%%%%%%%%%%%%%%%%%%%%%%%%%%%%%%%%%%%%%%%%%%%%%%%%%%%%%%%%%%%%%%%%%
%%%%%%%%%%%%%%%%%% Perturbation Theory in Open String Field Theory %%%%%%%%%%
%%%%%%%%%%%%%%%%%%%%%%%%%%%%%%%%%%%%%%%%%%%%%%%%%%%%%%%%%%%%%%%%%%%%%%%%%%%%%

\section{Perturbation theory in open string field theory}
We begin with a summary of Witten's formulation of open bosonic string field
theory (OSFT) \cite{Witten:1985cc} with an emphasis on perturbation theory.
For general reviews of OSFT 
see~\cite{lpp,Gaberdiel-Zwiebach,Thorn:1988hm,Zwiebach:nj}.
The classical field for OSFT is a ghost number one
state, $\Psi$,  in the free open string Fock space.
The string field,
$\Psi$, has a natural expansion in terms of the open string fields
\begin{equation}
\Psi = \int \, d^{26}p \left[
\phi(p) c_1 |0;p\rangle + A_{\mu}(p) \alpha_{-1}^{\mu} c_1 |0;p\rangle
 + \psi(p) c_0 |0;p\rangle + \ldots\right],
\end{equation}
where $\phi$ is the open string tachyon, $A_{\mu}$ is the gauge field and
$\psi$ is an auxiliary field.  The vacuum $|0\rangle$ denotes the
$SL(2,{\bf R})$ vacuum.  The classical action is given by
\begin{equation}\label{eq:WittenAction}
  S(\Psi)  = \frac{1}{2}\int \Psi \star Q_B \Psi 
            +\frac{g}{3} \int \Psi \star \Psi\star \Psi.
\end{equation}
The definitions for the $\star$-product and string integration are
given in \cite{lpp,Gross:1986ia,Gross:1986fk,cst,Samuel,Ohta,rsz} in terms
of both oscillator expressions and conformal field theory correlators.
We will also make use of two-string and three-string vertices,
$\langle V_2|$ and $\langle V_3 |$, which are defined by
\begin{eqnarray}
\langle V_2 | |\Psi_1\rangle |\Psi_2\rangle 
  &=& \int \Psi_1 \star \Psi_2,\\
  \langle V_3 | |\Psi_1\rangle |\Psi_2\rangle |\Psi_3\rangle
  &=& \int \Psi_1 \star \Psi_2 \star \Psi_3,
\end{eqnarray}
and are elements of the two-string and three-string Fock spaces
respectively.

The theory has a large gauge
group.  Infinitesimally the gauge transformations are given by
\begin{equation}
  \Psi \to \Psi + Q_B \Lambda + g(\Psi \star \Lambda - \Lambda \star \Psi),
\end{equation}
where $\Lambda$ is any ghost number $0$ field.

To quantize the theory we must fix a gauge.  The standard choice for
gauge fixing is Feynman-Siegel (FS) gauge fixing which imposes the 
condition $b_0 \Psi =0$.  
It is straightforward to perform tree-level calculations in this gauge,
but some care is required when trying to
impose this condition on path integrals.  Roughly speaking,
it turns out that if one
tries to introduce Fadeev-Poppov ghosts to fix $b_0 \Psi =0$, 
the ghosts themselves suffer from a gauge redundancy similar
to the gauge redundancy of the original action.  To fix this new
gauge redundancy one must introduce ghosts for ghosts.  As the new ghosts
have their own redundancy, this process proceeds forever, creating 
an infinite tower of ghost fields.  Happily, at the end of the
day this entire procedure can be summarized as follows~\cite{Thorn:1988hm}:
\begin{enumerate}
\item The field $\Psi$ is fixed by $b_0 \Psi = 0$.
\item The ghost number of $\Psi$ is allowed to range over all ghost numbers,
not just ghost number~1.  The fields of ghost numbers other than one are
all ghost fields.
\item $\Psi$ is a grassmann odd field.  To define what this means, 
suppose the states 
$\{|s\rangle \}$ form a basis for the open string Fock space such 
that each $|s\rangle$ has definite ghost number. Then if we write $\Psi$ in a 
Fock space expansion as $\Psi = \sum_s |s\rangle \psi_s$, then $\psi_s$
has the opposite grassmannality of $|s\rangle$.
\end{enumerate}
The form of the action remains the same as in equation~(\ref{eq:WittenAction}).
Using the FS gauge condition of $\Psi$ we can simplify the kinetic term:
\begin{equation}\label{eq:FSAction}
S_{FS}(\Psi) =   \frac{1}{2}\int \Psi \star c_0 L_0 \Psi 
+\frac{g}{3} \int \Psi \star \Psi \star \Psi.
\end{equation}
Given the gauge fixed action we can now develop the Feynman rules for 
perturbation theory. We can do this in two ways, which we will refer
to as the conformal field theory method and the oscillator method.

In the conformal field theory method, the Feynman rules are given in terms
of rules for sewing strips of world-sheet together.  Amplitudes
may be evaluated by conformally mapping the resulting diagrams to
the upper half plane for genus 0 or the cylinder for genus 1.

In the oscillator method
the Feynman rules are calculated directly from the action  using the usual
methods from field theory but summing over the infinite number of fields.
For any amplitude this gives rise to correlators which can be evaluated
using squeezed state methods.

In the next two sections we consider each of these two methods in turn.

%%%%%%%%%%%%%%%%%%%%%%%%%%%%%%%%%%%%%%%%%%%%%%%%%%%%%%%%%%%%%%%%%%%%%%%%%%%
%%%%%%%%%%%%%%%%%%%%%%%%%%%%%%%%%%%%%%%%%%%%%%%%%%%%%%%%%%%%%%%%%%%%%%%%%%%
%                        Conformal Field Theory Methods                   %
%%%%%%%%%%%%%%%%%%%%%%%%%%%%%%%%%%%%%%%%%%%%%%%%%%%%%%%%%%%%%%%%%%%%%%%%%%%
%%%%%%%%%%%%%%%%%%%%%%%%%%%%%%%%%%%%%%%%%%%%%%%%%%%%%%%%%%%%%%%%%%%%%%%%%%%

\section{Evaluation of the  tadpole using conformal field theory}
\label{S:ConformalAnalysis}

In this section we calculate the one-point function using conformal
field theory methods.  We begin the calculation with the assumption
that we are working with the theory on a D$25$-brane.  In section
\ref{s:maptocylinder} we describe how the diagram can be computed by
constructing a map from the original Witten diagram to the cylinder.
In section \ref{s:Tsmall}, we specialize to the limit where the
internal loop of the diagram is small and study the divergences in
this limit. In section \ref{s:Interpretingdivs} we discuss the origin
of these divergences from the propagation of closed string modes over
long distances (these divergences are discussed further in Section 6).
Finally, in section \ref{s:plessthan25} we discuss how the calculation
differs for the theory on a D$p$-brane with $p\ne25$.

%%%%%%%%%%%%%%%%%%%%%%%%%%%%%%%%%%%%%%%%%%%%%%%%%%%%%%%%%%%%%%%%%%%%%%%%%%
        \subsection{Mapping the Witten diagram to the cylinder}
%%%%%%%%%%%%%%%%%%%%%%%%%%%%%%%%%%%%%%%%%%%%%%%%%%%%%%%%%%%%%%%%%%%%%%%%%%
\label{s:maptocylinder}
We begin with a brief discussion of the world-sheet interpretation of
FS gauge fixed OSFT.  The derivation of this interpretation is
given in \cite{Giddings:1986bp,Giddings:1986wp,Thorn:1988hm,Zwiebach:1990az}.  
The Feynman rules consist of one propagator and one vertex.

The propagator is given by an integral over world-sheet strips of fixed width.
By convention the strips are of width $\pi$, and length $T$, 
where $T$ is integrated from $0$ to $\infty$.
To ensure the right measure on moduli space $b(\sigma)$ is integrated
across the strip \cite{Giddings:1986bp}.
\FIGURE{
 \epsfig{file=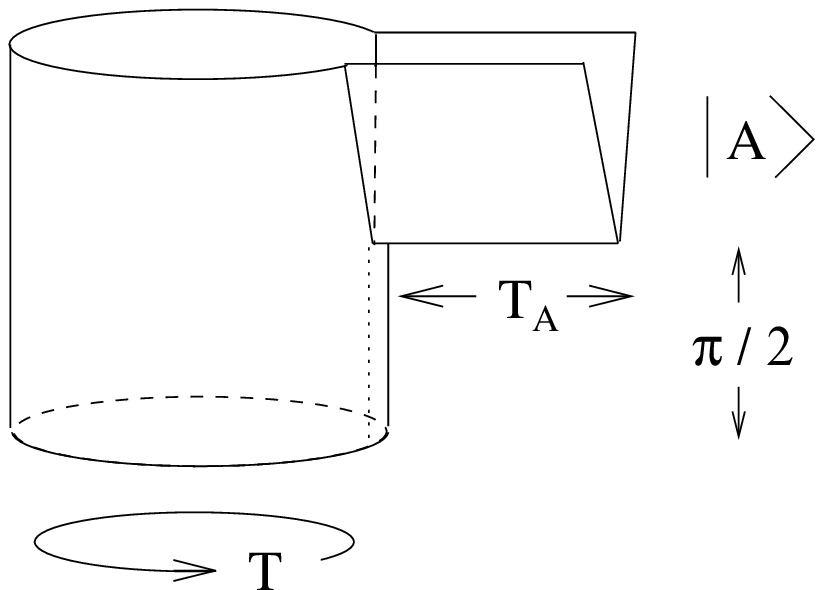,width=6cm}
 \caption{\footnotesize The world-sheet diagram for the one-point function.}
 \label{f:TadpoleDiagram}
}
The only vertex in the theory is a prescription for gluing three strips
together.  The right half of the first strip is glued to the left half
of the second and similarly for the second and third strips and 
the third and first strips.

Using these rules we can construct the one-point amplitude.  
We start with an external state $|A\rangle$ which propagates along
a strip of length $T_A$.
We then  take a second strip of length $T$ and glue both ends of
it and the end of the first propagator together using the vertex.
The resulting diagram is pictured in figure~\ref{f:TadpoleDiagram}.

We wish to study this diagram using conformal field theory methods.
To do this we use the methods of \cite{Freedman:fr} to map the diagram
to a cylinder.  For another approach to this conformal mapping
problem, see \cite{Zemba:rf}.  Taking the limit $T_A \to \infty$, we
can map the external state, $|A\rangle$, to a puncture at the boundary
of the cylinder.

It is convenient to flatten the diagram by cutting along the folded
edge of the external propagator in figure~\ref{f:TadpoleDiagram} and
cutting the internal propagator in half.  The resulting diagram is
displayed in figure~\ref{f:Torus} a. 
We will let $\rho$ be the coordinate on the
Witten diagram and $u$ be the coordinate on the cylinder.  To enforce
Neumann boundary conditions along the boundaries of the diagram, we 
use the doubling trick. Since the double of the cylinder is a torus, we may
use the theory of elliptic functions to determine $\rho(u)$.  

Consider the image of the Witten diagram under $\rho\to u$ shown in 
figure~\ref{f:Torus} b.
\FIGURE{
\epsfig{file=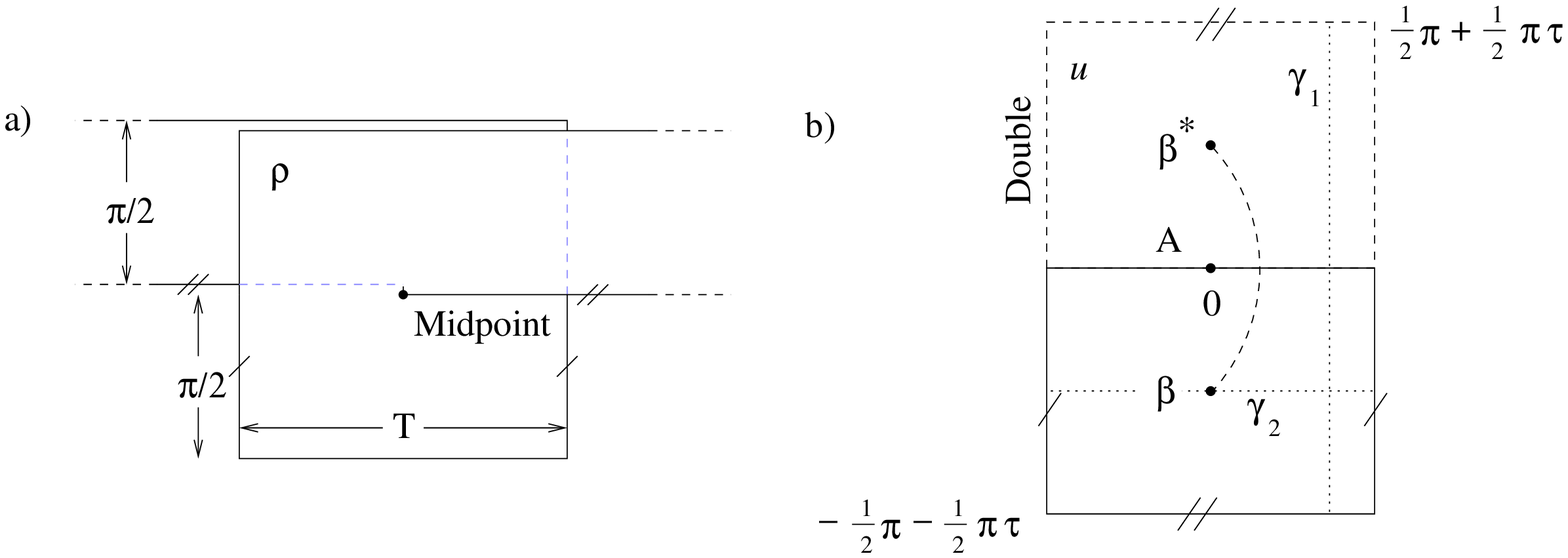,width=16cm}
\caption{\footnotesize 
a) The tadpole diagram laid flat by cutting
the external propagator along its middle and cutting the internal 
propagator in half.  The edges to be identified are indicated by dashes.
b) The doubled image of the tadpole under the conformal map $u(\rho)$.  
The left
and right sides of the image are identified as well as the top and bottom.
The image of the midpoint of the vertex is denoted by $\beta$.
}
\label{f:Torus}
}
The top and bottom of the diagram are identified as well as the left and
right edges.  The external state, $|A\rangle$, is mapped to a puncture
at point $A$ which we choose to be at $u=0$.  
The midpoint is mapped to point $\beta$.  By the symmetry
in the diagram, we can set the real part of $\beta$ to zero.
 Since the vertex
in the original Witten diagram had an angle of $3\pi$, the function $\rho(u)$
must behave as $\rho(u)-\rho(\beta)\sim (u-\beta)^{3/2}$ near $\beta$.  This
implies that $d\rho/du$ has a branch cut.  The height of the torus is
given by a purely imaginary parameter $\tau$ to be determined later.
The integral around the puncture $A$
 of $d\rho/du$ corresponds\footnote{
One has to be careful that the contour does not cross the branch cut so that,
 in the $\rho$ coordinates, the contour is continuous.
},
in the original Witten diagram, to the total width of the external propagator
plus its double, which is fixed to be $2 \pi i$.  
This implies that $d\rho/du$ has a simple pole at $A$ of residue one.

We now define the quadratic differential $\phi(u) = (\frac{d\rho}{du})^2(u)$.
{}From the form of $d\rho/du$ near $\beta$ we see that $\phi(u)$ has no branch
cut, just a simple zero at $\beta$.  In order to preserve Neumann boundary
conditions we must also include a zero at the image of $\beta$ under the 
doubling.  Since we have put the top of the diagram along the real axis, we 
just get a second zero at $\beta^{*}$.  The only other piece of analytic
structure we need is that since $d\rho/du$ had a simple pole at $A$, $\phi(u)$
has a double pole at $A$.

Now since $\phi(u)$ is a meromorphic function on a torus we may determine it
using $\vartheta$-functions:
\begin{equation}
\phi(u)  = 
   C \frac{\vartheta_1(u-\beta,q)
    \vartheta_1(u-\beta^{*},q)}{\vartheta_1^2(u,q)},
\end{equation}
where $q = e^{i \pi \tau}$.
The constant $C$ is determined from the condition that $\sqrt{\phi(u)}$ has
a pole of residue one at $A$:
\begin{equation}
  C = \frac{(\vartheta_1^{\prime }(0))^2}{\vartheta_1(-\beta)
   \vartheta_1(-\beta^{*})}.
\end{equation}
The two constants $\beta$ and $\tau$ can be determined
from the height and width of the diagram by integrating $d\rho/du$ along
the curves $\gamma_1$ and $\gamma_2$.  We have
\begin{eqnarray}
\oint_{\gamma_1} du \sqrt{\phi(u)} &=& 2 \pi i,\nonumber\\
\oint_{\gamma_2} du \sqrt{\phi(u)} &=& T.\label{eq:constraints}
\end{eqnarray}

\noindent
In general these relations cannot be solved analytically, but it
is straightforward to solve them numerically and thus to determine
$\tau$ and $\beta$ as functions of $T$.  

At this point one could, in principle, evaluate the diagram for any
given $A$.  If we suppose that the state $A$ is defined by a vertex
operator inserted on a half-disk with coordinates $v$, we can easily
compute the map $\rho (v)$ from the half-disk to the tadpole diagram.
The diagram at a fixed modular parameter is then computed by
evaluating
\begin{equation}\label{eq:TorusCorr}
  \langle ( u (\rho) \circ \rho(v) \circ A) 
( u(\rho)\circ  \frac{1}{2 \pi i}\int\, d\rho\, b(\rho)) \rangle_{\text{torus}}
\end{equation}
where the contour of integration runs across the internal propagator.
This correlator implicitly defines a fock space state,
$|\mathcal{T}(T)\rangle$, given by
\begin{equation}\label{eq:TofTDef}
  \langle A | \mathcal{T}(T)\rangle \equiv \langle ( u (\rho) \circ
 \rho(v) \circ A) ( u(\rho)\circ \frac{1}{2 \pi i}\int\, d\rho\,
 b(\rho)) \rangle_{\text{torus}}.
\end{equation}
Note that the state $|\mathcal{T}(T)\rangle$ is a function of the 
modular parameter $T$.  The full tadpole diagram is given by
integrating over this modular parameter.  We thus define the full
tadpole state $|\mathcal{T}\rangle$ by
\begin{equation}
  |\mathcal{T}\rangle = \int_{0}^{\infty} dT\,\,
   |\mathcal{T}(T)\rangle.
\label{eq:tadpole-integral}
\end{equation}
The expression (\ref{eq:TorusCorr}) can only be evaluated numerically,
since we do not know $\rho(u)$ explicitly (only its derivative) and we
cannot analytically solve the constraints (\ref{eq:constraints}).
Furthermore, there are several types of divergences in the integral
over $T$ in (\ref{eq:tadpole-integral}).  The integrand
(\ref{eq:TofTDef}) diverges as $T \rightarrow \infty$ due to the open
string tachyon.  While this divergence is difficult to treat in the
CFT approach, its physical origin is clear and is quite transparent in
the oscillator approach, where this divergence can be treated by a
suitable analytic continuation.  We discuss the open string tachyon
divergence further in Sections 4 and 6.  In addition to the the
divergence as $T \rightarrow \infty$, there are further divergences as
$T \rightarrow 0$ arising from the closed string.  These divergences
are much more subtle, as closed strings are not explicitly included
among the degrees of freedom in OSFT, but arise as composite states of
highly excited open strings.  We thus seek to evaluate the tadpole
diagram in an expansion around $T = 0$, where much of the interesting
physics in the diagram is hidden.

%%%%%%%%%%%%%%%%%%%%%%%%%%%%%%%%%%%%%%%%%%%%%%%%%%%%%%%%%%%%%%%%%%
\subsection{The $T\to 0$ limit}\label{s:Tsmall}
%%%%%%%%%%%%%%%%%%%%%%%%%%%%%%%%%%%%%%%%%%%%%%%%%%%%%%%%%%%%%%%%%%

We now focus on the region of moduli space near $T=0$.  
Unfortunately the 
map $\sqrt{\phi(u)}$  cannot easily be expanded around this limit.  
To get around this we use a trick.  It 
turns out that the conformal map greatly simplifies if, instead
of fixing the integral along $\gamma_1$ to be $2 \pi i$, 
we set it equal to some parameter $H$ and take $H \to i \infty$ holding $T$
fixed.  This is equivalent to gluing  a semi-infinite cylinder
to the bottom of the tadpole.  Later we will see how to
replace this long cylinder with a boundary state to reduce back to the
finite length cylinder case, but for the moment we just consider the conformal
map in this limit.

By solving the constraints (\ref{eq:constraints}) numerically, one can verify
 that as $H \to \infty$ with $T$ fixed, 
$\beta$ limits to 
a constant $\beta_0$, while $\tau \to i \infty$.  
Recalling that $q = e^{i \pi \tau}$,
we see that since $\tau$ is pure imaginary, $q \to 0$ as $H\to\infty$.  Thus we
may set $q = 0$ in our map to get
\begin{equation}
  \frac{d \rho}{du}\Big|_{q=0} = \sqrt{\phi(u)}\,\Big|_{q=0} =
  \sqrt{\cot^2(u)-\cot^2(\beta_0)}.
\end{equation} 
We can now solve for T in terms of $\beta_0$ by performing the integral
along $\gamma_2$:
\begin{equation}
  T = \oint_{\gamma_2} du \sqrt{\cot^2(u)-\cot^2(\beta_0)}  = 
    -\frac{i\pi}{\sin(\beta_0)}.
\end{equation}
Using this relation, we can eliminate $\beta_0$ from the definition
of our conformal map:
\begin{equation}
  \lim_{H\to\infty}\left(\frac{d \rho}{du}\right)  
  = \sqrt{1+\left(\frac{T}{\pi}\right)^2+\cot^2(u)}.
\end{equation}
This function may even be integrated analytically although the resulting
expression is cumbersome.  For notational simplicity we now consider
$d\rho/du$ only in the limit of $H\to \infty$ and we will
assume that the tadpole diagram has an infinitely long tube extending
from the bottom. 

Before we consider the effect of replacing the long tube at the
bottom of the diagram with a boundary state, we consider the effect
of the map $\rho(u)$ on the external state $A$ as $T\to 0$.
Note that if we take the limit that $T\to 0$, $d\rho/du$
 simplifies even further.
\begin{equation}
 \lim_{T\to 0}   \left(\frac{d \rho}{du}\right) =
  \sqrt{1+\cot^2(u)}  = \frac{1}{\sin(u)},
\end{equation}
where one must be careful about the interpretation of the branch cuts.
Integrating this function yields
\begin{equation}
  \int \, du \frac{1}{\sin(u)}  = 
   \log\left(-\tan\left(\frac{u}{2}\right)\right).
\end{equation}
While this may not seem like a familiar map, it is actually a representation
of the identity state.  Putting
\begin{equation}\label{eq:h}
 h(z) = \frac{1+i z}{1-iz},
\end{equation}
we consider the circle of conformal maps pictured in 
figure~\ref{f:ConformalMapCircle}.
\FIGURE{
\epsfig{file=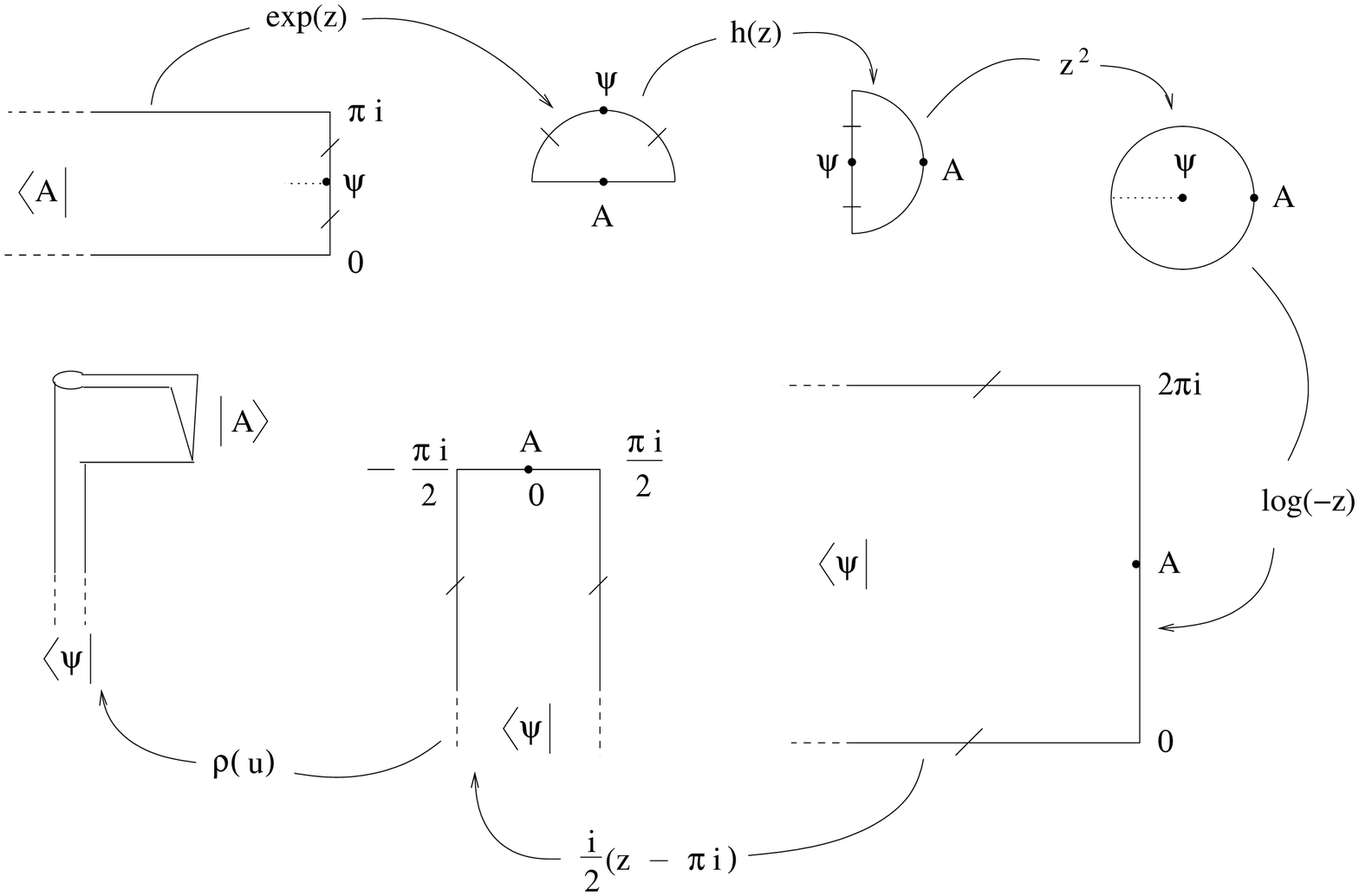,width=15cm}
\caption{\footnotesize
A circle of conformal maps showing the equivalence of the two
prescriptions for the identity state.  In the limit that $T\to 0$
traversing the diagram clockwise from the surface in the upper
left to the tadpole in the bottom left is equivalent to the trivial
map $f(z)  =z$.
}
\label{f:ConformalMapCircle}
}
One can verify that traversing the diagram counterclockwise (starting from
the vertical strip at the bottom and proceeding to the representation
of the identity in the upper left), gives the same map as $\rho(u)$ when
$T\to0$.
This implies that the tadpole diagram with a long tube attached to the
bottom is conformally equivalent to the identity state with an operator 
inserted in the corner in the limit $T\to 0$.  
Such states are known  as Shapiro-Thorn states~\cite{Shapiro:gq,Shapiro:ac}.

We now must deal with the fact that the original diagram had a tube of finite 
length extending from the bottom.  We thus consider the effect of 
replacing the infinite tube at the
bottom of the tadpole with 
the closed string boundary state with Neumann boundary 
conditions.  
The boundary state in disk coordinates
for Neumann boundary conditions can be written
\begin{equation}
\langle \mathcal{B} |  = 
\langle 0| (\tilde{c}_{-1}+c_{1})(\tilde{c}_{0}+c_{0})(\tilde{c}_{1}+c_{-1})
  \exp\left( \sum_{m > 1} b_m \tilde{c}_m +\tilde{b}_m c_m \right)
  \exp\left( \sum_{n\geq 1} \frac{1}{n} \alpha_n \cdot \tilde{\alpha}_n \right)
\end{equation}
where the oscillators are the usual  closed string oscillators and
$\langle 0 |$ is the closed string $SL(2,\bf{R})$ vacuum.  When we map
to cylinder coordinates, the disk becomes a semi-infinite tube with the
boundary state being propagated in from infinity.  By rescaling
the size of the cylinder we can make it the same circumference as
the long tube at the bottom of the tadpole.

Consider taking the long tube that extends from the bottom of the diagram
and cutting it off a distance $\pi/2$ below the vertex.  We can then attach
the boundary state, in cylinder coordinates, onto the bottom of the diagram.
Before adding the boundary state, the topology of the tadpole diagram is a
annulus with one vertex operator  (representing the external state)
inserted on the outer boundary.   Attaching the boundary state 
plugs the hole in the annulus, changing the topology to a disk.
The cost of doing this is that there is now an additional vertex
operator, representing the boundary state, inserted in the
interior of the disk.  For general values of the modulus $T$, the
complexity of the boundary state makes this replacement impractical,
but, for small $T$, only a few terms in the boundary state become relevant.

After this modification, we can
map the tadpole diagram to the unit disk with the image of the boundary state
at $z=0$ and the image of the state $|A\rangle$ at $z=1$,
where $z$ is the coordinate on the disk.
We call the coordinates on the boundary state disk $w$.
Since the boundary state is not conformally invariant, it will be mapped to 
$z(w)\circ \mathcal{B}$.  
As before, we denote by $v$ the coordinates on the half-disk where
the vertex operator for the external state $A$ is defined and
we let $z(v)$ be the map that takes the external state $A$ into the 
disk coordinates.  We can then write down a formal expression for the
tadpole diagram
\begin{equation}
  \langle (z(v) \circ A )(z(\rho) \circ \frac{1}{2 \pi i}\int d\rho\, b(\rho))
( z(w)\circ \mathcal{B}) \rangle_{\text{disk}}.
\end{equation}
Mapping everything to the upper-half plane using the
map $h(z)$ defined in (\ref{eq:h}), we can write this as
an inner product between $A$ and the  ket $|\mathcal{T}(T)\rangle$
defined in  (\ref{eq:TofTDef}).  
If $| A\rangle$
is the external state in half-disk coordinates, the diagram
at fixed modular parameter is given by
\be \label{eq:Tdef}
  \langle A| \,U_{h(z)\circ z(v)}^{\dagger} 
(h(z)\circ z(\rho) \circ \frac{1}{2 \pi i}\int d\rho\, b(\rho))
\left(h(z)\circ z(w)\circ \mathcal{B} 
\right)|0\rangle
\ee
which implies that 
\begin{equation}\label{eq:TofTdef2}
  |\mathcal{T}(T)\rangle = \,U_{h(z)\circ z(v)}^{\dagger} 
(h(z)\circ z(\rho) \circ \frac{1}{2 \pi i}\int d\rho\, b(\rho))
\left(h(z)\circ z(w)\circ \mathcal{B} 
\right)|0\rangle,
\end{equation}
where the operator $U_{f}$ is
defined by its action on local operators
\begin{equation}
   U_{f} \mathcal{O} U_{f}^{-1} = f \circ \mathcal{O}.
\end{equation}
and $U_{f}^{\dagger}$ is the BPZ dual of $U_{f}$.  
Such operators have been considered in 
\cite{LeClair:1988sj,Rastelli:2000iu,Schnabl:2002gg,Schnabl:2002ff}.

Since, as we've already discussed, the map $z(v)$ just limits to
a representation of the identity state as $T \to 0$, the 
operator $U_{h(z)\circ z(v)}^{\dagger}$ has a well behaved limit as  $T \to 0$.
Thus, when we analyze the small $T$ limit, we only need to focus
on the behavior of $z(w) \circ \mathcal{B}$ and 
$z(\rho) \circ \int d\rho\, b(\rho)$.
As it turns out, it is computationally easier to 
calculate $w(z)$ rather than $z(w)$.
Pictorially,  $w(z)$ is given in figure~\ref{f:wMapPicture}.

\FIGURE{
\epsfig{file=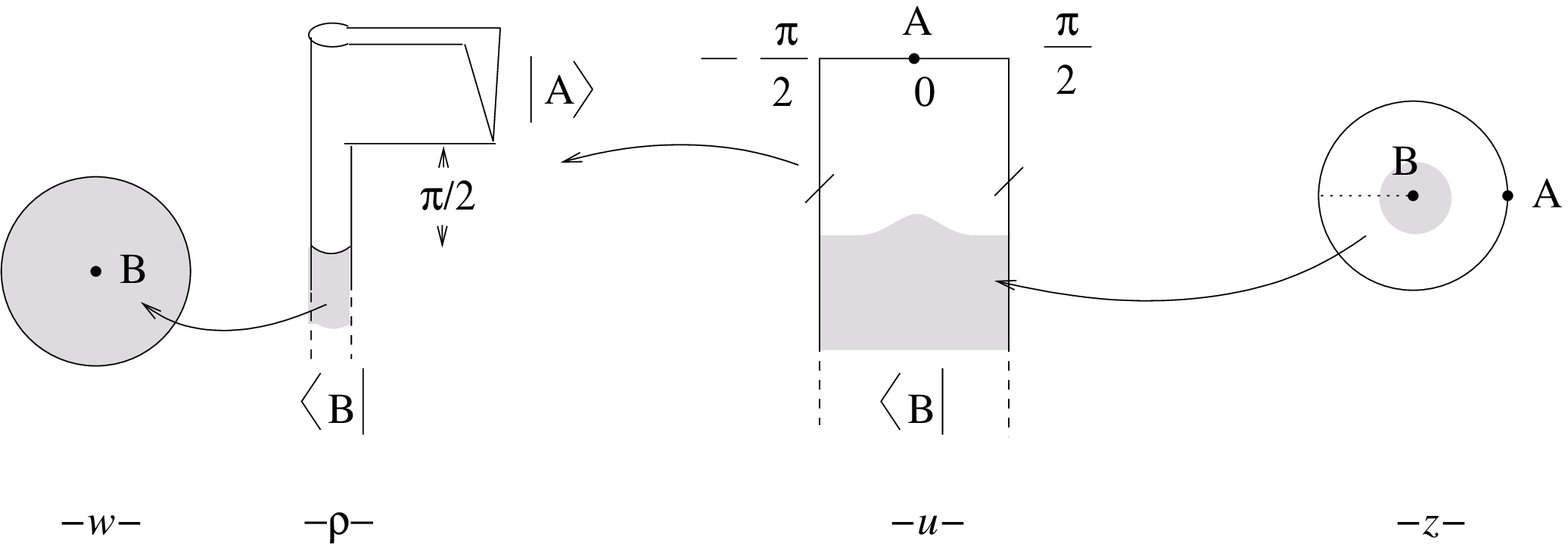,width=15cm}
\caption{\footnotesize
The map $w(z)$ is shown as a series of maps.  The shaded regions show
the images of the boundary state.
}
\label{f:wMapPicture}
}

The map from the disk to the vertical strip is given by
\begin{equation}
 u(z) = \frac{i}{2} \log(z).
\end{equation}
Since we already know the derivative of the map from the vertical strip to
the Witten diagram, $d \rho/du $, we can easily compute the derivative
of the map from the disk to the Witten diagram. To eliminate excess factors
of $\pi$, we put $W= T / \pi$.  We then have
\begin{equation}
 \frac{d \rho}{dz} = \frac{d \rho}{d u} (u(z)) \frac{d u}{dz}  =
  \frac{i}{2}\frac{\sqrt{W^2(z-1)^2-4z}}{z(z-1)}.
\end{equation}
Integrating this function gives
\begin{multline}
  \rho(z) = -\frac{i}{2} 
\Bigl(
 2 \tan^{-1} (   \frac{1+z}{  \sqrt{W^2(z-1)^2-4z}  }   )-W\log(z)\\
  + W \log(-2z +(1+z) W \sqrt{W^2(z-1)^2-4z}+W^2 (1+z^2)) 
\Bigl)+g(W),
\end{multline}
where $g(W)$ is the constant of integration.  By fixing the image of the
midpoint, we can determine $g(W)$ to be
\begin{equation}
  g(W)  =\frac{i}{2} \Bigl( \pi+2 \cot^{-1} (W) 
   + i \log \Bigl( \frac{W+i}{W-i}\Bigl) + W \log(2+2 W^2) \Bigl).
\end{equation}
Finally, we map the bottom of the Witten diagram to the disk with the
boundary state in the center.  This map is given by
\begin{equation}
  w(\rho) = \exp\left( \frac{\pi-2i\rho}{W}\right).
\end{equation}
We can thus calculate $w(\rho(z)) = w(z)$.  Since the full expression
is quite complicated we only display the series
expansion:
\begin{equation}
  w(z) = e^{ \pi/W+i\log \left( \frac{W+i}{W-i}\right)/W} 
\Biggl[
-\frac{1+W^2}{W^2} z+2 \frac{1+W^2}{W^4} z^2 +
\frac{2 W^4+W^2-1}{W^6}z^3+\mathcal{O}(z^4)
\Biggl].
\end{equation}
We can now find $z(w)$ as a power series in w.  Putting
\begin{equation}
  k(W) = e^{- \pi/W-i\log \left( \frac{W+i}{W-i}\right)/W},
\end{equation}
we have
\begin{equation}\label{eq:zofw}
z(w) = -k \frac{W^2}{1+W^2} w+2 k^2 \frac{W^2}{(1+W^2)^2} w^2
-k^3 \frac{W^2(7+2 W^2)}{(1+W^2)^3} w^3+\mathcal{O}(w^4).
\end{equation}
As one might expect, the unit disk is mapped to smaller and 
smaller regions as $W\to 0$.  This suggests that the boundary state
might be mapped to some local operator.
In fact, all the positive weight parts of $|B\rangle$ will be suppressed
in the small $W$ limit.
Thus the only relevant terms from the boundary state are the 
weight zero and weight $(-1,-1)$ fields.  These are given by
\begin{eqnarray}
  |\mathcal{B}\rangle  
    &=& c_{1}(c_0+\tilde{c}_0)\tilde{c}_1|0\rangle\nonumber \\
&& + \alpha_{-1} \cdot \tilde{\alpha}_{-1} {c}_{1} (c_0+\tilde{c}_0)
    \tilde{c}_1 
|0\rangle
- (c_1 c_{-1}+\tilde{c}_{-1}\tilde{c}_1) (c_0+\tilde{c}_0)
|0\rangle\nonumber\\
& & + \text{higher weight states}.\label{eq:BoundaryState}
\end{eqnarray}
Since the term $c_{1}(c_0+\tilde{c}_0)\tilde{c}_1|0\rangle$  
is a  weight $(-1,-1)$ 
primary it picks up the 
coefficient
\begin{equation}
  \left(\frac{1+W^2}{kW^2}\right)^2
\end{equation}
under the $z(w)$ map.  Similarly, the term $\alpha_{-1} \cdot
\tilde{\alpha}_{-1} \tilde{c}_{1} (c_0+\tilde{c}_0) c_1 |0\rangle$ is
a weight $(0,0)$ primary and thus is unchanged by the $z(w)$ map.

Unfortunately, the term, $- (c_1 c_{-1}+\tilde{c}_{-1}\tilde{c}_1)
(c_0+\tilde{c}_0) |0\rangle$, is not a primary and is mapped to
\begin{align}
  - (c_1 c_{-1}+\tilde{c}_{-1}\tilde{c}_1) (c_0+\tilde{c}_0)
             |0\rangle
  -\frac{8}{W^4}\,c_{1}(c_0+\tilde{c}_0)\tilde{c}_1|0\rangle 
\nonumber\\
  +\frac{4}{W^2}\, c_1(c_{-1}+\tilde{c}_{-1}) \tilde{c}_1 |0\rangle
+\frac{2}{W^2} (c_1 -\tilde{c}_1) c_0 \tilde{c}_0.
\end{align}
As can be seen in the second term, this state mixes with the weight
$(-1,-1)$ state $c_{1}(c_0+\tilde{c}_0)\tilde{c}_1|0\rangle$ under
conformal maps. This mixing will play a role later.

To fully account for the behavior of the tadpole near
$W \to 0$, we must take care of the
insertion of the $b_0$.  Since the conformal map from the disk to
the tadpole has already been found, the transformation of $b_0$ to
the disk coordinates is straightforward.
Calling the resulting operator $B$, we get
\begin{equation}
B = \frac{1}{\pi} \int dz \,\frac{z(z-1)}{\sqrt{W^2(z-1)^2-4z}}\, b(z) - 
\text{c.c.},
\end{equation}
where the contour runs along the real axis from $-1$ to $z(-1)$. 
Note that as $W\to 0$, $z(-1) \to 0$ also.  Thus we can get
collisions between $b(z)$ and the various $c$'s in the boundary state.
We can expand $B$ in terms of the modes found in equation 
(\ref{eq:BoundaryState}), keeping only the most divergent terms for
each mode $b_n$:
\begin{equation}
  B \sim  \ldots  -\frac{4 }{3 \pi} \,b_{-1} + \frac{1}{W^2}\, b_{0} 
  -\frac{1}{\pi W^3 k}\, b_{1} +\ldots - \text{c.c.}
\end{equation}
We can now let $B$ act on the conformally transformed boundary state.
We keep only the most divergent terms from each state and drop all finite 
terms:
\begin{eqnarray}
  B\,( z(w) \circ |\mathcal{B}\rangle) &=& 
      -\frac{2}{k^2 W^6} c_1 \tilde{c}_1|0\rangle  \nonumber\\
      & & -\frac{1}{k^2 W^4}:c_1 (c_0 +\tilde{c}_0) \tilde{c}_1 B:
        |0\rangle  
      -\frac{4 }{3 \pi  k^2 W^4}
             (c_0 +\tilde{c}_0)(\tilde{c}_1 -c_1)|0\rangle
  \nonumber\\
      & & + \frac{2}{W^2} \alpha_{-1} \cdot \tilde{\alpha}_{-1}
          \tilde{c}_1 c_1 |0\rangle  \nonumber\\
       & & - \frac{2}{W^2} (c_1 c_{-1}  +\tilde{c}_{-1} \tilde{c}_1)|0\rangle  
       -\frac{16}{3\pi W^2} (c_{-1}+\tilde{c}_{-1})(\tilde{c}_1-c_1)
       |0\rangle
\nonumber\\
       & & -\frac{8}{3\pi W^2} c_{0} \tilde{c}_0 |0\rangle
+\ldots.
\label{eq:FullLimit}
\end{eqnarray}
To see directly how these terms cause the tadpole to diverge, consider
for example the term $-\frac{2}{k^2 W^6} c_1 \tilde{c}_1|0\rangle$,
which is the most divergent term in the expansion
(\ref{eq:FullLimit}).  Plugging this term into the expression for
$|\mathcal{T}(T)\rangle$ given in equation (\ref{eq:TofTdef2})
gives
\begin{align}\label{eq:Tfrommostdivergent}
  |\mathcal{T}(T)\rangle &\sim -\frac{2}{k^2 W^6} U_{h(z)\circ z(v)}^{\dagger}
 (h(z)\circ c(0) \,\tilde{c}(0))|0\rangle\nonumber\\
& \propto  \frac{ e^{2 \pi^2 /T}}{ T^6}
U_{h(z)\circ z(v)}^{\dagger} c(i) \,\tilde{c}(i) |0\rangle.
\end{align}
As discussed above, as $T\to 0 $ the map $h(z)\circ z(v)$ limits to the
map representing the identity state.  Thus if we are only interested in
$|\mathcal{T}(T)\rangle$ near $T = 0$ we can replace
\begin{equation}
  h(z)\circ z(v) = f(v) = \frac{2v}{1-v^2},
\end{equation}
where $f(z)$ is the usual expression for the identity state in the
upper-half plane geometry.
Thus we can write, for small $T$,
\begin{equation}\label{eq:leadingTdivergence}
  |\mathcal{T}(T)\rangle \sim \frac{ e^{2 \pi^2 /T}}{ T^6}\,\,
 U_{f(v)}^{\dagger} c(i) \,\tilde{c}(i) |0\rangle.
\end{equation}
The full tadpole state is then given by the integral of this state
over $T$.  Since the state $U_{f(v)}^{\dagger} c(i) \,\tilde{c}(i)
|0\rangle$ does not depend on $T$, the small $T$ region of the
integral is determined by the integral over the function $e^{2 \pi^2
/T} /T$, which diverges.  One can also consider what we would get if we
take, for example, the term $\frac{2}{W^2} \alpha_{-1} \cdot
\tilde{\alpha}_{-1} \tilde{c}_1 c_1 |0\rangle$ from the expansion
(\ref{eq:FullLimit}).  A similar calculation yields
\begin{equation}\label{E:masslessdivergence}
  |\mathcal{T}(T)\rangle_{\alpha_{-1} \cdot \tilde{\alpha}_{-1}
          \tilde{c}_1 c_1 |0\rangle} \sim
\frac{1}{T^2}\,\, U_{f(v)}^{\dagger} c(i)\partial X^{\mu}(i)\,\tilde{c}(i)
 \bar{\partial} X_{\mu}(i) |0\rangle,
\end{equation}
where the subscript indicates that this is only the behavior of the 
terms in $|\mathcal{T}(T)\rangle$ arising from term
$\alpha_{-1} \cdot \tilde{\alpha}_{-1}
          \tilde{c}_1 c_1 |0\rangle$
in the expansion (\ref{eq:FullLimit}).
Since one cannot integrate $1/T^2$ near zero, this also leads to
a divergence in $|\mathcal{T}\rangle$.

For comparisons with the results in section~\ref{s:Oscillators}, it is
useful to note that the expression for the leading divergence of the
$|\mathcal{T}(T)\rangle$, given in (\ref{eq:leadingTdivergence}),
can be rewritten as
\begin{equation}
\label{eq:OscillatorFormOfSmallT} |\mathcal{T}(T)\rangle \sim \frac{
  e^{2 \pi^2 /T}}{ T^6}\,\, \exp \left( -\frac{1}{2} a^{\dagger}_m
C_{mn} a^{\dagger}_n - c^{\dagger}_m C_{mn} b^{\dagger}_n \right) c_0
c_1 |0\rangle, 
\end{equation}
 where $a^{\dagger}_m$, $c^{\dagger}_m$ and $b^{\dagger}_m$ are the
usual open string oscillators and $C_{mn} = (-1)^m \delta_{mn}$.
Equation~(\ref{eq:OscillatorFormOfSmallT}) may be verified using the
methods of \cite{lpp}.

%%%%%%%%%%%%%%%%%%%%%%%%%%%%%%%%%%%%%%%%%%%%%%%%%%%%%%%%%%%%%%%%%%%%%%%%%%%
\subsection{Interpretation of the divergences}
%%%%%%%%%%%%%%%%%%%%%%%%%%%%%%%%%%%%%%%%%%%%%%%%%%%%%%%%%%%%%%%%%%%%%%%%%%%
\label{s:Interpretingdivs}
The divergences in (\ref{eq:FullLimit}) may be interpreted as arising from
the
propagation of tachyonic and massless closed string modes down a long
cylinder.  The tube at the bottom of the tadpole is a tube of
length $\frac{\pi}{2}$ and circumference $T$.  Rescaling this tube
by a factor of $\frac{2\pi}{T}$ gives a tube of length $\frac{\pi^2}{T}$
and constant circumference $2 \pi$. These are the standard lengths for
closed string theory.  If we think of the boundary state at the end of
the tube as a closed string state, we may propagate it along the
length of the tube using the operator
\begin{equation}
  \exp\left(-\frac{\pi^2}{T}(L_0+\tilde{L}_0)\right).
\end{equation}
For the term in the boundary state given by 
$c_{1}(c_0+\tilde{c}_0)\tilde{c}_1|0\rangle$, which
we may think of as coupling to the closed string tachyon, 
this gives a prefactor
of 
\begin{equation}
 e^{2 \pi^2/T} \sim \frac{1}{k^2}.
\end{equation}
For the weight zero terms from the boundary state there is no term
picked up from the propagation.  However we must account for the
measure on the moduli.  Taking the modulus to be the length of the
cylinder $s = \frac{\pi^2}{T}$, the usual measure on a cylinder is
just $ds = -\pi^2 \frac{dT}{T^2}$.

Of course the full tadpole diagram is not just a cylinder.
Complicated conformal factors can and do arise from the specific
manner in which the cylinder at the bottom of the diagram is attached
to the external open string state.  Furthermore since the boundary
state is not a conformal primary, conformal transformations can mix
the behavior of the various terms.  This is seen in
equation~(\ref{eq:FullLimit}).  The first three terms diverge as $\sim
\frac{1}{k^2}$ suggesting that these divergences come from the
propagation of the closed string tachyon over the long cylinder.  The
next four terms diverge as $\sim \frac{1}{T^2}$ suggesting that these
terms arise from the massless closed string sector fields.  The term
$\alpha_{-1} \cdot \tilde{\alpha}_{-1} \tilde{c}_1 c_1 |0\rangle$ is
of the right form to correspond to the graviton/dilaton.  Since this
field does not mix with any of the other fields under the conformal
map $z(w)$, there is no ambiguity that this divergence arises purely
from the massless sector.  The other fields diverging as $1/W^2$,
correspond to auxiliary fields.  Because these states are not
conformal primaries, they mix with the states coupling to the tachyon
field. Thus, these divergences may be due in part to the closed string tachyon

The tachyon divergence may be partially treated by an analytic
continuation.  If we take the weight of the state $c_1
(c_0+\tilde{c}_0) \tilde{c}_1 |0\rangle$ to be $(h_1,h_1)$, we can try
to perform the integrals over the modular parameter with the
assumption that $h_1>0$.  Since the term, $-(c_1 c_{-1}
+\tilde{c}_{-1}\tilde{c}_{-1})(c_0+\tilde{c}_0)|0\rangle$, mixes with
$c_1(c_0+\tilde{c}_0) \tilde{c}_1 |0\rangle$ under conformal maps we
must also take this state to have some arbitrary weight $(h_2,h_2)$.
We can then substitute $h_{1} = -1$ and $h_{2} = 0$ at the end of the
calculation.  This prescription works for all the terms containing
powers of $1/k$.  Unfortunately there are subleading terms which
contain no factors of $1/k$, but still diverge as badly as $1/W^5$.

These divergences, which mix with the massless divergences, seem to be
an unfortunate consequence of the geometry of the Witten diagram.  In
other versions of string field theory, such as the one discussed in
section \ref{ss:OCSFT}, these divergences do not arise.  It is
possible that a more sophisticated method for treating these
divergences would eliminate them and that we are merely limited by our
inability to correctly identify the tachyonic degrees of freedom,
since they are encoded in a highly nontrivial fashion in terms of the
fundamental open string degrees of freedom.
It is also possible,  since we are dealing with off-shell
physics, that spurious divergences are arising.  Such unexpected
off-shell physics was found in \cite{Freedman:fr}, where extra poles
were found in an internal closed string propagator.
Fortunately, however, we can eliminate these $1/W^n$ divergences by
considering lower dimensional D$p$-brane backgrounds, which is the
subject of the next section.

%%%%%%%%%%%%%%%%%%%%%%%%%%%%%%%%%%%%%%%%%%%%%%%%%%%%%%%%%%%%%
          \subsection{Lower dimensional D$p$-branes}
%%%%%%%%%%%%%%%%%%%%%%%%%%%%%%%%%%%%%%%%%%%%%%%%%%%%%%%%%%%%%
\label{s:plessthan25}

Having examined the case of the D$25$-brane, we turn to the general
case of the D$p$-brane for $p<25$.  The only change in the analysis we
need to make is to replace the boundary state, $|\mathcal{B}\rangle$,
with the correct boundary state for a D$p$-brane. This state is given
by
\begin{equation} \label{eq:GeneralB}
  |\mathcal{B}_p\rangle = \exp \left( \sum_{n = 1}^{\infty}\, \mp
\frac{1}{n} \alpha_{-n}\cdot \tilde{\alpha}_{-n} - b_{-m}
\tilde{c}_{-m} - \tilde{b}_{-m} c_{-m} \right) \int d^{25-p} q_{\bot}
e^{-i q_{\bot}\cdot y_{\bot}} |q_{\bot},q_{\parallel} =0\rangle,
\end{equation}
where $y_{\bot}$ is the location of the brane, $q_{\bot}$ is the
momentum transverse to the brane, and $q_{\parallel}$ is the momentum
parallel to the brane. We will set $y_{\bot}^{\mu} = 0$ for
convenience. The minus sign in front of the $\alpha$'s is chosen for
Neumann boundary conditions and the plus sign for Dirichlet.

We can now study the divergence structure of the tadpole diagram by
studying  $B(z(w) \circ |\mathcal{B}_p\rangle)$.  Since the sign changes
in (\ref{eq:GeneralB}) do not affect the divergence structure we need
only consider the effect of the momentum dependence.
Because it is simpler, we begin with the massless sector.  Consider the
contribution from the graviton/dilaton.
\begin{equation}
  B(z(w) \circ |\mathcal{B}_p\rangle) = \ldots - 
  \int d^{25-p}q_{\bot}\,
\left(\frac{k W^2}{1+W^2}\right)^{q_{\bot}^2} \frac{2}{W^2}
   \alpha_{-1} \cdot\tilde{\alpha}_{-1} c_{1}\tilde{c}_{1} |q_{\bot}\rangle
+ \ldots
\end{equation}
Note that the factor of $\left(\frac{k W^2}{1+W^2}\right)^{q_{\bot}^2}$
kills the $W\to 0$ divergence if $q_{\bot}^2>0$.  To see whether the point
$q_{\bot} =0 $ contributes a divergence in the integral, we map this
state to a local operator at $i$ in the upper-half plane using the map
$h^{-1}(z)$.  This gives the operator
\begin{equation}
  \int d^{25-p}q_{\bot}\,
(2)^{q_{\bot}^2}
\left(\frac{k W^2}{1+W^2}\right)^{q_{\bot}^2} \frac{2}{W^2}
  \, c\partial X(i)\, \tilde{c} \bar{\partial} X(i)
e^{iq_{\bot}\cdot X(i)} .
\end{equation}
In the upper-half plane geometry, the external state is mapped to
some local operator at the origin.  By Lorentz invariance, this
operator cannot have any momentum dependence.  

We can now evaluate the tadpole in this geometry using the Green's
function relevant to the Dirichlet/Neumann boundary conditions. The term
$ e^{iq_{\bot}\cdot X(i)}$ produces a factor of $(2)^{-k^{2}_{\bot}}$ when
it contracts with itself.  There can also be additional momentum dependent
factors when $ e^{iq_{\bot}\cdot X(i)}$ contracts with other $X$'s in
the boundary state and in the external state.  These will produce additional
factors of $q_{\bot}^2$.  As we will see below, these factors will only
make things more convergent, so we can ignore them.  If we take just the
momentum-dependent and $W$-dependent terms we get
\begin{equation}\label{eq:MomentumDependentMassless}
  \int d^{25-p}q_{\bot}\, \left(\frac{k W^2}{1+W^2}\right)^{q_{\bot}^2}
  \frac{2}{W^2}
 = \text{Const} \times \int_{0}^\infty dr \, r^{25-p-1}
\left(\frac{k W^2}{1+W^2}\right)^{r^2}
  \frac{2}{W^2}.
\end{equation}
Dropping the constant from the angular integral, this gives
\begin{equation}
  \frac{2}{W^2}\left[-\log\left( \frac{k W^2}{1+W^2}
  \right)\right]^{\tfrac{1}{2}(p-25)}.
\end{equation}
Expanding this function around $W =0$ gives
\begin{equation}\label{eq:SmallWIntegrand}
   \left(\frac{W}{\pi}\right)^{\tfrac{25-p}{2}} \frac{1}{W^2}
+\mathcal{O}(W^{\tfrac{25-p-2}{2}}).
\end{equation}
We are interested in when this can be integrated near $W = 0$.
Additional factors of $W$ could arise from the external state  map but these
will only aid the convergence.  From the expansion 
(\ref{eq:SmallWIntegrand}) we see that
we can integrate (\ref{eq:MomentumDependentMassless}) with respect to
$W$ near $W=0$ if
\begin{equation}
\label{E:pcontraint}
   p \le 22.
\end{equation}
Thus, for sufficiently many transverse dimensions, there are no
divergences from the massless sector.  As we discuss in Section 6,
this constraint has a natural interpretation in terms of the long
range behavior of the massless fields around the brane.

Now let's look at the tachyon sector.  From the calculation in the
massless sector, we can see that the only effect of momentum
dependence is to give an extra power of $W^{ (25-p)/2}$.
Unfortunately, this is not enough to suppress the factor of $1/k^2$ so
one must still resort to some form of analytic continuation as we did
in the D$25$-brane theory.  Unlike the D$25$-brane case, however, this
analytic continuation can now be used to make the diagram completely
finite provided $p$ is small enough.  Recall that the terms which
still diverged after analytic continuation were at worst of the form
$1/W^5$.  Thus to make the tadpole integrable around $W = 0$, we must
choose $p \leq 16$. While we have a natural interpretation for the
constraint (\ref{E:pcontraint}), this additional restriction of the
dimensionality of the branes seems to be an artifact of the interplay
between our somewhat ad hoc choice of analytic continuation and the
details of the Witten OSFT.  We do not believe that there is any
universal significance to this constraint; indeed, it is not evident
in the alternative string field theory discussed in
section~\ref{ss:OCSFT}.

As in the D$25$-brane case, it is instructive to compare these results
with the propagation of the boundary state along a closed string tube
of length $\frac{\pi}{W}$ and circumference $2 \pi$.  
As before, the propagation
of the boundary state is  represented by
\begin{equation}\label{eq:ClosedStringVersion}
  \exp\left(-\frac{\pi}{W}(L_0+\tilde{L}_0)\right) |\mathcal{B}_p \rangle.
\end{equation}
Consider decomposing the boundary state into a sum over states of
definite weight, $(h,h)$, and momentum, $q_{\bot}$,
\begin{equation}
  |\mathcal{B}_p \rangle = \sum_{h=-1}^{\infty} \int d^{25-p}q_{\bot} 
  |\mathcal{B}_p(q_{\bot},h)\rangle.
\end{equation}
We can then write (\ref{eq:ClosedStringVersion}) as
\begin{equation}
  \sum_{h=-1}^{\infty} \int d^{25-p}q_{\bot}
  \exp\left(-\frac{\pi}{W}(2h + k^2_{\bot})\right)
  |\mathcal{B}_p(q_{\bot},h)\rangle.
\end{equation}
Now consider the different terms in the sum over $h$.  For $h>0$, the
term in the exponent $2h + k^2_{\bot}$ is always greater than zero.
Thus the limit as $W \to 0$ is well defined.  For $h = 0 $ the limit
$W \to 0$ is well defined for the region of integration where
$q_{\bot}^2 >0$.  As we saw above, the region of the integral around
$q_{\bot}^2 = 0$ is also defined for sufficiently small $p$.

For the tachyon, however, we have $h= -1$.  Now, whenever $q_{\bot}^2
< 2$, the limit as $W \to 0$ is divergent.  Thus the added momentum
dependence does nothing to help the tachyon divergence and we must
 resort to analytic continuation.

%%%%%%%%%%%%%%%%%%%%%%%%%%%%%%%%%%%%%%%%%%%%%%%%%%%%%%%%%%%%%%%%%%%%%%%%%%%
%%%%%%%%%%%%%%%%%%%%%%%%%%%%%%%%%%%%%%%%%%%%%%%%%%%%%%%%%%%%%%%%%%%%%%%%%%%

\section{Evaluation of the tadpole using oscillator methods}

%%%%%%%%%%%%%%%%%%%%%%%%%%%%%%%%%%%%%%%%%%%%%%%%%%%%%%%%%%%%%%%%%%%%%%%%%%%
%%%%%%%%%%%%%%%%%%%%%%%%%%%%%%%%%%%%%%%%%%%%%%%%%%%%%%%%%%%%%%%%%%%%%%%%%%%
\label{s:Oscillators}

Having examined the one-point function using conformal field theory
methods, we now evaluate it using the oscillator approach of
\cite{Taylor:2002bq}. In this section we primarily specialize to the
D$25$-brane.  We will comment on the lower dimensional branes at the
end of the section.  In section \ref{s:squeezed}
we review the oscillator form of the two- and three-string vertices
and use squeezed state methods to compute the one-loop tadpole
in terms of infinite matrices of Neumann coefficients.  In
sections~\ref{s:SmallT} and \ref{s:MandR}, we analyze the results
using numerical and analytical methods and compare with our results
from section~\ref{S:ConformalAnalysis}.

%%%%%%%%%%%%%%%%%%%%%%%%%%%%%%%%%%%%%%%%%%%%%%%%%%%%%%%%%%%%%%%%%%%%%%%%%%%
%                  \subsection{The Feynman rules for OSFT}
%%%%%%%%%%%%%%%%%%%%%%%%%%%%%%%%%%%%%%%%%%%%%%%%%%%%%%%%%%%%%%%%%%%%%%%%%%

%%%%%%%%%%%%%%%%%%%%%%%%%%%%%%%%%%%%%%%%%%%%%%%%%%%%%%%%%%%%%%%%%%%%%%%%%%%%%
\subsection{Oscillator description of the one-loop tadpole}
%%%%%%%%%%%%%%%%%%%%%%%%%%%%%%%%%%%%%%%%%%%%%%%%%%%%%%%%%%%%%%%%%%%%%%%%%%%%%

\label{s:squeezed}
We begin by writing an oscillator description for the tadpole diagram,
following~\cite{Thorn:1988hm,Taylor:2002bq}.
The oscillator expressions for the the two- and three-string
vertices $|V_2\rangle$ and $|V_3\rangle$ are squeezed states
in the two-fold and three-fold tensor product of the string Fock space
with itself
\cite{lpp,Gross:1986ia,Gross:1986fk,cst,Samuel,Ohta,rsz}.  
Explicit formulae for these vertices are given below.
In Feynman-Siegel gauge, $b_0 | \Psi \rangle = 0$, and the propagator
is given by $b_0/L_0$, which we can represent using a Schwinger
parameter
\be
\label{eq:prop}
\frac{b_0}{L_0} = b_0\int_0^{\infty} dT \, e^{-T L_0}.
\ee
We can use the vertices and the propagator to write the tadpole as
\begin{equation}
\label{eq:firststep}
\ket{{\mathcal T}} = -g\int_0^{\infty} dT \,
\left._{1,2}\bra{\tilde{V}_2}b_0^{2} \; e^{-\onehalf T (L_0^{(1)} +
L_0^{(2)}) } \ket{V_3}_{1,2,3} \right. .
\end{equation}
%\begin{equation} % changed
%e^{-T L_0^{(1)}} |\tilde{V}_2 \rangle 
%= e^{-\frac{T}{2}( L_0^{(1)}+L_0^{(2)})} 
%|\tilde{V}_2 \rangle.
%\end{equation} % changed
The three-string vertex is given by
\begin{eqnarray}
\label{eq:v3}
 \ket{V_3} &=& \int d^{26}p_1 d^{26}p_2 \, \exp 
\left( - \onehalf a^{i\dag}_{n} V^{ij}_{nm}         
a^{j\dag}_m - a^{i\dag}_n V^{ij}_{n0} p_j - \onehalf p_i V^{ij}_{00} p_j 
                   \right) \\ \nonumber
 && \phantom{spa} \times \, 
\exp \left(- c^{i\dag}_n X^{ij}_{nm} b^{j\dag}_m \right)\;
                c_0^1 \ket{\hat{0};p_1}_1 \;c_0^2\ket{\hat{0};p_2}_2 
\;c_0^3 
\ket{\hat{0};p_3 = - p_1 - p_2}_3 ,
\end{eqnarray}
where sums on all repeated indices are understood.  The vacuum
$|\hat{0}\rangle = c_{1}\ket{0}$, where as before $\ket{0}$ is the
$SL(2,{\bf R})$-invariant vacuum. The Neumann coefficients
$V^{ij}_{nm}$ and $X^{ij}_{nm}$ are given by standard formulae
\cite{Gross:1986ia,Gross:1986fk,rsz} and are tabulated in \cite{wt2}
among other places.  The two-string vertex $\ket{\tilde{V}_2}$ is
related to the usual two-string vertex $| V_2 \rangle$ through
\begin{equation}
|\tilde{V}_2\rangle = (-1)^{N_g^{(1)}} |V_2\rangle
\label{eq:extra-sign}
\end{equation}
and
has the following oscillator representation \be
\label{eq:v2}
  \ket{\tilde{V}_2} = \int d^{26}p \, 
(c_0^1+c_0^2)\exp \left( - a^{1\dag}_n C_{nm} a^{2\dag}_{m} - c^{1\dag}_{n}
        C_{nm} b^{2\dag}_{m} - c^{2\dag}_{n} C_{nm} b^{1\dag}_{m} \right) 
\ket{\hat{0};p}_1
       \ket{\hat{0};-p}_2,
\ee
where
\begin{equation} % changed
C_{nm} = \delta_{nm} (-1)^n.
\end{equation} % changed
The extra sign in (\ref{eq:extra-sign}) arises from the fact that the
$b_0$ in the propagator
anticommutes with Grassmann-odd states in $| V_2 \rangle$
\cite{Thorn:1988hm}.
Note that in (\ref{eq:firststep}) we have used the property
\begin{equation} % changed
e^{-T L_0^{(1)}} |\tilde{V}_2 \rangle = e^{-\frac{T}{2}(
  L_0^{(1)}+L_0^{(2)})} |\tilde{V}_2 \rangle
\end{equation} % changed
to make the expression more symmetric.

We may represent the action of the propagator
on the vertex by multiplying each term in the vertex by the
appropriate function of its conformal weight.  That is, since $[L_0^{(k)},
\, a^{i\dag}_n V^{ij}_{nm} a^{j\dag}_m]  = (n \delta_{ik} +
m\delta_{jk}) a^{i\dag}_n V^{ij}_{nm} a^{j\dag}_m $, we have
\begin{eqnarray}
\lefteqn{ e^{-TL^{(k)}_0} \exp \left( -\onehalf a^{i\dag}_n V^{ij}_{mn} 
a^{j\dag}_m \right)
c_0^1 \ket{\hat{0};0}_1\;c_0^2 \ket{\hat{0};0}_2\;c_0^3 \ket{\hat{0};0}_3 }
 \\
  &  &\hspace{1in} =   e^T \exp \left( -\onehalf a^{i\dag}_n
    V^{ij}_{mn} a^{j\dag}_m e^{-T(n\delta_{ik}+
          m\delta_{jk})} \right) 
c_0^1 \ket{\hat{0};0}_1\;c_0^2 \ket{\hat{0};0}_2\;c_0^3 \ket{\hat{0};0}_3
, \nonumber
\end{eqnarray}
with an analogous result in the ghost sector.  As in \cite{Taylor:2002bq}, we
denote a vertex which has absorbed a propagator in this
fashion by a hat, so that 
\begin{equation} % changed
 \hat{V}^{i_kj_l}_{nm}(T_k, T_l) \equiv
 e^{-nT_k/2} \,V^{i_kj_l}_{nm}\, e^{-mT_l/2} .
\end{equation} % changed

Using the explicit representations (\ref{eq:prop}, \ref{eq:v3},
\ref{eq:v2}) of the vertices and propagators, the evaluation
of $\ket{\mathcal{T}}$ at fixed modular parameter $T$ reduces
to computing the inner product of two squeezed states.
The formula for squeezed state expectation values is given by \cite{kp}
\be
\label{eq:sqz1}
\bra{0} e^{-\onehalf a S a} e^{-\mu \adag-\onehalf \adag V \adag} \0 =
  \det (1-SV)^{-1/2} e^{-\onehalf \mu (1-SV)^{-1} S \mu} ,
\ee
where $a$ and $\adag$ are understood to be vectors, and matrix multiplication 
is implicit.  We also need the corresponding fermionic formula
\begin{equation} % changed
\label{eq:sqz2}
\bra{0} e^{c S b} 
e^{-\lambda_c \bdag - \cdag\lambda_b - \cdag X \bdag} \ket{0}
  = \det (1-SX) e^{-\lambda_c (1-SX)^{-1} S \lambda_b} 
\end{equation} % changed
where we have suppressed the ghost zero mode dependence.

We now apply these formulas to the tadpole.  
Since the diagram factorizes into separate matter and ghost
portions, we will discuss the matter and ghost parts in turn.
The matter portion of the expectation value in the integrand of \er{firststep}
is given by
\be
\label{eq:foo1}
\int d^{26}q \, e^{-T (q^2 - 1) - q^2 V^{11}_{00} } 
     \left._{1,2}\bra{0} e^{-\onehalf a S a} e^{-\onehalf \adag \tV \adag - 
\mu \adag -\onehalf a^{3\dag} V^{11}a^{3\dag} } 
     \ket{0}_{1,2,3} \right. ,
\ee
where we have defined
\barray
S &=& \left( \begin{array}{cc} 0 & C \\ C & 0 \end{array} \right), \\
\tV &=& \left( \begin{array}{cc} \hV^{11}(T,T) & \hV^{12}(T,T) \\ 
\hV^{21}(T,T) & \hV^{11}(T,T)
         \end{array}\right)\vphantom{\Biggr)^{1}_{1_2}} 
, \\
\mu & = & \left( \begin{array}{c} \hV^{21}(T,0) \\ 
   \hV^{12}(T,0) \end{array} \right) a^{3\dag} +  
   \left( \begin{array}{c} \hV^{11}_{n0}(T,0) - \hV^{12}_{n0} (T,0) \\ 
\hV^{21}_{n0}(T,0)- 
              \hV^{11}_{n0} (T,0) \end{array} \right) q .
\earray
Using the squeezed state formula \er{sqz1} and carrying out the resulting 
Gaussian
integral over momentum (with the appropriate analytic continuation),
equation~\er{foo1} becomes 
\begin{equation} % changed
 e^T \, \left(\frac{2\pi}{Q \det (1-S\tV)} \right)^{13}  \exp \left(
         -\onehalf \adag_n  M_{nm} \adag_m \right) \0 ,
\end{equation} % changed
where
\begin{eqnarray}
 M_{nm} & = & Q_{nm} - \frac{Q_nQ_m}{Q} , \nonumber \\
 Q_{nm} & = & V^{11}_{nm} + \left( \begin{array}{c} \hV^{12}_{n\cdot}(0,T) \\ 
\hV^{21}_{n\cdot}(0,T) \end{array} 
       \right)^T \frac{1}{1-S\tV} S \left( \begin{array}{c} \hV^{21}_{\cdot m}
(T,0) \\ 
   \hV^{12}_{\cdot m}(T,0) \end{array} \right)  ,\nonumber \\
 Q_n    & = & V^{13}_{0n} - V^{23}_{0n} + \left( \begin{array}{c} 
\hV^{11}_{0\cdot}(0,T) - 
           \hV^{21}_{0\cdot} (0,T) \\ \hV^{12}_{0\cdot}(0,T)- 
\hV^{11}_{0\cdot}(0,T) \end{array} \right)^T 
      \frac{1}{1-S\tV} S \left( \begin{array}{c} \hV^{21}_{\cdot n}(T,0) \\ 
\hV^{12}_{\cdot n}(T,0) \end{array} 
       \right)  , \\
\label{eq:mattermatrix}
 Q      & = & 2 V^{11}_{00} + 2 T + \left( \begin{array}{c} \hV^{11}_{0\cdot}
(0,T) - 
           \hV^{21}_{0\cdot} (0,T) \\ \hV^{12}_{0\cdot}(0,T)-
           \hV^{11}_{0\cdot}(0,T) \end{array} \right)^T 
           \frac{1}{1-S\tV} S \left( \begin{array}{c}
           \hV^{11}_{\cdot 0}(T,0) - \hV^{12}_{\cdot 0} (T,0) \\ 
           \hV^{21}_{\cdot 0}(T,0)- 
              \hV^{11}_{\cdot 0} (T,0) \end{array} \right) .  \nonumber
\end{eqnarray}

Meanwhile, the ghost portion of the expression \er{firststep}
is
\be
\label{eq:foo2}
 \left._{1,2}\bra{0} e^{ c S b} e^{- \cdag \tX \bdag - 
\lambda_c \bdag- \cdag 
      \lambda_b-c^{3 \dag} X^{11}b^{3\dag}} \0_{1,2,3} \right. ,
\ee
with
\begin{eqnarray}
 \tX &=& \left( \begin{array}{cc} \hX^{11}(T,T)\quad & \hX^{12}(T,T) \\  
\nonumber
                                 \hX^{21}(T,T)\quad & \hX^{11}(T,T)  
\end{array}
\right) , \\ 
 \lambda_c &=&c^{\dag 3} \left( \begin{array}{cc} \hX^{12} (0,T) & \hX^{21} 
(0,T) \end{array}\right), \\
 \lambda_b &=& \left( \begin{array}{c} \hX^{21} (T,0) \\ \hX^{12} (T,0) 
\end{array} \right) b^{\dag 3} 
+ \left( \begin{array}{c} \hX^{21}_{n0} (T,0) \\ \hX^{12}_{n0} (T,0) 
\end{array} \right) b^{3}_0. 
  \nonumber
\end{eqnarray}
Using the squeezed state formula \er{sqz2}, we find
\begin{equation} % changed
 \det(1-S\tX) \, \exp \left( -c^{\dagger}_nR_{nm}b_m^{\dagger}
 \right) c_0 \ket{\hat{0}}, 
\end{equation} % changed
where we have defined
\be
\label{eq:ghostmatrix}
 R = X^{11} + \left( \begin{array}{cc}  \hX^{12} (0,T) & \hX^{21} (0,T) 
      \end{array}\right) \frac{1}{1 - S\tX} S \left( \begin{array}{c} 
           \hX^{21} (T,0) \\ \hX^{12} (T,0) \end{array} \right).
\ee
Up to an overall constant, then, $|\mathcal{T}\rangle$ is given by,
\be
\label{eq:tadpole}
\ket{{\mathcal T}} = \int_0^{\infty} dT \, e^T 
\frac{ \det (1 - S\tX)}{( Q \det (1-S\tV) )^{13} }
       e^{-\onehalf \adag M \adag - c^{\dagger} R b^{\dagger} }
       c_0\ket{\hat{0}} . 
\ee

%%%%%%%%%%%%%%%%%%%%%%%%%%%%%%%%%%%%%%%%%%%%%%%%%%%%%%%%%%%%%%%%%%%%%%%%%%%%
                \subsection{Divergences in the tadpole}
%%%%%%%%%%%%%%%%%%%%%%%%%%%%%%%%%%%%%%%%%%%%%%%%%%%%%%%%%%%%%%%%%%%%%%%%%%%%
\label{s:SmallT}

We are now interested in evaluating the tadpole integral
(\ref{eq:tadpole}).  At any finite value of $T$, the integrand is a
state in the string Fock space with finite coefficients for each
zero-momentum state.  The matrices $X^{rs}_{nm}$ and
$V^{rs}_{nm}$ are infinite-dimensional, and while we have expressions
for each matrix element, we cannot analytically compute the integrand
of (\ref{eq:tadpole}).  Nonetheless, by truncating the matrices $X$
and $V$ at finite oscillator level, we can numerically estimate the
value of the integrand.  Empirical evidence indicates that the
integrand converges exponentially quickly as the level of truncation
is increased, where the rate of convergence depends on the modular
parameter $T$.  For large $T$, we can also consider expanding
(\ref{eq:tadpole}) as a function of $e^{-T}$.  The contribution at
order $e^{(1-k) T}$ represents the portion of the tadpole arising from
the propagation of open string states at level $k$ around the loop.

It is immediately clear from (\ref{eq:tadpole}) that the integral for
the tadpole diverges as $T \rightarrow \infty$ due to the term of
order $e^{T}$.  This term arises from the propagation of the open
string tachyon around the loop.  In an expansion in terms of the level
of the field propagating around the loop, it is easy to see how this
divergence can be removed by analytic continuation.  Unlike the
divergence from the closed string tachyon considered in the CFT
calculation, the field causing this divergence is explicitly
considered as one of the fundamental degrees of freedom in OSFT, and
thus the associated tachyon divergence can easily be removed by
analytic continuation.  This analytic continuation is most transparent
when we do the calculation level by level in fields.  The method of
level truncation by fields has previously been used in many OSFT
calculations, and is probably the most effective way of dealing with
the open string tachyon; because we are more interested in closed
string physics here, however, we will not pursue this approach further
and we will instead approximate (\ref{eq:tadpole}) by truncation in
oscillator level, which leads to a simpler calculation of the
integrand at small $T$.

\FIGURE{
\epsfig{file=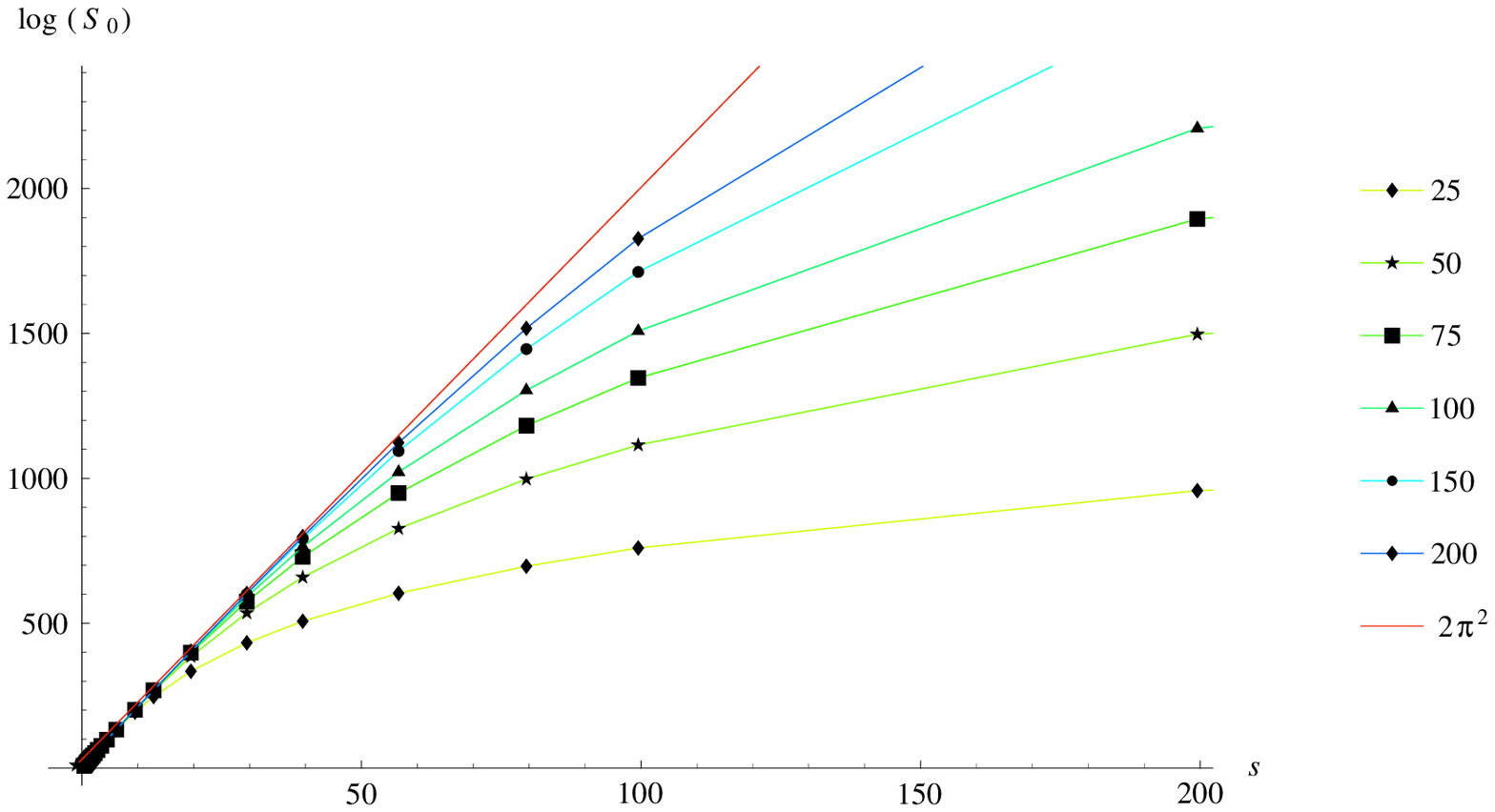,width=16cm}
\label{fig:tachyon}
\caption{$\log S_0$ vs. $s = 1/T$.  The straight line plotted has a
  slope of $2\pi^2$, which is the analytical prediction from section 3.}
}

{}From the CFT analysis, we expect to see a divergence in the integrand
of (\ref{eq:tadpole}) as $T\rightarrow 0$.  Indeed, we see numerical
evidence for such a divergence.  Let us first consider the scalar
portion of the state \er{tadpole}, \be
\label{eq:scalar}
  S_0(T) = e^T \frac{\det (1- S\tX )}{\det (1 - S \tV )^{13} Q^{13} }.
\ee 

Our numerical computations of $S_0$ are plotted in
figure~\ref{fig:tachyon}.  We find that as $T\rightarrow 0$, $S_0$
first increases exponentially, and then falls off, taking a finite
value at $T = 0$.  Both aspects of this behavior have simple
explanations.  The exponential increase is the divergence arising from
the negative mass-squared of the closed string tachyon, and takes the
form $S_0 \propto e^{B/T}$.  We can compute the coefficient $B$
numerically, as we discuss below.  The fall-off occurs because as
$T\rightarrow 0$ we consider smaller and smaller world-sheet
distances, which require higher and higher oscillator modes to
resolve.  Level truncation essentially acts as a UV cutoff, rendering
the divergence finite as $T\to 0$.

{}From the conformal field theory analysis, we expect that if we
could compute $S_0(T)$ to infinite level we would find
\begin{equation} % changed
  S_0(T)  \sim e^{2 \pi^2 /T}.
\end{equation} % changed
In figure~\ref{fig:tachyon} we plot the quantity $\log S_0$ versus $s =
1/T$ at successively higher levels.  The straight line with slope 
$2 \pi^2$ is the CFT prediction for the infinite level behavior in
the region $s \gg 1$.  As is evident from the plot, our numerical data
is in good accord with this prediction.  As expected, as the level
increases the region of linear behavior becomes larger, while the
fall-off becomes less rapid, so that successively better estimations of
$B$ are obtained.  

Our data, considered as a function of level, converges exponentially
quickly; that is, to a good approximation, we may write the amplitude
obtained at finite level as \be\label{eq:LogS} \log S_0^{(L)} (T) =
\log S_0^{({\mathrm exact})} (T) - A e^{-r(T) L}, \ee 
where $L$ is the
level.  The rate of convergence $r(T)$ is approximately
 given by 
\be\label{eq:rT}
r(T) \approx 0.002 + 0.949 \,\,T.  \ee 
This approximation to the function $r(T)$ is obtained by numerically
determining the rate of convergence for several points at fixed values
of $T$ using the ansatz $r(T) = A+B\, T$.  Since the first terms
dropped in the level $L$ truncation are of order $e^{-LT}$, we indeed
expect $r (T) = T$, in close agreement with our empirical data.  These
results are consistent with the general empirical observation that,
while integrated amplitudes converge polynomially in $1/L$ (when
finite), the integrand itself converges exponentially in $L$ with a
rate that depends on the modular parameter \cite{Taylor:2002bq}.

Looking at the region $T > .007$, we find the best fit given by
equations (\ref{eq:LogS}) and (\ref{eq:rT}) gives
\begin{equation} % changed
   B\approx.9911 \times 2 \pi^2.
\end{equation} % changed
  This is within less than 1\% of the value $2\pi^2$ found
analytically in section 3, so we have a very strong agreement between
our two methods of calculation for the leading rate of divergence.

%%%%%%%%%%%%%%%%%%%%%%%%%%%%%%%%%%%%%%%%%%%%%%%%%%%%%%%%%%%%%%%%%%%%%%
                   \subsection{The matrices $M$ and $R$}
%%%%%%%%%%%%%%%%%%%%%%%%%%%%%%%%%%%%%%%%%%%%%%%%%%%%%%%%%%%%%%%%%%%%%%

In this subsection, we consider the form of the matrices $M$ and $R$
in the exponential part of (\ref{eq:tadpole}).  We show that in the
limit $T \rightarrow 0$, these matrices reduce to $C_{nm}$, so that
just as in the CFT calculation, the leading divergence is associated
with a Shapiro-Thorn state.  Furthermore, we compute the first
subleading terms in $M$ and $R$.  We show that these subleading terms
agree with what is expected from the CFT calculation.

\label{s:MandR}
In the small $T$ limit, the loop of the tadpole diagram reduces to an
identification of the left and right halves of the incoming string.
Therefore, the matrices $M_{nm} = Q_{nm} + Q_n Q_m/Q$ and $R_{nm}$
should both limit to $C_{nm}$ in order to describe the identification
of the sides of the string.  For any finite level,
we can demonstrate analytically that this
is indeed the case for the matrix $M$ of \er{mattermatrix} and $R$ of
\er{ghostmatrix}.  Consider the identity
\be
\label{eq:matrix-id}
  \left( \begin{array}{cc} J_2 & J_3 \end{array}\right) \left(
\begin{array}{cc} 1 - C J_3 & -CJ_1 \\ -CJ_1 & 1 - CJ_2 \end{array}
\right)^{-1} = \left( \begin{array}{cc} C & C \end{array} \right) 
\ee
which holds for any matrices $J_i$ that satisfy $J_1 + J_2 + J_3 = C$.

In level truncation, we can simply apply the identity \er{matrix-id}
at $T = 0$, with $J_i$ equal to $X ^ {1 i}$ and $V^{1i}$ in the
ghost and matter sectors respectively.  This gives us the 
result
\begin{align}\label{eq:RandMareC}
   R_{nm} (T=0) &=C_{nm}  \nonumber \\
    M_{nm}(T=0) &=C_{nm}  .  
\end{align} 
Since the sum condition needed to demonstrate this result is linear,
we find that $M (T = 0) $and $ R (T = 0) $ are equal to $C$ level by
level when we calculate them in level truncation.  This reproduces the
formula we found in equation~(\ref{eq:OscillatorFormOfSmallT}).
Unfortunately, one can show that there is no
analogue of  equation (\ref{eq:matrix-id}) without level truncation.   
This makes the verification of (\ref{eq:RandMareC})
without truncating the matrices quite difficult.  
We discuss some of the subtleties of comparing the
level truncated analysis and the infinite dimensional matrix analysis in
appendix \ref{A:InfiniteLevel}.

We can also look at the first order corrections to $M_{nm}$ and
$R_{nm}$.  Doing a linear fit  near $T=0$ to the various coefficients
gives for the first five diagonal elements at level $25$,
\begin{eqnarray}
M_{11} &\approx& -0.9999999 + 1.0000031 \; (T)\nonumber \\
M_{22} &\approx& 0.9999994 -  1.0000095 \; (2 T)\nonumber \\
M_{33} &\approx& -0.9999982 + 1.0000265 \; (3 T)\nonumber \\
M_{44} &\approx& 0.9999953 -  1.0000349 \; (4 T)\nonumber \\
M_{55} &\approx& -0.9999912 + 1.0000694 \; (5 T).
\end{eqnarray}
The off diagonal elements are consistent with being zero to $\mathcal{O}(T)$.
We find similar behavior in the ghost sector.  At level $150$ we
find the following coefficients for a linear fit near $T=0$

\begin{eqnarray}
    R_{11} &\approx& -0.999995 + 0.999775\, \left( T \right) \nonumber \\
    R_{22} &\approx& 0.999989 - 0.999642\, \left(2T \right)  \nonumber\\
    R_{33} &\approx& -0.999985 + 0.999761\, \left( 3T \right)\nonumber \\
    R_{44} &\approx& 0.999979 - 0.999627\, \left( 4T \right) \nonumber\\
    R_{55} &\approx&  -0.999975 + 0.999741\, \left( 5T \right).
\end{eqnarray}
As with the matter sector, the off diagonal elements are approximately
zero to $\mathcal{O}(T)$.

These coefficients suggest that the first order corrections to 
$M_{nm}$ and $R_{nm}$ are given by
\begin{align}\label{eq:InfiniteLevel}
   M_{nm} &= C_{nm}- m C_{nm} \,T +\mathcal{O}(T^2)\nonumber\\
   R_{nm} &= C_{nm}- m C_{nm} \,T +\mathcal{O}(T^2).
\end{align}
It might seem that this formula should be easy to derive by just Taylor
expanding the expressions for $M_{nm}$ and $R_{nm}$ to first order in
$T$.  In level truncation, this expansion is straightforward and yields
the result
\begin{align}\label{eq:LevelTruncated}
  M_{nm}^{L} &= C_{nm}- 2 m C_{nm} \,T+\mathcal{O}(T^2)\nonumber\\
   R_{nm}^{L} &= C_{nm}- 2 m C_{nm} \,T+\mathcal{O}(T^2),
\end{align}
where we have put the superscript $L$ in the matrices to 
emphasize that this expression is only valid in level truncation.
Note that equations (\ref{eq:InfiniteLevel}) and (\ref{eq:LevelTruncated})
are off by a factor of two!  This disagreement stems from the fact that
one must take the level to infinity and then expand around $T=0$.
Expanding around $T=0$ and then taking the level to infinity gives
incorrect results.  For a more complete discussion of this issue
see appendix \ref{A:InfiniteLevel}.

We can now compare (\ref{eq:InfiniteLevel}) with our results from the
conformal field theory method.  As it turns out the linear correction
comes entirely from the map that acts on the external state.  The
corrections that arise from the map that acts on the boundary state
only enter at $\mathcal{O}(T^2)$.  Using the notation of
section~\ref{s:Tsmall} we can write the external state map as $f(z) =
h(z)\circ z(v)$.  At $T=0$, $f(z)$ is just the map corresponding to
the identity state
\begin{equation} % changed
  f(z)\Big|_{T=0} = \frac{2z}{1-z^2}.
\end{equation} % changed
As it turns out, the first order correction in $T$ takes a
simple form
\begin{equation} % changed
  f(z)  = f(z)\Big|_{T=0}\circ\left(\frac{z}{1+T/2}\right)+\mathcal{O}(T^2).
\end{equation} % changed
Thus to  $\mathcal{O}(T)$ we can write
\begin{equation} % changed
  f(z) \circ |A\rangle  = f(z)\Big|_{T=0}\circ \left(\frac{z}{1+T/2}\right)
\circ |A\rangle.
\end{equation} % changed
Since we also have
\begin{equation} % changed
  \left(\frac{z}{1+T/2}\right)
  \circ |A\rangle  = (1+T/2)^{-L_0} |A\rangle,
\end{equation} % changed
the change in the tadpole state from this correction to the external
state map can be accounted for by just taking
\begin{equation} % changed
  |\mathcal{T}\rangle\Big|_{T=0} \to  
(1+T/2)^{-L_0} \left(|\mathcal{T}\rangle\Big|_{T=0}\right)+
\mathcal{O}(T^2),
\end{equation} % changed
where we are only setting $T=0$ in the map from the external state to
the geometry.  Acting with $(1+T/2)^{-L_0}$ on an arbitrary state is
straightforward since one just takes, for example,
\begin{equation} % changed
  a^{\dagger}_m  \to (1+T/2)^{-m} a^{\dagger}_m,
\end{equation} % changed
and similarly for the ghosts.  Thus the terms in the exponent
become
\begin{multline}
   -\frac{1}{2}a^{\dagger}_m C_{mn} a^{\dagger}_n
   - c_m C_{mn} b_n \to 
-\frac{1}{2} (1+T/2)^{-2 m}a^{\dagger}_m C_{mn} a^{\dagger}_n
   -(1+T/2)^{-2 m} c^{\dagger}_m C_{mn} b^{\dagger}_n\nonumber \\
 = -\frac{1}{2}  a^{\dagger}_m (C_{mn} - mT\,C_{mn})a^{\dagger}_n
   -c^{\dagger}_m (C_{mn} - mT\,C_{mn}) b^{\dagger}_n +\mathcal{O}(T^2),
\end{multline}
which reproduces what we found in equations (\ref{eq:InfiniteLevel}).

%%%%%%%%%%%%%%%%%%%%%%%%%%%%%%%%%%%%%%%%%%%%%%%%%%%%%%%%%%%%%%%%%%%%%%%%%%%%
                 \subsection {Summary of oscillator calculation}
%%%%%%%%%%%%%%%%%%%%%%%%%%%%%%%%%%%%%%%%%%%%%%%%%%%%%%%%%%%%%%%%%%%%%%%%%%%%

We have given an analytic expression for the tadpole in terms of
infinite-dimensional matrices of Neumann coefficients.  A divergence
associated with the open string tachyon arises for large Schwinger
parameter $T$, and can be dealt with by straightforward analytic
continuation when the amplitude is expanded in the level of the open
string field propagating in the loop.  We have numerically analyzed
the behavior of the tadpole integrand as $T \rightarrow 0$.  
Our numerical approximations have reproduced to a high degree of
accuracy
the leading divergence in this limit.  Near $T = 0$ the tadpole takes the form
\be
 \ket{{\mathcal T}} \sim\int_0 dT e^{2\pi^2/T}
e^{-\onehalf \adag C \adag - c^{\dagger} C b^{\dagger}} \0 + \ldots 
\ee
This leading term
is the zero-momentum Shapiro-Thorn state $\ket{\Phi} = e^{-\onehalf
\adag C \adag - c^{\dagger} C b^{\dagger}} \0 $ that describes the
closed-string tachyon 
\cite{Shapiro:gq,Shapiro:ac}.  The oscillator calculation thus agrees
with the conformal calculation of section 3.

We have also identified the leading corrections (in $T$) to the limit $M,R
= C$.  These corrections are linear in $T$ and do not
represent Shapiro-Thorn states for massless closed string fields.  In
terms of the conformal calculation discussed in the previous section,
these corrections may be understood as coming from the conformal
transformation of the incoming string.

Ideally we would like to be able to see the Shapiro-Thorn states for
the graviton and dilaton.  These could arise from terms in $M$ that
vanish as $e^{-2\pi^2/T}$ as $T\rightarrow 0$; such terms can give a
finite contribution to the amplitude because of the exponential
divergence in the scalar portion of the amplitude $S_0$.  However,
such exponentially dying corrections are not only many orders of
magnitude smaller than the leading corrections, but also many orders
of magnitude smaller than the error introduced by level truncation.
Without analytic control over the matrices $\tV$ and $\tX$ or some
way of explicitly removing the divergence due to the closed string
tachyon, we cannot examine the subleading terms directly in numerical
experiments.

Finally, we note that it would be easy to redo the calculations for
the lower-dimensional branes. We would simply introduce a few minus
signs for the Dirichlet coordinates and eliminate the momentum
integrals for the transverse directions.  Without some new approach to
the calculation, however, all we would find is the same divergence from the
closed string tachyon.  We would not be able to observe that the
massless sector was no longer contributing to the divergence.

%%%%%%%%%%%%%%%%%%%%%%%%%%%%%%%%%%%%%%%%%%%%%%%%%%%%%%%%%%%%%%%%%%%%%%%%%%%
%%%%%%%%%%%%%%%%%%%%%%%%%%%%%%%%%%%%%%%%%%%%%%%%%%%%%%%%%%%%%%%%%%%%%%%%%%%
  \section{The open string tadpole in open-closed string field theory}
%%%%%%%%%%%%%%%%%%%%%%%%%%%%%%%%%%%%%%%%%%%%%%%%%%%%%%%%%%%%%%%%%%%%%%%%%%%
%%%%%%%%%%%%%%%%%%%%%%%%%%%%%%%%%%%%%%%%%%%%%%%%%%%%%%%%%%%%%%%%%%%%%%%%%%%

\label{ss:OCSFT}
\FIGURE{
\epsfig{file=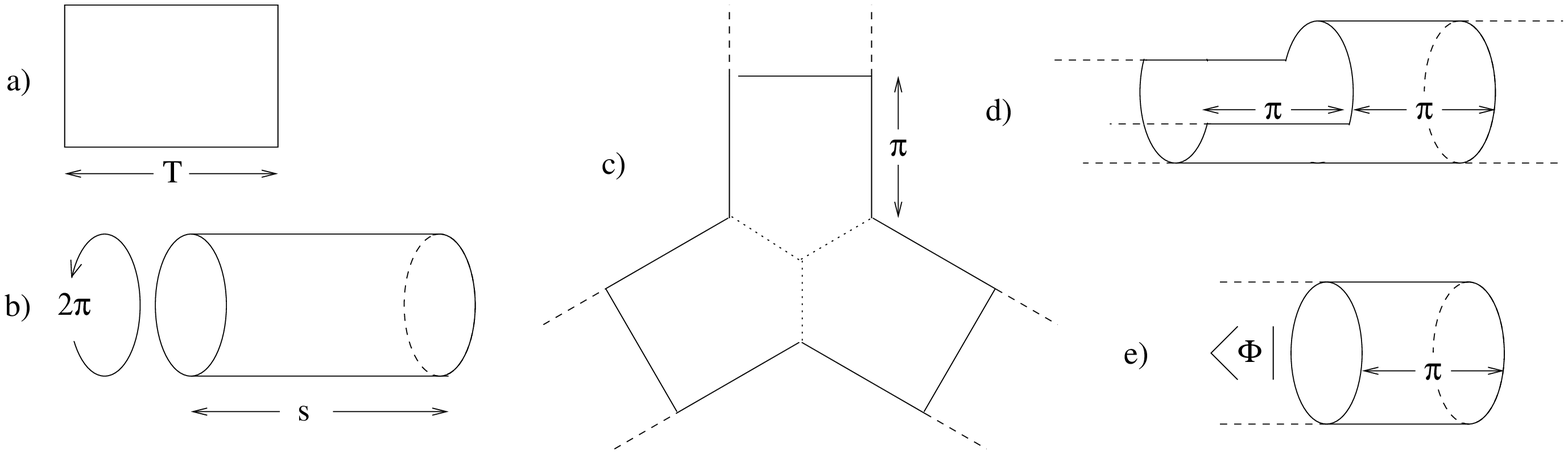,width=15cm}
\caption{\footnotesize
The relevant vertices in OCSFT. a) The open string propagator is the same
as in OSFT. b) The closed string propagator is a tube of circumference $2 \pi$.
Its length, $s$, is integrated from $0$ to $\infty$.  c) The open string
vertex is similar to the OSFT vertex but world-sheet stubs of length $2\pi$
are added on each side.  d) The open-closed string transition vertex represents
an open string which turns in into a closed string.
e) The vertex representing a closed string being absorbed by the brane.
}
\label{f:OCSFTVertices}
} 
To understand the physics hidden in the OSFT tadpole, it is useful to
study the same diagram in a version of string field theory which
includes closed strings explicitly.  This open-closed string field
theory (OCSFT), due to
Zwiebach \cite{Zwiebach:1997fe,Zwiebach:1990qj}, has the nice property
that it divides up moduli space so that there are never any
pinched-off world-sheets.  This
is convenient for us, since it implies that in this theory the tube at
the bottom of the OSFT tadpole will be explicitly written as a closed
string propagator.  For a related discussion in a different version
of string field theory, see \cite{Green:pf}.

OCSFT is complicated by the fact that its action is non-polynomial
but, since we are working only to order $g$, we will only need to
consider a few terms.  In fact since we are only interested in the
divergent part of the tadpole, we truncate the theory to just the
terms in the action relevant to the $T \to 0$ part of the tadpole
moduli space.  These terms may be summarized as follows
\begin{enumerate}
\item
  Start with the usual OSFT action but modify the cubic term by adding
to the vertex strips of length $\pi$ to each of the three sides.  The
resulting vertex is shown in figure~\ref{f:OCSFTVertices}~c.
The stubs added to the Witten vertex eliminate the region of the
tadpole near $T \to 0$.
\item
  Add to the theory a set of closed string fields $\Phi$ with kinetic
term $\langle \Phi| (c_0-\tilde{c}_0) (Q_B+\tilde{Q}_B) |\Phi\rangle$.
The field $\Phi$ is assumed to satisfy $(b_0 - \tilde{b}_0)\Phi=0$.
We also impose the analogue of FS gauge $(b_0 + \tilde{b}_0) \Phi = 0$.
The propagator of the theory is given by integrating over closed string
tubes of length $s$ with insertions of $b_0 \tilde{b}_0$.
\item
Add the open-closed string vertex shown in figure~\ref{f:OCSFTVertices}~d.
We will denote this vertex $\langle V_{\text{OC}}||\Psi \rangle|\Phi\rangle$. 
\item Add a vertex in which a closed string is absorbed by the brane.
This is given by 
$\langle \Phi | (c_0 -\tilde{c}_0)
e^{-\pi (L_0+\tilde{L}_0)}| \mathcal{B} \rangle$.
Pictorially this vertex is shown in figure~\ref{f:OCSFTVertices}~e.
\end{enumerate}

We can now consider the OCSFT version of the open string tadpole.  Using
the new open string vertex in figure~\ref{f:OCSFTVertices} c, 
and the open string propagator one can construct
a tadpole diagram that looks just like the one we considered on OSFT.  The
only difference is that because of the stubs, the loop cannot have length
$T < 2\pi$.  The part of the tadpole moduli space near $T\to 0$
is covered by a new diagram
formed by gluing an open string propagator to an open-closed vertex, 
then attaching a closed string propagator and finally capping the diagram 
with the vertex in figure~\ref{f:OCSFTVertices}~e.  The resulting diagram
is pictured in figure~\ref{f:OCSFTTadpole}.
\FIGURE{
\epsfig{file=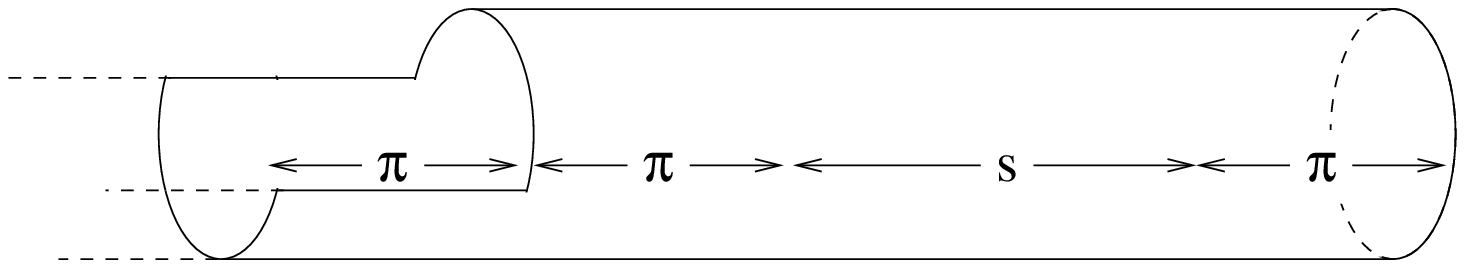,width=10cm}
\caption{\footnotesize
The OCSFT representation of the open string tadpole.
}
\label{f:OCSFTTadpole}
}

Note that the dependence on $s$ is
quite simple since it is just the propagation of the boundary state
a distance $ s$.  Thus everything to the right of the
open-closed vertex can be represented as
\begin{equation}
(b_0 +\tilde{b}_0)
\int_{ \pi}^{\infty} ds' \, e^{-s' (L_0+\tilde{L}_0)}
|\mathcal{B}\rangle.
\end{equation}
As in the OSFT version of the tadpole, this integral diverges because
of the weight $(-1,-1)$ field in $|\mathcal{B}\rangle$.  As in OSFT
we can treat this divergence by analytic continuation in the weight
of the tachyon field.  This analytic continuation is equivalent to the
replacement
\begin{equation} % changed
  (b_0 +\tilde{b}_0)
\int_{ \pi}^{\infty} ds' \, e^{-s' (L_0+\tilde{L}_0)}|\mathcal{B}\rangle
 = \frac{b_0 +\tilde{b}_0}{L_0+\tilde{L}_0} e^{-\pi (L_0+\tilde{L}_0)}
|\mathcal{B}\rangle
\end{equation} % changed
where the operator $(L_0+\tilde{L}_0)^{-1}$ is defined
on all states that are not weight $(0,0)$.  This analytic continuation
is closer in spirit to the analytic continuation used in Section 4 to
remove the open string tachyon divergence than the analogous analytic
continuation used in Section 3 to deal with the closed string tachyon,
since the closed string tachyon is explicitly included among the
fundamental degrees of freedom in OCSFT.  We can also have divergences
from the massless sector, but as we showed for the OSFT tadpole, 
these are only relevant for the D$25$, D$24$ and D$23$-branes and 
we postpone discussion of these cases until the end of the section.

Now consider our truncated action (we drop the open string three-vertex since
it is not relevant here)
\begin{eqnarray}
  S[\Psi,\Phi] = \frac{1}{2}\int \Psi \star Q_B \Psi 
  & & + \frac{1}{2}\langle \Phi| (c_0 -\tilde{c}_0) (Q_B +\tilde{Q}_B)
|\Phi\rangle
\nonumber\\ & & +
                 g\,\langle V_{\text{OC}} || \Psi \rangle | \Phi \rangle
           +\,\langle \Phi |(c_0-\tilde{c}_0) e^{-\pi (L_0+\tilde{L}_0)}|
             \mathcal{B} \rangle.
\end{eqnarray}
Note that the closed string tadpole appears even at the classical level.
We can try to shift the closed string field by $\Phi \to \Phi +  \delta \Phi$
to cancel this tadpole.  This gives
\begin{equation} % changed
  \delta S[\Psi,\Phi] = 
g\,\langle V_{\text{OC}}|| \Psi \rangle| \delta \Phi \rangle 
+\langle \Phi| (c_0 -\tilde{c}_0) (Q_B +\tilde{Q}_B)|\delta\Phi\rangle
+ \text{const.}
\end{equation} % changed
Notice that if we can find a $\delta \Phi$ such that
$(Q_B +\tilde{Q}_B) |\delta \Phi \rangle = -e^{-\pi (L_0+\tilde{L}_0)}|
             \mathcal{B} \rangle$,
then the closed string tadpole will be canceled.
We then get the new action
\begin{eqnarray}
  S[\Psi,\Phi+\delta\Phi] = \frac{1}{2}\int \Psi \star Q_B \Psi 
& & + \frac{1}{2}\langle \Phi| (c_0 -\tilde{c}_0) (Q_B 
+\tilde{Q}_B)|\Phi\rangle
\nonumber\\
& & + g\,\langle V_{\text{OC}} | | \Psi \rangle | \Phi \rangle +
 g \,\langle V_{\text{OC}}| | \Psi \rangle |\delta \Phi \rangle.
\end{eqnarray}
Notice that the closed string tadpole is now eliminated, but there is
now a new contribution to the open string tadpole.  

Let's recalculate the open string tadpole in the shifted theory.  The
original diagram that we calculated before is now gone because the
closed string tadpole (which made up the right half of the diagram)
has been canceled.  However, there is a new diagram coming from the
term $g \langle V_{\text{OC}} | \Psi \rangle |\delta \Phi \rangle$.
This diagram may be thought of as the original open string tadpole,
but with the closed string propagator chopped off and the state
$-|\delta \Phi \rangle$ stuck onto the end.  Note that since this
diagram has no closed string modulus to integrate, it is finite as
long as $|\delta \Phi \rangle$ is finite.  This fact will be useful
for us when we discuss the cases where we have divergences from
the massless sector in the original tadpole.

Now, as discussed in \cite{Polchinski:jq}, the equation
$(Q_B +\tilde{Q}_B) |\Phi\rangle = -e^{ \pi (L_0 +\tilde{L}_0)} |\mathcal{B}
\rangle $ is equivalent to the linearized Einstein's equations in 
the background of the brane.  Thus the new open string tadpole
represents a coupling between the closed string background and the
the open strings.

For OSFT this may not seem important since there is no closed string
background to shift.  However, what makes this discussion relevant for
OSFT is that the open string tadpole after the shift in the closed
string background is actually equal to the open string tadpole before we
made the shift.  This is seen by solving the equation for $|\Phi\rangle$ to
get
\begin{equation} % changed
  |\Phi\rangle = -\frac{b_0 + \tilde{b}_0}{L_0+\tilde{L}_0}
   e^{ \pi (L_0 +\tilde{L}_0)} |\mathcal{B}
\rangle.
\end{equation} % changed 
Thus, replacing the closed string propagator with $-|\Phi\rangle$ 
is equivalent to doing nothing.  The implication for the 
open string tadpole in OSFT is that as long as the diagram is finite
it naturally incorporates the linearized shift in the closed string
background.

This leads to the question: what can we say about the tadpole diagrams
which diverge because of the massless sector?  For the finite diagrams
the closed string propagator essentially represents the inverse of the
BRST operator. For the divergent diagrams this representation is not
defined when acting on the boundary state.  To cure this problem one
must invert the BRST operator by hand.  To see how this is done it is
useful to note how BRST invariance is maintained in the shifted and
unshifted theory.  
Recall that in string theory BRST invariance for
scattering diagrams reduces to the fact that exact states should
decouple from on-shell states.  For the tadpole diagram this simply
implies that the tadpole should be annihilated by the BRST operator.

In the unshifted theory, when $Q_B+\tilde{Q}_B$ is pulled through the
OCSFT tadpole diagram, it picks a contribution from the closed
string propagator given by
\begin{align}
 \int_{ \pi}^{\infty} ds' \, \{Q_B+\tilde{Q}_B,b_0 +\tilde{b}_0\}
 \, e^{-s' (L_0+\tilde{L}_0)} 
|\mathcal{B}\rangle
&=
\int_{ \pi}^{\infty} ds' \, (L_0+\tilde{L}_0) \, e^{-s' (L_0+\tilde{L}_0)} 
|\mathcal{B}\rangle
\nonumber\\
&= -\int_{ \pi}^{\infty} ds' \, \frac{\partial}{\partial T}
     \, e^{-s' (L_0+\tilde{L}_0)} |\mathcal{B}\rangle
\nonumber\\
&= -
     \, e^{-s' (L_0+\tilde{L}_0)}\big|_{s=\pi}^{\infty}|\mathcal{B}\rangle.
\label{eq:SurfaceTerm}
\end{align}
As with the open string, we only pick up contributions at the
endpoints of integration.  The contribution at $s = \pi$ 
cancels with a surface term from a part of the tadpole moduli space
that we haven't included.  

In the shifted theory, we have replaced the closed string propagator
with a surface term.  Acting on the surface term with
$Q_B+\tilde{Q}_B$ we get
\begin{equation}\label{eq:QOnCounterTerm}
  -(Q_B+\tilde{Q}_B)  |\delta\Phi\rangle
 = e^{-\pi (L_0+\tilde{L}_0)} |\mathcal{B}\rangle,
\end{equation}
which is the same surface term at $s = \pi$ that we found in
(\ref{eq:SurfaceTerm}). Note, though, that (\ref{eq:QOnCounterTerm})
is the same equation that we used to find $|\Phi\rangle$ in
the first place.  Thus the condition for BRST invariance 
essentially determines the surface term.

This fact is quite useful for studying the tadpole for the cases where
it diverges.  If we take one of the OSFT tadpoles which diverges, we
can remove the small $T$ region of integration and replace it with a
surface term.  By the above discussion, this surface term is
mostly determined by BRST invariance.  We will employ this fact in
section \ref{sec:regulating} where we study the physics hidden
in the divergent diagrams.

%%%%%%%%%%%%%%%%%%%%%%%%%%%%%%%%%%%%%%%%%%%%%%%%%%%%%%%%%%%%%%%%%%
%%%%%%%%%%%%%%%%%%%%%%%%%%%%%%%%%%%%%%%%%%%%%%%%%%%%%%%%%%%%%%%%%%
             \section{Divergences and closed strings}
%%%%%%%%%%%%%%%%%%%%%%%%%%%%%%%%%%%%%%%%%%%%%%%%%%%%%%%%%%%%%%%%%%
%%%%%%%%%%%%%%%%%%%%%%%%%%%%%%%%%%%%%%%%%%%%%%%%%%%%%%%%%%%%%%%%%%

In this section we discuss the various divergences which arise in the
one-loop tadpole as $T \rightarrow 0$, and the role which closed
strings play in the structure of the tadpole.  In subsection
\ref{sec:continuation} we discuss the leading divergence in the
tadpole and the closed string tachyon which is responsible for this
divergence.  In subsection \ref{sec:massless} we discuss the massless
closed string modes and the piece of the tadpole arising from them.
In subsection \ref{sec:regulating} we study the physics hidden in the
divergent diagrams using BRST invariance.  Finally, in subsection
\ref{sec:two-loops} we discuss some further problems which arise at
two loops in OSFT.

\subsection{The leading divergence and the closed string tachyon}
\label{sec:continuation}

In both the conformal field theory and oscillator calculations, we
found a leading divergence in the open string tadpole which arises from
the region of the modular integration near $T = 0$.  In terms of the
dual parameter $s = \pi^2/T$, the divergence arises from an integral
of the form
\begin{equation}
\int^\infty ds \; \left[ e^{2s} \exp \left(
-\frac{1}{2}a^{\dagger} \cdot C \cdot a^{\dagger} -c^{\dagger}\cdot
C\cdot b^{\dagger} 
\right)| 0 \rangle + {\rm subleading\ terms} \right]\,.
% \label{eq:}
\end{equation}
This divergent type of integral is a standard problem when we have a
Schwinger parameter associated with a tachyonic state.  In principle,
we would like to simply analytically continue the integral, using
\begin{equation}
\int_\lambda^\infty e^{as} ds \rightarrow
-\frac{1}{a}  e^{a \lambda}.
% \label{eq:}
\end{equation}
Since the closed string channel associated with the divergent integral
is not included explicitly in OSFT, however, it is rather subtle to
carry out an analytic continuation of this type.  In the CFT
calculation, we can do this explicitly once we have expanded around $T
= 0$ as in (\ref{eq:FullLimit}).  As suggested in Section 3, the terms
in the small $T$ expansion associated with the tachyon component of
the boundary state can be separately analytically continued.
Even here, however, we run into difficulties, and have to resort
to considering lower dimensional branes to ensure a completely
finite diagram.

Moreover, the expansion around $T = 0$ is a difficult starting point
for any exact or approximate calculation of the complete tadpole
diagram.  In order to get a numerical approximation to the tadpole
amplitude including the analytic continuation for the closed string
tachyon, it is necessary to break the modular integral into several
parts.  The integral for $T > T_0$ is finite (after the open string
tachyon is analytically continued as discussed in Section 4), and can
in principle be computed approximately using level truncation on
fields in the oscillator formalism.  The integral for $T < T_0$ can be
approximately computed when $T_0$ is small using the small $T$
expansion and an explicit analytic continuation of the tachyon term.
While this approach allows us to deal with the divergence from the
closed string tachyon ``by hand'' in the particular diagram considered
here, we should emphasize that this method is not general, and for
higher-loop diagrams it is much more difficult to see how analogous
divergences can be controlled.

In the oscillator approach, it is even less clear how the divergence
from the closed string tachyon can be treated.  Since level
truncation regulates the divergence, the region of the modular
integral near $T = 0$ is softened, and the divergent piece cannot be
precisely isolated.  It seems that in the level-truncated theory,
unlike in the CFT picture, there is no way to implement by hand an
analytic continuation to deal with the closed string divergence.  For
this reason, we cannot use the oscillator approach to study other
parts of the tadpole, such as the finite part and the part depending
on massless closed strings.  While this is unfortunate, this problem
is an artifact of the closed string tachyon.  It seems likely that in
superstring field theory, this problem would not occur, so that the
oscillator method would be a much more useful approach for analyzing
detailed features of loop amplitudes.  

It is also worth pointing out at this point that a small modification
of the usual formulation of OSFT might make the closed string tachyon
divergence much more tractable.  If we were to explicitly formulate
OSFT in terms of a Lorentzian world-sheet, the real Schwinger
parameterization used here could be replaced by an integral with an
imaginary exponent.  Unlike the divergent integral $\int e^{as}$, the
complex integral $\int e^{ias}$ is oscillatory.  With such an
oscillatory integrand, level truncation should simply suppress the
integrand at large values of $s$, effectively regulating the theory
and giving a finite result for integrals which diverge in the
Euclidean formulation.  Unlike the {\it ad hoc} approach used to
implement analytic continuation in the CFT calculation of the tadpole,
this approach would immediately generalize to all diagrams, and would
in one stroke deal with the open string tachyon as well as the closed
string tachyon.  We will not pursue this approach further here, but it
is an interesting possible avenue for further investigations.

%%%%%%%%%%%%%%%%%%%%%%%%%%%%%%%%%%%%%%%%%%%%%%%%%%%%%%%%%%%%%%%%%%%%%
   \subsection{Tadpole contributions from massless closed strings}
%%%%%%%%%%%%%%%%%%%%%%%%%%%%%%%%%%%%%%%%%%%%%%%%%%%%%%%%%%%%%%%%%%%%%
\label{sec:massless}

Let us now turn our attention to the massless closed string modes.  As
mentioned in the previous subsection, the terms in the tadpole arising
from these modes cannot be seen in the oscillator calculation without
a new formulation of the theory.  These terms do appear explicitly,
however, in the small $T$ expansion of the tadpole
(\ref{E:masslessdivergence}).

To understand the structure of the terms associated with massless
closed strings it is helpful to consider the schematic form of the
OCSFT calculation from Section 5.  In OCSFT, it is clear that the open
string tadpole arises directly from the closed string tadpole.  The
closed string tadpole in turn encodes the structure of the D-brane as
a source for the closed string fields.  The D-brane boundary state
$|{\cal B} \rangle$ only couples to the closed string fields.  If we
forget about the open strings and solve the equations of motion for
the closed string, the linearized equation of motion corresponds to
the linearized gravity equations in the presence of the brane.
For a D$p$-brane source, the linearized gravitational equations take
the schematic form
\begin{equation}
\partial^2 \phi (x) = \delta^{25-p} (x_\bot) \,.
\label{eq:linearized-phi}
\end{equation}
(We ignore here the details of the tensor structure of the full
gravity multiplet in order to elucidate the underlying physics; the
exact equations for the gravitational fields are written in the
following subsection.)  In momentum space, the solution of
(\ref{eq:linearized-phi}) is
\begin{equation}
\phi (k) = \frac{1}{k_\bot^2}  \delta (k_\parallel).
\label{eq:linearized-momentum}
\end{equation}
In position space, the solution is just the usual
\begin{equation}
\phi (x) \propto r^{p-23}, \;  p \neq 23; \;\;\;\;\;
\phi (x) \propto \ln r, \; p = 23 \,.
% \label{eq:}
\end{equation}
As discussed in Section 5, the open string tadpole in OCSFT is
unchanged if we explicitly shift the closed string background to
cancel the tadpole.  This corresponds to turning on a linearized
gravity background of the form (\ref{eq:linearized-momentum}).  In
OCSFT, the open string tadpole then arises by acting on this closed
string background with the open-closed interaction vertex.  Thus, in
OCSFT, we can naturally think of the open string tadpole as coming
directly from the closed string background arising from the D-brane.

In OSFT, the analysis in Section 3 of the small $T$ expansion of the
one-loop tadpole reveals a highly parallel structure to that just
described.  The terms in (\ref{eq:FullLimit}) associated with massless
fields are those with a $T$ dependence of the form $1/T^2$.  The
integration measure $dT/T^2$ is proportional to the measure $ds$ for
the dual (closed string) modular parameter.  Unlike the closed string
tadpole, the open string tadpole only has support at vanishing
momentum $p = 0$.  As discussed in Section 3, the integral over
$q_\bot$ in the closed string boundary state gives extra powers of $T$
in the modular integral, so that when $p \leq 22$ the modular integral
is convergent.  This is analogous to smoothing out the solution
(\ref{eq:linearized-momentum}) by integrating with the measure $\int
d^{25-p}k_\bot \sim \int k_\bot^{24 -p}dk_\bot$ in the vicinity of the
singular point $k_\bot$.  For $p \leq 22$ the resulting integral is
convergent, while for $p \geq 23$ the integral diverges.

This somewhat schematic discussion indicates that
part of the open string tadpole can be seen as arising from the closed
string background associated with the D$p$-brane.  For $p \leq 22$,
this piece of the tadpole is finite.   In the following subsection, we
consider D-branes for which this part of the tadpole diverges, and we
develop the line of reasoning just described in more detail.

%%%%%%%%%%%%%%%%%%%%%%%%%%%%%%%%%%%%%%%%%%%%%%%%%%%%%%%%%%%%%%%%%%%%
\subsection{Divergent diagrams and BRST invariance}
%%%%%%%%%%%%%%%%%%%%%%%%%%%%%%%%%%%%%%%%%%%%%%%%%%%%%%%%%%%%%%%%%%%%
\label{sec:regulating}

Having analyzed the case where the tadpole diagram is finite, we now
turn to the cases where the diagram diverges\footnote{We would like to
thank Eva Silverstein for very useful discussions on the issues in
this section and in Appendix A.1.  A world-sheet discussion of issues
related to the divergences we describe in this section will be given
in \cite{McGreevy-Silverstein}}.  While we have no satisfactory way
of regulating these diagrams, we can still extract some interesting
physics from them.  What we attempt to show is that BRST invariance
mostly determines the physics hidden in the divergent part of the
modular integral.

Recall from section \ref{ss:OCSFT} that,
for the tadpole diagram, BRST invariance amounts to checking that
$Q_B$ annihilates the tadpole state, $|\mathcal{T}\rangle$.
Since $Q_B$ is a total derivative on moduli space, we will only
pick up contributions at the boundaries of moduli space.
In other words, we can only pick up surface terms at $T = 0$ and
$T = \infty$.  Such surface terms can, in fact, arise because of 
the divergences in the tadpole and are discussed in 
appendix~\ref{s:BRSTanomaly}.

Since the $T\to 0$ limit of the tadpole diagram is divergent, we
must introduce some sort of regulator.  A very simple choice is
to just cut off the modular integral at some minimum value $T_0$.
In other words, we replace $|\mathcal{T}\rangle$ with the regulated
state
\[
  |\mathcal{T}\rangle_{T_0}  =\int_{T_0}^{\infty} |\mathcal{T}(T)\rangle.
\]
where we have included the subscript $T_0$ to explicitly indicate that
this is a regulated form of the tadpole.  Since the subscript $T_0$ is
cumbersome, we drop it in subsequent equations, but, throughout this
section, we will assume that $|\mathcal{T}\rangle$ is regulated in this
way.  Introducing the cutoff, $T_0$, explicitly breaks the BRST
invariance of the diagram.  Acting on $|\mathcal{T}\rangle$ with $Q_B$
gives a surface term at $T_0$.  This is seen in the following
calculation
\begin{eqnarray} \label{eq:Qidentity1}
  \langle \mathcal{T}| Q_B
&=& 
\int_{T_0}^\infty dT\,
\langle V_3 |b_0^{(2)} e^{-T L_0} |\tilde{V}_2\rangle Q_B\nonumber\\
&=& 
\int_{T_0}^\infty dT\, \frac{d}{dT}
\langle V_3 | e^{-T L_0}  |\tilde{V}_2\rangle\nonumber\\
&=& 
\langle V_3 | e^{-T_0 L_0} |\tilde{V}_2\rangle. \label{eq:QT}
\end{eqnarray}
Note that the last line of (\ref{eq:Qidentity1}) is similar to the
original tadpole diagram except that we have fixed the modular
parameter at $T_0$ and dropped the insertion of $b_0$.
Thus in the conformal field theory method, we can represent the
surface term by just taking $T\to T_0$ and dropping the integral
of $b(z)$ across the world-sheet.  

For the tadpole diagram to be BRST invariant, we would require that 
$Q_B$ acting on the region of integration $T< T_0$ should 
cancel this surface term.  Unfortunately, for the divergent diagrams,
we can not define this region of integration. Thus we consider
replacing the $T<T_0$ region of integration with a new diagram.
 We then try to understand what restrictions the condition of
BRST invariance imposes.  This condition will not fix the new
diagram completely but will tell us something about the physics
of the $T<T_0$ region.

To construct the new diagram, we start with the surface term we
found from acting on $|\mathcal{T}\rangle$ with $Q_B$ and
modify it.  As we discussed for the tadpole diagram, we
may replace the boundary at the bottom of the surface term with a
boundary state without changing the diagram.  
Instead of doing this, however, we replace the bottom
of the diagram with some arbitrary closed string field $\Phi$.
Call this new diagram $|\mathcal{T}_{\Phi}\rangle$.  The surface
term we found from acting on $|\mathcal{T}\rangle$ with $Q_B$ 
can then be written as $|\mathcal{T}_{\mathcal{B}}\rangle$.

We now propose replacing the region of integration $T<T_0$ with the
surface term $|\mathcal{T}_\Phi \rangle$.  To ensure that this choice
of surface term is valid, one would have to show that any open string
state can be written in the form $|\mathcal{T}_{\Phi}\rangle$ for some
closed string state $\Phi$.  While, from the conformal field theory
expression for $|\mathcal{T}_{\Phi}\rangle$, this seems likely, we will
content ourselves with the fact that this choice of surface term is
general enough for our purposes.  With this caveat in mind, we can then
ask: what condition does BRST invariance impose on $\Phi$?
To check this we need to consider the action of $Q_B$ on
$|\mathcal{T}_\Phi\rangle$. 

We now show that $Q_B |\mathcal{T}_{\Phi}\rangle =
|\mathcal{T}_{(Q_B+\tilde{Q}_B) \Phi}\rangle$.  This can be shown by a
contour pulling argument.  We start with the BRST current contour
running across the external leg of the diagram and pull it to the
right.  Since the Witten vertex is BRST invariant, we can freely slide
the contour over it.  If we slide the two ends of the contour all the
way to the right of the diagram, they eventually meet each other and
we can join them together to form a closed contour.  This contour can
then be pulled down the tube at the bottom of the diagram to act on
$|\Phi\rangle$.  This process is pictured in
figure~\ref{f:ContourArgument}.  \FIGURE{
  \epsfig{file=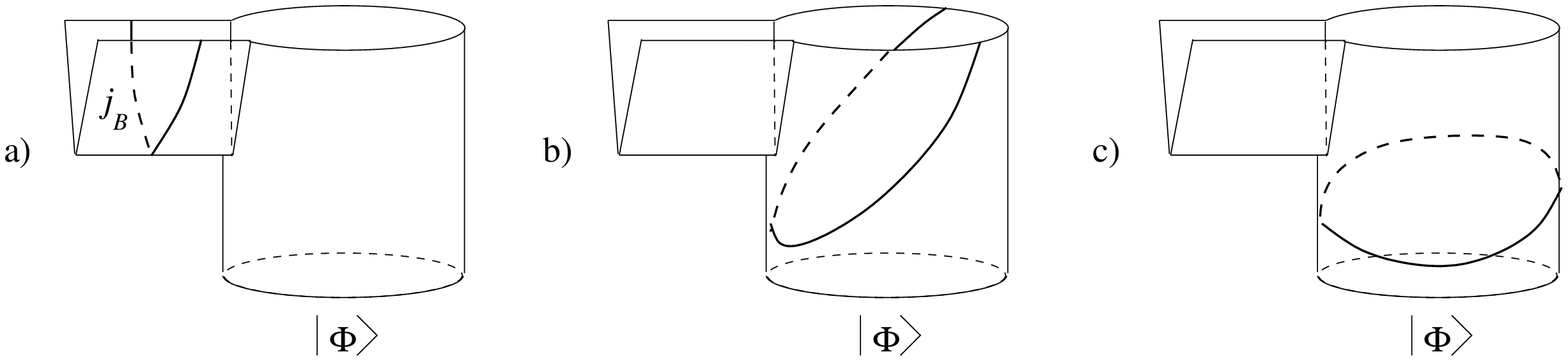,width=14cm}
\caption{\footnotesize
Contour pulling argument showing that 
$Q_B |\mathcal{T}_{\Phi}\rangle = |\mathcal{T}_{Q_B \Phi}\rangle$.
a) The contour of the BRST current $j_B$ starts on the external leg.
b) Pulling the contour to the right we can freely pass it over
the Witten vertex.
c) The two ends of the contour are joined together.  The contour can
now be thought of as acting on $|\Phi\rangle$.
}
\label{f:ContourArgument}
}

Note that in the notation we are
using we can write, 
$Q_B |\mathcal{T}\rangle = |\mathcal{T}_{\mathcal{B}}\rangle$.  Thus, if
we modify the tadpole diagram by 
$|\mathcal{T}\rangle \to |\mathcal{T}\rangle + |\mathcal{T}_{\Phi}\rangle$,
the condition for BRST invariance becomes
\begin{equation} 
 0 = Q_B(|\mathcal{T}\rangle + |\mathcal{T}_{\Phi}\rangle) =  
  |\mathcal{T}_{\mathcal{B}}\rangle +
|\mathcal{T}_{(Q_B+\tilde{Q}_B) \Phi}\rangle  = 
|\mathcal{T}_{\mathcal{B}+(Q_B+\tilde{Q}_B) \Phi}\rangle.
\end{equation} 
So to cancel the BRST anomaly we should choose 
\be \label{eq:QPhiandB}
   (Q_B+\tilde{Q}_B)| \Phi \rangle =  - |\mathcal{B}\rangle.
\ee 
Notice that this  is essentially the same
closed string field we used when we shifted the OCSFT action.

%%%%%%%%%%%%%%%%%%%%%%%%%%%%%%%%%%%%%%%%%%%%%%%%%%%%%%%%%%%%%%%%%%%%%%%%%%%%%
%\subsection{Solving the equation $(Q_B+\tilde{Q}_B)| \Phi \rangle 
%= - |\mathcal{B}\rangle$}
%%%%%%%%%%%%%%%%%%%%%%%%%%%%%%%%%%%%%%%%%%%%%%%%%%%%%%%%%%%%%%%%%%%%%%%%%%%%%

\label{s:SolvingQinverseB}
We now try to solve for $|\Phi\rangle$.  We restrict discussion
to the D$25$-brane case since it is the simplest.
  Equation~(\ref{eq:QPhiandB})  was
already solved in \cite{Polchinski:jq} and we follow the discussion
there.
First, note that equation~(\ref{eq:QPhiandB}) requires that  
the boundary state is BRST-closed and BRST-exact.  While it is true
that $(Q_B+\tilde{Q}_B) |\mathcal{B}\rangle = 0$, 
it is not true that $|\mathcal{B}\rangle$
is exact.  This can be seen by looking at the expansion of
$|\mathcal{B}\rangle$ in equation~(\ref{eq:BoundaryState}). 
We can resolve this issue by noting that 
the cohomology of $Q_B$ is  defined
using states with well-behaved momentum dependence.  For example one can check
that 
\begin{equation} % changed
  Q_B\, x_0^{\mu} |0\rangle  = c_1 \alpha_{-1}^{\mu} |0\rangle,
\end{equation} % changed
contrary to the fact that $c_1 \alpha_{-1}^{\mu} |0\rangle$ is in 
the cohomology of $Q_B$.
Thus, we proceed by just taking a general 
state of weight $(0,0)$ and $(-1,-1)$ fields, acting on it with 
$(Q_B+\tilde{Q}_B)$ and
comparing it with $|\mathcal{B}\rangle$.  We use the same parametrization
for $|\Phi\rangle$
as  \cite{Polchinski:jq}
\begin{multline}
  |\Phi\rangle = \Bigl(A(p) c_1 \tilde{c}_1 
     - \alpha_{-1}^{\mu} \tilde{\alpha}_{-1}^{\nu} h_{\mu \nu}(p) 
              c_1 \tilde{c}_1  \\
     + (c_1 c_{-1} -\tilde{c}_1 \tilde{c}_{-1})(\Phi(p) -h^{\mu}_{\mu}(p) /2 )
+(c_0+\tilde{c}_0)(c_1 \alpha_{-1}^{\mu} -\tilde{c}_1 
\tilde{\alpha}_{-1}^{\mu})
  i \zeta_{\mu}(p) \Bigr)|0\rangle.
\end{multline}
We then compute
\begin{eqnarray}
(Q_B+\tilde{Q}_B) | \Phi \rangle &=& 
\left(\tfrac{1}{2} p^2-1\right)
A(p)c_{1} (c_0+ \tilde{c}_0)\tilde{c}_1 |0\rangle 
\label{eq:QPhi}\\
& & +\left[
  -\tfrac{1}{2} p^2 h_{\mu \nu}(p) + i p_{\mu} \zeta_{\nu}(p)
    +i p_{\nu} \zeta_{\mu}(p)\right]\alpha_{-1}^{\mu} \tilde{\alpha}_{-1}^{\nu}
    (c_0+\tilde{c}_0)c_1\tilde{c}_1|0\rangle 
\nonumber\\
  & & + \left[ \tfrac{1}{2} p^2\bigl(\Phi(p)-h_{\mu}^{\mu}(p)/2\bigr)+ i 
p^{\mu}
   \zeta_{\mu}(p) \right] 
    (c_0+\tilde{c}_0)(c_{1}c_{-1}-\tilde{c}_1 \tilde{c}_{-1})|0\rangle
\nonumber\\
  & & + \left[ -p^{\mu} h_{\mu \nu}(p)- p^{\mu}\bigl(\Phi(p)-h_{\mu}^{\mu}(p)/2
    \bigr)
    +2i\zeta_{\mu}(p)\right]
       (\tilde{\alpha}_{-1}c_{-1}+ \alpha_{-1} \tilde{c}_{-1}) 
     c_1\tilde{c}_1|0\rangle.\nonumber
\end{eqnarray}
Substituting equation~(\ref{eq:QPhi}) into 
$(Q_B+\tilde{Q}_B)|\Phi\rangle = -|\mathcal{B}\rangle$
gives
\begin{eqnarray} \label{eq:lingrav}
(\tfrac{1}{2}\partial^2+1)A(x)  &=& 1
\nonumber\\
 \tfrac{1}{2} \partial^2 h_{\mu\nu}(x) -\partial_{\mu} \zeta_{\nu}(x) 
 -\partial_{\nu} \zeta_{\mu}(x) &=& \eta_{\mu \nu}
\nonumber\\
 \tfrac{1}{2}\partial^2 \Phi(x) - \tfrac{1}{4}\partial^2 h_{\mu}^{\mu}(x)
 +\partial^{\mu}\zeta_{\mu}(x) &=& 1
\nonumber\\
 \tfrac{1}{2}\partial^{\nu} h_{\mu \nu}(x)
 +\tfrac{1}{2}\partial_{\mu} \Phi(x) 
 -\tfrac{1}{4}\partial_{\mu} h_{\nu}^{\nu}(x) &=&  \zeta_{\mu}(x).
\end{eqnarray}
If we eliminate $\zeta_{\mu}(x)$, these equations are equivalent to the
linearized gravity equations in the background of the brane 
\cite{Polchinski:jq}.
Note that the closed string tachyon shift is simply given by $A(p)=1$.

To generalize to arbitrary $p$ one must replace the R.H.S. of equation
(\ref{eq:lingrav}) with the appropriate source terms from the lower
dimensional boundary states.  These source terms will have
$\delta$-functions for the transverse dimensions.  Solving these
equations one can check that the long range behavior of the fields
described in section~\ref{sec:massless} is reproduced.

Since the terms in $\ket{\mathcal{B}}$
of higher weight are BRST-exact we can act on them with
\begin{equation} % changed 
-\frac{b_0+\tilde{b}_0}{L_0+\tilde{L}_0} 
\end{equation} % changed
to solve for the rest of
$\ket{\Phi}$.

We have thus shown that BRST invariance of the tadpole implies that
the region of integration $T<T_0$ should, for consistency, represent a
closed string background.  The fact that this region of integration
actually diverges, even when the solutions to the equations
(\ref{eq:lingrav}) are finite, arises from the specific prescription
which OSFT uses to define the inverse of the BRST operator in terms of
a closed string propagator.

One might ask why we cannot simply adopt this construction as a way of
defining the tadpole.  These are a number of problems with this idea.
First, there is an ambiguity in the choice of $\Phi$ under the shift
$\Phi \to \Phi + Q_B \delta \Phi$.  This implies that if we were to
use this scheme for one of the finite diagrams, we would, in general,
get a different answer than the answer we would get from using
analytic continuation.  There is also the problem that we have
introduced an unphysical parameter $T_0$.  One can check that this is
not much of a problem since a shift in $T_0$ can be accommodated by a
corresponding BRST exact shift of $\Phi$.  Perhaps the most serious
problem with this somewhat arbitrary division of moduli space is that
it would lead to a breakdown in unitarity since for $T<T_0$ one can
no longer consider the tadpole to be made up of an open string loop.

Finding a sensible way of systematically dealing with the divergent
massless closed string tadpoles encountered in this section seems to
be an important problem for string field theory.  Although these
divergences are only associated with D$p$-branes having codimension 3
or less, such branes are important to understand in string theory.
While it seems difficult to make sense of the theory in, for example,
a space-filling brane background, we know from the recent work on the
Sen conjectures that string field theory in such a background contains
within it other backgrounds corresponding to lower-dimensional
D-branes and empty space-time.  Since the theory is well defined in
these other backgrounds, it seems that the technical problems
encountered in the presence of high-dimensional D-branes should have
some robust technical solution which does not require a significant
modification of the theory.  On the other hand, it may be possible
that the theory does not have a well-defined perturbation series in
all backgrounds.  This question is particularly pressing for
superstring field theory, where we might expect the theory to be
completely well-defined with finite tadpoles for D$p$-branes with $p <
7$, while D$p$-branes with $p \geq 7$ would have similar divergences
to those discussed in this section.

%%%%%%%%%%%%%%%%%%%%%%%%%%%%%%%%%%%%%%%%%%%%%%%%%%%%%%%%%%%%%%%%%%%%%%%
    \subsection{Beyond one loop}
%%%%%%%%%%%%%%%%%%%%%%%%%%%%%%%%%%%%%%%%%%%%%%%%%%%%%%%%%%%%%%%%%%%%%%%
\label{sec:two-loops}
Having examined the divergences in the tadpole diagram we discuss the
situation for higher-loop diagrams.  Although we have no explicit
computations, we predict that there will be additional divergences of
a new kind.  For example at the two-loop level one can consider a
torus diagram with a hole.  If we put an open string state on the edge
of the hole we have a contribution to the two-loop tadpole diagram.
In the conformal frame natural to OSFT this diagram is pictured in
figure \ref{f:2loop}.
\FIGURE{
\epsfig{file=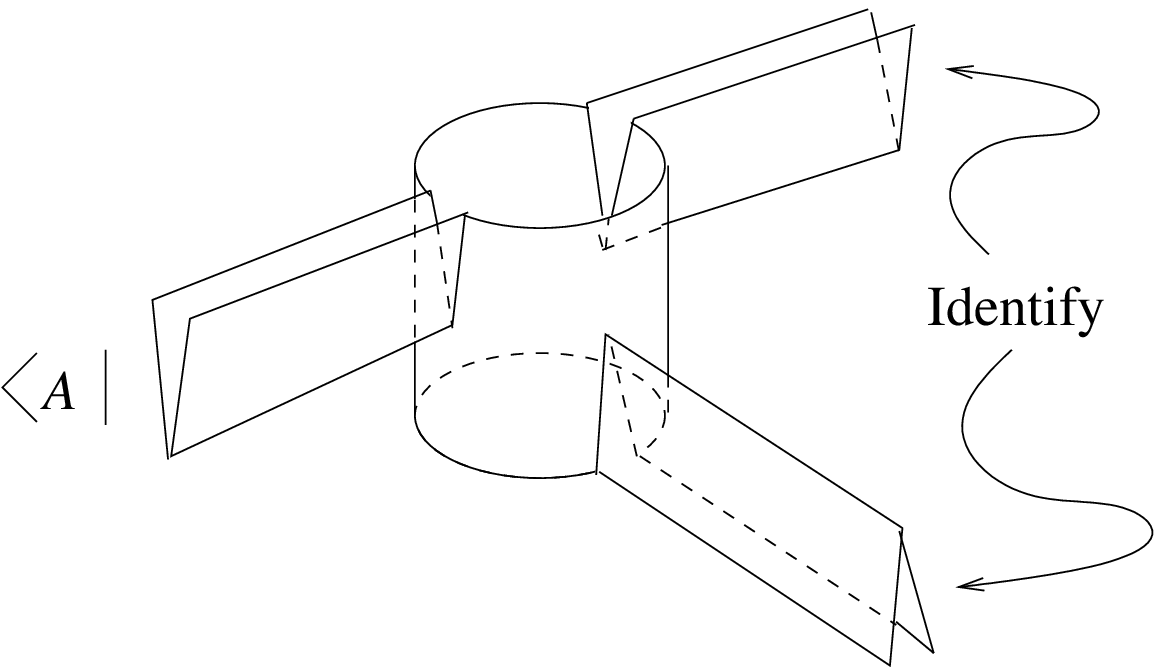,width=8.5cm}
\caption{\footnotesize
  The two-loop non-planar tadpole diagram contributing to
a breakdown in BRST invariance.
}
\label{f:2loop}
}
There is a region of moduli space in which the hole is separated from
the rest of the diagram by a long tube.  In this limit the diagram may
be viewed as an open string which turns into a closed string which
propagates over a long tube and then ends in a torus.  This long tube
leads to a divergence and associated BRST anomaly in world-sheet
string theory and represents a shift in the cosmological constant
\cite{Polchinski:jq}. The divergence occurs both because of the closed
string tachyon and the massless fields and thus cannot be treated
by a simple analytic continuation. 

We expect a similar divergence and BRST anomaly in OSFT.  Just as the
the one-loop open string tadpole could be thought of as a coupling between
the open strings and the closed string tadpole, here the two-loop open
string tadpole can be thought of as a coupling between the open
strings and the one-loop closed string tadpole.  The closed string
tadpole we studied in the context of the one-loop open string tadpole
arose because the brane was a source for the closed string fields.
The one-loop closed string tadpole occurs even in pure closed string
theory in the absence of the brane.  Thus we expect that the
divergence we find will be independent of the number of transverse
dimensions.

At higher loops still more diagrams with divergent long tubes appear.
For every divergent closed string tadpole we will find a corresponding
set of divergences in OSFT.  Without some general framework, each of
these divergences must be treated individually.  To some extent, this
same problem arises in ordinary world-sheet string theory, but one
might have hoped that in OSFT the problem would be more tractable.

%%%%%%%%%%%%%%%%%%%%%%%%%%%%%%%%%%%%%%%%%%%%%%%%%%%%%%%%%%%%%%%%%%%%%%%%%%%
%%%%%%%%%%%%%%%%%%%%%% Conclusion and further directions %%%%%%%%%%%%%%%%%%
%%%%%%%%%%%%%%%%%%%%%%%%%%%%%%%%%%%%%%%%%%%%%%%%%%%%%%%%%%%%%%%%%%%%%%%%%%%

\section{Discussion}

In this paper we have explicitly calculated the one-loop open string
field theory tadpole using both conformal field theory and oscillator
methods and found that the calculations agree.  There is a divergence
in the loop diagram due to the propagation of an open string tachyon
around the loop; this divergence is easily dealt with by analytic
continuation in an expansion of the diagram in the level of the open
string field in the loop using the oscillator approach to OSFT.  We
also find, however, that the one-loop tadpole diagram diagram diverges
due to the propagation of the closed string tachyon over a long closed
string tube.  This divergence arises only because we are working in
bosonic string theory, and can also in principle be treated by
analytic continuation.  In practice, we can perform this analytic
continuation for the one-loop diagram considered in this paper by
explicitly using our knowledge of the role of closed strings in this
diagram, but this is much more subtle than in the case of the open
string tachyon, as the closed strings are not explicitly included as
degrees of freedom in OSFT.  The analytic continuation we use does not
deal with the closed string tachyon divergence successfully for
low-codimension branes.  This procedure is not easily generalized to
more complicated diagrams, and we do not have any systematic way of
dealing with such divergences other than order-by-order in
perturbation theory.

Once we have treated the leading divergences in the one-loop tadpole
by hand, the tadpole becomes finite as long as we work in a D-brane
background with sufficiently many codimensions.  For D-branes of
dimension $p \geq 23$, we find divergences from the propagation of
massless modes over long distances.  For the D$25$-brane theory, this
divergence also contributes to a BRST anomaly.  It does not seem to be
possible to treat these divergences without stepping outside the
framework of OSFT.  Indeed, recent work indicates that even the
Fischler-Susskind mechanism for dealing with such divergences in the
world-sheet formalism may have previously unexpected subtleties
\cite{ams,McGreevy-Silverstein}.  We describe, however, a procedure
for using BRST invariance that reveals much of the physics hidden in
these divergent diagrams.  Achieving a better understanding of these
divergences seems to be an important unsolved problem for string field
theory.

We have found that the open string tadpole essentially arises from the
coupling between the closed string background and the open strings.
The closed string has a tadpole in the presence of a D-brane,
expressing the linearized massless fields produced by the D-brane
source.  The fact that this shift in the closed string background is
seen in the one-loop tadpole diagram considered here demonstrates that
OSFT actually captures the running of the background geometry, and
thus contains a highly nontrivial aspect of closed string physics.

At two loops, OSFT runs into further difficulties with divergences and
probably BRST anomalies as well.  We believe that these problems,
which are already well-known from perturbative string theory, doom any
attempt to construct a complete quantum theory from bosonic open
string field theory, without the introduction of some fundamental new
idea.  It is, of course, possible that some as-yet unknown mechanism
stabilizes the closed string tachyon and turns bosonic string theory
into a consistent theory.  But if this is the case, this mechanism is
no more apparent from the point of view of OSFT than from the
traditional perturbative approach to the theory.  In this sense, our
study of quantum effects in bosonic OSFT has not led to any surprising
new insights.

The positive results we have found here are, however, firstly that
OSFT does not have any new problems or complications in the definition
of the quantum theory which could not be predicted based on known
issues in the perturbative theory, second that the one-loop tadpole in
bosonic OSFT can be understood in a physically sensible fashion, and
third that the OSFT tadpole naturally contains information about the
shift in the closed string background due to the D-brane with respect
to which the open string theory is originally defined.  

While bosonic string theory has been a nice toy model with which to
explore a variety of phenomena in string field theory, such as tachyon
condensation, it remains unclear whether this theory is connected in
any fundamental way with supersymmetric string theory and M-theory,
and even whether this theory can be defined in a complete and
consistent fashion quantum mechanically.  In order to make further
progress in understanding how far string field theory can take us in
probing the fundamental nature of string theory, when we begin to
consider seriously the construction of a nonperturbative quantum
theory, it is probably necessary to work with a supersymmetric string
field theory.  The results of this paper seem to give positive support
to the hypothesis that supersymmetric open string field theory, if it
exists classically, may naturally extend to a sensible quantum theory.
Two possible candidates for SUSY OSFT are the Berkovits theory
\cite{Berkovits:1995ab} and the Witten theory
\cite{Witten:1986qs,Preitschopf:1989gp,Arefeva:1989cm}.  If one of
these theories can be shown to be completely consistent and to
correctly incorporate the Ramond as well as the NS sector of open
strings, it seems likely that an analogous calculation to the one in
this paper will, for D$p$-branes with $p < 7$, give rise to a finite
one-loop tadpole which encodes the linearized gravitational fields
from the D-brane source.  Such a tadpole should not suffer from any of
the divergences encountered in this paper.  In the supersymmetric
theory, there seems to be no reason why higher loop calculations
should not continue to incorporate the higher-order gravitational
effects of the D$p$-brane, so that the full theory should completely
encode the D$p$-brane geometry seen by closed strings.

There is clearly a significant amount of work remaining to be done to
substantiate this story, but if this picture can be realized
explicitly, it could open several exciting new directions for
progress.  This would give a new nonperturbative quantum definition of
string theory.  In this theory, if it is unitary, closed strings
should arise as composite states.  This would give a new more general
open string model exhibiting open-closed string duality, from which
the CFT description of strings in AdS space would arise as a special
case in the usual decoupling limit.  Because OSFT can be defined
nonperturbatively, it is possible to imagine using level truncation to
compute numerical approximations to finite nonperturbative quantities,
thus potentially accessing new nonperturbative features of string
theory.  Because, as we have seen in this paper, the closed string
background is encoded in the quantum open string diagrams, it is
possible that changes in the closed string background might be studied
completely in terms of OSFT, once a full definition of the quantum
theory exists.  At a more pragmatic level, if the methods of this
paper are applied to SUSY OSFT, it should be possible to study
directly the production of closed strings in dynamical open string
tachyon condensation \cite{Sen-rolling}.  As was shown in recent work
on this subject, a full physical understanding of the rolling tachyon
requires treating the back reaction of the radiated closed strings on
the rolling open string tachyon (see \cite{Leblond:2003db,Lambert:2003zr} for a
recent discussion and further references).  String field theory seems
like the natural context in which to study this process, and quantum
computations in OSFT like the one in this paper may, even in the
bosonic theory, shed light on this puzzle.

At a more technical level, we have seen in this paper that the two
different approaches to computing OSFT diagrams, the oscillator
approach and the CFT approach, give somewhat orthogonal information
about the structure of the theory.  The CFT approach has the advantage
of giving analytic expressions for amplitudes.  From these analytic
expressions, it is possible to take the limit where open string
modular parameters become small, which in some situations corresponds
to the limit where closed string physics plays an important role.  In
this paper, as in \cite{Freedman:fr}, it was possible to analytically
study the contribution from closed strings by expanding around this
limit.  On the other hand, the CFT approach is  only tractable for simple
diagrams.  For any diagram at genus $g > 1$, the CFT approach is not
easily applicable.  Furthermore, even for computing finite diagrams at
genus $g \leq 1$, the CFT approach gives a complicated integral
expression in terms of implicitly defined functions, which can only be
numerically approximated.  The oscillator approach, unlike the CFT
approach, can be applied to arbitrary diagrams of OSFT.  It is even
possible to imagine truncating OSFT at finite level and including a
momentum cutoff so that even nonperturbative quantities in OSFT can be
approximated by finite-dimensional integrals.  While closed strings
are not included exactly when OSFT is truncated at finite oscillator
level, as we found in this paper the effects of closed strings can be
clearly seen when the oscillator cutoff is sufficiently high; in the
level-truncated theory, the closed strings act more like resonances
than like asymptotic states.  Although we found here that the closed
string tachyon divergence made it difficult to extract other physical
effects in the oscillator calculation, in a theory without closed
string tachyons, like the superstring, it seems that the oscillator
method should be a viable approach for approximating any loop diagram.
Even for the bosonic theory, a general method for analytically
continuing or otherwise taming the tachyon divergence---such as a
Lorentzian world-sheet formulation---would allow us to extract useful
physics from the oscillator method.  Summarizing these observations,
it seems likely that in further developments, such as in a systematic
formulation of a quantum supersymmetric open string field theory, both
these approaches to computations will play useful roles.

%%%%%%%%%%%%%%%%%%%%%%%%%%%%%%%%%%%%%%%%%%%%%%%%%%%%%%%%%%%%%%%%%%%%%%%%%%%%
                         \section*{Acknowledgments}
%%%%%%%%%%%%%%%%%%%%%%%%%%%%%%%%%%%%%%%%%%%%%%%%%%%%%%%%%%%%%%%%%%%%%%%%%%%%

We would like to thank Erasmo Coletti, Bo Feng, Yang-Hui He, Hong Liu,
Hirosi Ooguri, Martin Schnabl, Ilya Sigalov, Eva Silverstein and
Barton Zwiebach for useful discussions; we would like to thank Barton
Zwiebach in particular for comments on the draft.  Thanks to Eva
Silverstein for sharing with us a preliminary version of
\cite{McGreevy-Silverstein}.  The numerical computations in this work
were done using {\em Mathematica}.  This work was supported by the DOE
through contract $\#$DE-FC02-94ER40818.

\appendix

%%%%%%%%%%%%%%%%%%%%%%%%%%%%%%%%%%%%%%%%%%%%%%%%%%%%%%%%%%%%%%%%%%%%%%%%%%%%%
%%%%%%%%%%%%%%%%%%%%%%%%%%%%%%%%%%%%%%%%%%%%%%%%%%%%%%%%%%%%%%%%%%%%%%%%%%%%%
              \section{The BRST anomaly in the D$25$-brane theory}
%%%%%%%%%%%%%%%%%%%%%%%%%%%%%%%%%%%%%%%%%%%%%%%%%%%%%%%%%%%%%%%%%%%%%%%%%%%%%
%%%%%%%%%%%%%%%%%%%%%%%%%%%%%%%%%%%%%%%%%%%%%%%%%%%%%%%%%%%%%%%%%%%%%%%%%%%%%
\label{s:BRSTanomaly}
We now study the BRST invariance of the tadpole in the D$25$-brane theory.
In any BRST-quantized string theory, a basic condition for 
gauge invariance is that BRST-exact states decouple from on-shell
diagrams.  For the tadpole diagram, this condition is simply that
$Q_B$ should annihilate the tadpole.
  
In sections~\ref{s:BRSTanomalyCFT} and \ref{s:BRSTanomalyOsc} we use
conformal field theory and oscillator methods to compute the action of
the BRST operator on the tadpole and find that the tadpole is not BRST
closed.  This breakdown of gauge invariance is familiar from ordinary
string perturbation theory in the context of the Fischler-Susskind
mechanism \cite{Fischler:ci,Das:dy,Fischler:1987gz,Polchinski:jq,Green:pf}.
Our results differ from previous work on BRST invariance in OSFT.  In
\cite{Bogojevic:1987ma} it was argued that $Q_B |\mathcal{T}\rangle$
should vanish, while in \cite{Thorn:1988hm} it was suggested that it
might be possible to avoid a breakdown in BRST invariance in
Feynman-Siegel gauge.  Our results indicate that in fact $Q_B
|\mathcal{T}\rangle\neq 0$, even in Feynman-Siegel gauge.

%%%%%%%%%%%%%%%%%%%%%%%%%%%%%%%%%%%%%%%%%%%%%%%%%%%%%%%%%%%%%%%%%%%%%%%%%%%%
    \subsection{The BRST anomaly in the conformal field theory method}
%%%%%%%%%%%%%%%%%%%%%%%%%%%%%%%%%%%%%%%%%%%%%%%%%%%%%%%%%%%%%%%%%%%%%%%%%%%%

\label{s:BRSTanomalyCFT}
We now study the BRST invariance of the OSFT tadpole
for the D$25$-brane theory.  On the grounds
of gauge invariance, we expect that $Q_B |\mathcal{T}\rangle = 0$. To
check this identity we first need to render $|\mathcal{T}\rangle$ a
finite state by imposing a cutoff.

We choose to simply cut off the integrals over the modular parameter
$T$ so that it never gets smaller that some minimum value $T_0$.  We
can then evaluate $Q_B|\mathcal{T}\rangle$ using equation
(\ref{eq:Qidentity1}).  This gives
\begin{equation} 
  \langle \mathcal{T}| Q_B =  
\langle V_3 | e^{-T_0 L_0} |\tilde{V}_2\rangle. \label{eq:QT1}
\end{equation}
We can then examine the behavior as $T_0\to 0$.  Conveniently, we have
already done all the work for this calculation in section
\ref{S:ConformalAnalysis}.  Notice that the only difference between
equation~(\ref{eq:QT1}) and equation~(\ref{eq:firststep}), is that $T$
is evaluated at a specific point, $T_0$, and that the $b_0$ is
missing.  In the conformal field theory language we can accommodate
this change by simply fixing the length of the internal propagator to
be $T_0$ and dropping the insertion of $b_0$.  This gives
\begin{equation} % changed
  Q_B|\mathcal{T}\rangle = U_{h(z)\circ z(v)}^{\dagger} \bigg|_{T=T_0}
  (h(z)\circ z(w)\bigg|_{T=T_0} \circ \mathcal{B})|0\rangle.
\end{equation} % changed
To map the resulting diagram to the disk, we can just use the same map $z(w)$
given in equation~(\ref{eq:zofw}).  
Since there is no $b_0$ to worry
about, the divergence structure is actually much simpler.  Since
$U_{h(z)\circ z(v)}^{\dagger}$ is well behaved as $T_0 \to 0$ and
$h(z)$ is independent of $T_0$, we only need to 
consider $z(w)\circ \mathcal{B}$.
Using equation~(\ref{eq:BoundaryState}) we get that the boundary state is
mapped to
\begin{eqnarray}
  z\circ |\mathcal{B}\rangle 
    &=&\left(\frac{1+W_0^2}{k_0 W_0^2} \right)^2 c_{1}(c_0+\tilde{c}_0)
\tilde{c}_1|0\rangle\nonumber \\
& & - \alpha_{-1} \cdot \tilde{\alpha}_{-1} \tilde{c}_{1} (c_0+\tilde{c}_0) c_1
|0\rangle
-2 (c_1 c_{-1}+\tilde{c}_{-1}\tilde{c}_1) (c_0+\tilde{c}_0)
|0\rangle\nonumber\\
& & + \text{terms that vanish as }W_0\to0\label{eq:zonB},
\label{eq:anomaly}
\end{eqnarray}
where $W_0$ is $T_0/\pi$ and $k_0$ is $k(W_0)$.
Clearly as $T_0 \to 0$ this expression is non-vanishing.

For theories on D$p$ branes with $p<25$ we will still find an 
anomaly for the terms coming from the closed string tachyon but
we no longer find an anomaly in the massless sector.
For example, the graviton/dilaton term in (\ref{eq:anomaly})
in the D$p$-brane theory becomes
\begin{equation} % changed
\int d^{25-p}q_{\bot}\,
   \left(\frac{1+W_0^2}{k_0 W_0^2} \right)^{q_{\bot}^2}
\alpha_{-1} \cdot \tilde{\alpha}_{-1} \tilde{c}_{1} (c_0+\tilde{c}_0) c_1
|q_{\bot}\rangle.
\end{equation} % changed
Mapping to the upper-half plane this  becomes
\begin{equation} % changed
  \int d^{25-p}q_{\bot} \, (2)^{q_{\bot}} 
\left(\frac{1+W_0^2}{k_0 W_0^2} \right)^{q_{\bot}^2}
 :c\partial X(i)\cdot \tilde{c} \bar{\partial} X(i) e^{i q_{\bot}\cdot X(i)}:.
\end{equation} % changed
Now consider evaluating the correlation between this operator
and the external state.  When  $e^{i q_{\bot}\cdot X(i)}$ contracts
with itself we pick up a factor of $(2)^{-q_{\bot}}$.  The operator
$e^{i q_{\bot}\cdot X(i)}$ can also contract with other $X$'s but these
just pick up factors of $q_{\bot}^2$.  Such terms will go to zero 
as $W \to 0$.  Thus the largest contribution comes from the
momentum integral
\begin{equation} % changed
  \int d^{25-p}q_{\bot} \left(\frac{1+W_0^2}{k_0 W_0^2} \right)^{q_{\bot}^2}
\propto 
\left[-\log\left(\frac{1+W_0^2}{k_0 W_0^2}\right)\right]^{\frac{1}{2}(p-25)},
\end{equation} % changed
which vanishes as $W_0 \to 0$.

%%%%%%%%%%%%%%%%%%%%%%%%%%%%%%%%%%%%%%%%%%%%%%%%%%%%%%%%%%%%%%%%%%%%%%
         \subsection{The BRST anomaly in level truncation}
%%%%%%%%%%%%%%%%%%%%%%%%%%%%%%%%%%%%%%%%%%%%%%%%%%%%%%%%%%%%%%%%%%%%%%

\label{s:BRSTanomalyOsc}

It is also possible to demonstrate the BRST anomaly in level
truncation.  It is easiest to frame this discussion in terms of
truncation on field level rather than oscillator level.  Let us begin
by considering a level expansion of the regulated equation
(\ref{eq:Qidentity1})
\begin{equation} % changed
  Q_B \int_{T_0}^\infty dT
 \langle \widetilde{V}_{2}| b_0e^{-T L_0} | V_3 \rangle 
 =  \langle\widetilde{V}_{2}| e^{-T_0 L_0}|V_3 \rangle
= \sum_{n}^{} c_n (T_0) e^{(1-n)T_0},
\label{eq:qt-regulated}
\end{equation} % changed
where the term associated with the coefficient $c_n$ arises from
fields in the loop of level $n$.  We can deal with the open string
tachyon divergence through analytic continuation.  The coefficients
$c_n (T_0)$ are finite for all values of $n> 1, T_0 > 0$, and can be
directly calculated by computing the contribution to either the first
or the second expression in (\ref{eq:qt-regulated}) at each level.
Direct computation at low levels confirms that both ways of computing
$c_n (T_0)$ give the same result.  Truncating the theory at finite
field level reduces the summation in (\ref{eq:qt-regulated}) to a
finite sum, from which we can take the $T_0 \rightarrow 0$ limit in
the level-truncated theory.  Thus, we see that (\ref{eq:QT1}) holds in
level truncation, even as $T_0 \rightarrow 0$, so that level
truncation naturally regulates the divergences in this equation
arising from closed strings.

To show that (\ref{eq:qt-regulated}) does not vanish in the limit
$T_0\rightarrow 0$,
let us examine the action of $Q_B$ on the level 2
sector of $\ket{{\mathcal T}}$.  This calculation is straightforward,
as $Q_B$ conserves level.  The fields appearing in
$\ket{{\mathcal T}}$ at level 2 are 
\be
\ket{{\mathcal T}^{(2)}} = \left( \beta_{\mu\nu} a^{\dag\mu}_1 a^{\dag\nu}_1 
    + \gamma \cdag_2 b_0 + \delta \cdag_1 \bdag_1 \right) c_0 \ket{\hat{0}} .
\ee
Upon acting with $Q_B$ we find
\be
Q_B \ket{{\mathcal T}^{(2)}} = \left(\beta^{\mu}_{\mu} - \gamma - 3
\delta \right) \cdag_2 c_0 \ket{\hat{0}}.
\ee
The condition that $ \ket{{\mathcal T}^{(2)}} $ be $Q_B$-closed thus 
reduces to the 
condition 
\be
\label{eq:qcl}
\beta^{\mu}_{\mu} - \gamma - 3\delta =0
\ee
 on the coefficients of the component states.

The coefficients $\beta_{\mu \nu}$, $\gamma$ and $\delta$ are given by
\bea \label{eq:betaeq}
\beta_{\mu\nu} &=& \eta_{\mu\nu} \int_0^{\infty} dT\, S_0(T) 
(-\frac{1}{2} M_{11}(T)), \nonumber\\
\gamma &=& \int_0^\infty dT \, S_0(T)(-R_{20} (T)), \nonumber \\
\delta &=& \int_0^\infty dT \, S_0(T)(-R_{11}(T)).
\eea
As discussed above,
level truncation serves to
render the integrals in equation (\ref{eq:betaeq}) finite, 
so that they are calculable numerically.  
The integrands become very sharply peaked near
$T=0$, so numerical analysis becomes difficult for small $T$.
Nevertheless, it is clear that effects near $T=0$ give rise to a
violation of \er{qcl}.  At successively higher levels, the weight
factor $S_0$ becomes more and more sharply peaked at $T=0$, and the
integral is well approximated by its endpoint value $ -\frac{1}{2}
M_{11}(0) S_0(0) $.  As we have demonstrated above, 
$M_{11}(0) = R_{11}(0) =-1$,
while $R_{20} (0) = 0$; these values manifestly fail to satisfy
\er{qcl}.  Moreover, as we remove the regularization by sending the
level to infinity, the violation becomes infinite.

We have thus seen in the operator formalism the presence of a
breakdown in BRST invariance which can be directly ascribed to the
presence of short-distance divergences, interpretable in terms of the
closed string tachyon.

%%%%%%%%%%%%%%%%%%%%%%%%%%%%%%%%%%%%%%%%%%%%%%%%%%%%%%%%%%%%%%%%%%%%%%%%%%%
                   \section{The infinite level limit}
%%%%%%%%%%%%%%%%%%%%%%%%%%%%%%%%%%%%%%%%%%%%%%%%%%%%%%%%%%%%%%%%%%%%%%%%%%%

\label{A:InfiniteLevel}
In this section we look at some of the subtleties of the infinite
level limit of the matrices in the oscillator calculations.
In particular, we are interested in comparing how 
level-truncated calculations
near $T=0$  compare with what we expect 
if we take the level to infinity.

%Since in both level truncation and in the infinite level limit,
%the matrices $R_{nm}$ and $M_{nm}$ limit to $C_{nm}$ as $T\to0$,
%it is more interesting to look first at the $\mathcal{O}(T)$ correction
%to $R_{nm}$ and $M_{nm}$ around $T=0$.

For definiteness, we work in the ghost sector.  The calculation in the
matter sector is completely analogous.  We begin by expanding the
expression for $R_{nm}$, given in (\ref{eq:ghostmatrix}), around $T=0$
in level truncation.  The only tricky part of the calculation is the
small $T$ expansion of
\begin{equation}\label{eq:MatrixInverse}
  \frac{1}{1-S\tX}\,.
\end{equation}
It turns out, however, that it is sufficient to consider the expansion
of (\ref{eq:MatrixInverse}) acting on 
$S \left( \begin{smallmatrix} X^{21} \\ X^{12} \end{smallmatrix} \right)$,
which gives
\begin{equation} \label{eq:ExpansionIdentity}
    \frac{1}{1 - S\tX} S \left( \begin{matrix} 
           X^{21} \\ X^{12} \end{matrix} \right)
  =
\left( \begin{matrix} 
           \mathbf{1} \\ \mathbf{1} \end{matrix} \right)-
T\,\left(\frac{1}{1 - S \tX} \Bigg|_{T=0}\right)
\left(
\begin{matrix}
  \{G, C X^{21}+C X^{11}\} \\
  \{G, C X^{12}+C X^{11}\}
\end{matrix}
\right)
+\mathcal{O}(T^2),
\end{equation}
where we have defined the matrix
\begin{equation} % changed
  G_{mn} = \frac{m}{2} \delta_{mn}.
\end{equation} % changed
Using these formulas, one can work out the expansion
\begin{equation}\label{eq:FirstOrderG}
  R_{mn}  = C_{mn} - 2 T \{G,C\}_{mn}+\mathcal{O}(T^2) 
= C_{mn} -  2 m T \, C_{mn} +\mathcal{O}(T^2) .
\end{equation}
In fact since the only identity needed to prove this relation is
the sum rule $X^{11}+X^{12}+X^{21}=C$, equation(\ref{eq:FirstOrderG})
holds exactly at every fixed level.

We now make the following claim: In spite of the fact that the leading
order correction is the same at every finite level, if we did not truncate
the matrices at all, we would get instead
\begin{equation}\label{eq:RExpansion}
  R_{mn}  = C_{mn} -  m T \, C_{mn} +\mathcal{O}(T^2).
\end{equation}
To see how this happens, consider a single matrix element $R_{11}$.
We expect from level-truncated analysis that $R_{11}(T) = -1 +
2\,T+\mathcal{O}(T^2)$.  In figure~\ref{f:DerivativeofR11}, we plot
$-\frac{d}{dT} R_{11}(T)$ near zero at various levels.  We see that
if we go extremely close to zero,  $-\frac{d}{dT} R_{11}(T)$ always
eventually approaches $-2$ as predicted by (\ref{eq:FirstOrderG}).

\FIGURE{
\epsfig{file=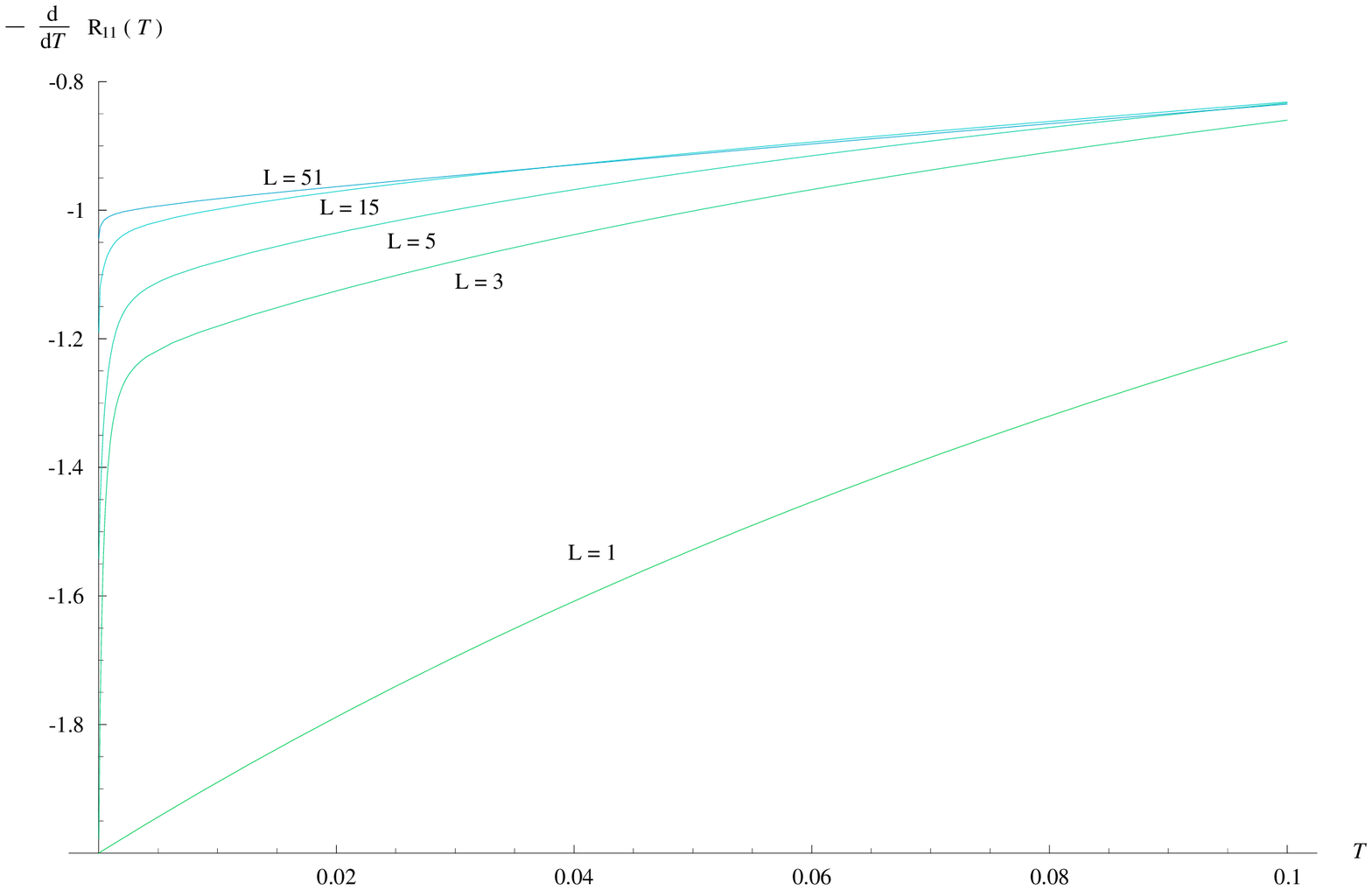,width=14cm}
\caption{\footnotesize
  Plot of $-\frac{d}{dT} R_{11}(T)$ at various levels, $L$.  We see that
near $T=0$, $-\frac{d}{dT} R_{11}(T)$ always approaches $-2$, but
at high levels comes arbitrarily close to hitting $-1$.
}
\label{f:DerivativeofR11}
}
As the level is increased, however, the region where $-\frac{d}{dT}
R_{11}(T)$ is close to $-2$ becomes arbitrarily small.  This falloff
near $T=0$ is analogous to the falloff we found when studying the
tachyon divergence in section \ref{s:SmallT}.  If we imagine taking
the level to infinity and then taking $T \to 0$, we can ignore the falloff
region and just take
\begin{equation} % changed
  \lim_{T\to 0}\left( -\frac{d}{dT} R_{11}(T)\right) = -1,
\end{equation} % changed
as predicted by the conformal field theory method.  

We can now ask: what went wrong with the calculation of equation 
(\ref{eq:FirstOrderG})?  To analyze this question, it is convenient
to work with the matrices $M^{ij} = C X^{ij}$.  In level truncation
these matrices satisfy one identity
\begin{equation} % changed
  M^{11}+M^{12}+M^{21} = 1.
\end{equation} % changed
If we do not truncate the matrices, they also satisfy \cite{spectroscopy}
\begin{equation} % changed
  (M^{11})^2 +(M^{12})^2 +(M^{21})^2  =1,
\end{equation} % changed
and commute.  These identities imply that at $T=0$, the matrix 
$1-S\tilde{X}$ has a non-vanishing kernel.  Thus the matrix
\begin{equation} % changed
  \frac{1}{1-S\tX} \Bigg|_{T=0}
\end{equation} % changed
is ill-defined.  This invalidates the expansion
(\ref{eq:ExpansionIdentity}). Even worse, the expression (\ref{eq:matrix-id}) 
that we used
to see that $R_{nm} \to C$ as $T \to 0$
is not defined for the untruncated matrices.  In fact one can check
that while it is true that
\begin{equation} % changed
  \lim_{\text{Level} \to \infty}\left\{
  \lim_{T\to 0} 
  \left(
    \begin{matrix}
      \hX^{12} & \hX^{21}
    \end{matrix}
  \right)
  \frac{1}{1-S \tX}\right\}
  =
  \left( 
    \begin{matrix}
       C & C
    \end{matrix}
   \right),
\end{equation} % changed
one finds that
\begin{equation} % changed
  \lim_{T\to 0} \left\{\lim_{\text{Level} \to \infty}
  \left(
    \begin{matrix}
      \hX^{12} & \hX^{21}
    \end{matrix}
  \right)
  \frac{1}{1-S \tX} \right\}
  \ne 
  \left( 
    \begin{matrix}
       C & C
    \end{matrix}
   \right)
\end{equation} % changed
in a similar manner to what we found for $-\frac{d}{dT} R_{11}(T)$.

In spite of this,  we find that $R_{nm} \to C$ in the
small $T$ limit unambiguously in numerical tests. 
At this time, we do not have enough control over the matrices
$\widetilde{M}$ and $\widetilde{X}$ to give an infinite level
proof of the identity \ref{eq:RExpansion}.

%%%%%%%%%%%%%%%%%%%%%%%%%%%%%%%%%%%%%%%%%%%%%%%%%%%%%%%%%%%%%%%%%%%%%%%%%%%%%
%%%%%%%%%%%%%%%%%%%%%%%%%%%%  Bibliography %%%%%%%%%%%%%%%%%%%%%%%%%%%%%%%%%%
%%%%%%%%%%%%%%%%%%%%%%%%%%%%%%%%%%%%%%%%%%%%%%%%%%%%%%%%%%%%%%%%%%%%%%%%%%%%%

\bibliographystyle{plain}
%\bibliography{papers}

\begin{thebibliography}{10}

%\cite{Witten:1985cc}
\bibitem{Witten:1985cc}
E.~Witten,
``Noncommutative Geometry And String Field Theory,''
Nucl.\ Phys.\ B {\bf 268}, 253 (1986).

%\cite{Sen:1999xm}
\bibitem{Sen:1999xm}
A.~Sen,
``Universality of the tachyon potential,''
JHEP {\bf 9912}, 027 (1999)
{\tt hep-th/9911116}.

%\cite{Ohmori:2001am}
\bibitem{Ohmori:2001am}
K.~Ohmori,
``A review on tachyon condensation in open string field theories,''
{\tt hep-th/0102085}.
%%CITATION = HEP-TH 0102085;%%

%\cite{DeSmet:2001af}
\bibitem{DeSmet:2001af}
P.~J.~De Smet,
``Tachyon condensation: Calculations in string field theory,''
{\tt hep-th/0109182}.
%%CITATION = HEP-TH 0109182;%%

%\cite{Zwiebach:nj}
\bibitem{Zwiebach:nj}
B.~Zwiebach,
``Is The String Field Big Enough?,''
Fortsch.\ Phys.\  {\bf 49}, 387 (2001).
%%CITATION = FPYKA,49,387;%%

%\cite{Taylor:2002uv}
\bibitem{Taylor:2002uv}
W.~Taylor,
``Lectures on D-branes, tachyon condensation, and string field theory,''
{\tt hep-th/0301094}.
%%CITATION = HEP-TH 0301094;%%

\bibitem{quantum}
% lookup 9905043
% lookup 0104164
% lookup 0105098
% lookup 0105227
% lookup 0105238
% lookup 0105312
% lookup 0109032
% lookup 0109187
%\bibitem{Asakawa:1999nc}
T.~Asakawa, T.~Kugo and T.~Takahashi,
``One-loop tachyon amplitude in unoriented open-closed string field  theory,''
Prog.\ Theor.\ Phys.\  {\bf 102}, 427 (1999)
{\tt hep-th/9905043};
%\bibitem{Alishahiha:2001tg}
M.~Alishahiha,
``One-loop correction of the tachyon action in boundary superstring field  
theory,''
Phys.\ Lett.\ B {\bf 510}, 285 (2001)
{\tt hep-th/0104164};
%\bibitem{Bardakci:2001ck}
K.~Bardakci and A.~Konechny,
``Tachyon condensation in boundary string field theory at one loop,''
{\tt hep-th/0105098};
%\bibitem{Craps:2001jp}
B.~Craps, P.~Kraus and F.~Larsen,
``Loop corrected tachyon condensation,''
JHEP {\bf 0106}, 062 (2001)
{\tt hep-th/0105227};
%\bibitem{Arutyunov:2001nz}
G.~Arutyunov, A.~Pankiewicz and B.~.~Stefanski,
``Boundary superstring field theory annulus partition function in the  
presence of tachyons,''
JHEP {\bf 0106}, 049 (2001)
{\tt hep-th/0105238};
%\bibitem{Minahan:2001pd}
J.~A.~Minahan,
``Quantum corrections in p-adic string theory,''
{\tt hep-th/0105312};
%\bibitem{Lee:2001jc}
T.~Lee, K.~S.~Viswanathan and Y.~Yang,
``Boundary string field theory at one-loop,''
J.\ Korean Phys.\ Soc.\  {\bf 42}, 34 (2003)
{\tt hep-th/0109032};
%\bibitem{Andreev:2001ak}
O.~Andreev and T.~Ott,
``On one-loop approximation to tachyon potentials,''
Nucl.\ Phys.\ B {\bf 627}, 330 (2002)
{\tt hep-th/0109187}.

%\cite{Thorn:1988hm}
\bibitem{Thorn:1988hm}
C.~B.~Thorn,
``String Field Theory,''
Phys.\ Rept.\  {\bf 175}, 1 (1989).
%%CITATION = PRPLC,175,1;%%

\bibitem{NGS}
D.\ J.\ Gross, A.\ Neveu, J.\  Scherk and J.\ H.\ Schwarz,
``Renormalization and unitarity in the dual resonance model,''
\PR {\bf D2}  (1970) 697.

\bibitem{Lovelace}
C.\ Lovelace, ``Pomeron form factors and dual Regge cuts,''
\PL {\bf B34} (1971) 500.

%\cite{Freedman:fr}
\bibitem{Freedman:fr}
D.~Z.~Freedman, S.~B.~Giddings, J.~A.~Shapiro and C.~B.~Thorn,
``The Nonplanar One Loop Amplitude In Witten's String Field Theory,''
Nucl.\ Phys.\ B {\bf 298}, 253 (1988).

%\cite{Shapiro:gq}
\bibitem{Shapiro:gq}
J.~A.~Shapiro and C.~B.~Thorn,
``BRST-Invariant Transitions Between Closed And Open Strings,''
Phys.\ Rev.\ D {\bf 36}, 432 (1987).

%\cite{Shapiro:ac}
\bibitem{Shapiro:ac}
J.~A.~Shapiro and C.~B.~Thorn,
``Closed String - Open String Transitions And Witten's String Field Theory,''
Phys.\ Lett.\ B {\bf 194}, 43 (1987).

%\cite{Zwiebach:1992bw}
\bibitem{Zwiebach:1992bw}
B.~Zwiebach,
``Interpolating string field theories,''
Mod.\ Phys.\ Lett.\ A {\bf 7}, 1079 (1992)
{\tt hep-th/9202015}.
%%CITATION = HEP-TH 9202015;%%

%\cite{Hashimoto:2001sm}
\bibitem{Hashimoto:2001sm}
A.~Hashimoto and N.~Itzhaki,
``Observables of string field theory,''
JHEP {\bf 0201}, 028 (2002)
{\tt hep-th/0111092}.
%%CITATION = HEP-TH 0111092;%%

%\cite{Gaiotto:2001ji}
\bibitem{Gaiotto:2001ji}
D.~Gaiotto, L.~Rastelli, A.~Sen and B.~Zwiebach,
``Ghost structure and closed strings in vacuum string field theory,''
{\tt hep-th/0111129}.
%%CITATION = HEP-TH 0111129;%%



\bibitem{a-Garousi}
% lookup 0201249
M.~Alishahiha and M.~R.~Garousi,
``Gauge invariant operators and closed string scattering in open string  field 
theory,''
Phys.\ Lett.\ B {\bf 536}, 129 (2002),
{\tt hep-th/0201249};
%\bibitem{Garousi:2002jt}
M.~R.~Garousi and G.~R.~Maktabdaran,
``Excited D-brane decay in cubic string 
field theory and in bosonic string  theory,''
Nucl.\ Phys.\ B {\bf 651}, 26 (2003)
{\tt hep-th/0210139}.
%%CITATION = HEP-TH 0210139;%%

\bibitem{closed}
%\bibitem{Strominger-closed}
A.\ Strominger,
``Closed strings in open string field theory,''
\PRL {\bf 58} 629 (1987);
M.\ Srednicki and R.\ Woodard,
``Closed from open strings in Witten's theory,''
Nucl.\  Phys.\  {\bf  B293}, 612 (1987),
{\tt hep-th/0201249};
% lookup 0005031
% lookup 0009061
% lookup 0010240
J.~A.~Harvey, P.~Kraus, F.~Larsen and E.~J.~Martinec,
``D-branes and strings as non-commutative solitons,''
JHEP {\bf 0007}, 042 (2000);
{\tt hep-th/0005031}
%\bibitem{Gibbons:2000hf}
G.~W.~Gibbons, K.~Hori and P.~Yi,
``String fluid from unstable D-branes,''
Nucl.\ Phys.\ B {\bf 596}, 136 (2001)
{\tt hep-th/0009061};
%\bibitem{Sen:2000kd}
A.~Sen,
``Fundamental strings in open string theory at the tachyonic vacuum,''
J.\ Math.\ Phys.\  {\bf 42}, 2844 (2001)
{\tt hep-th/0010240};
%\bibitem{Gerasimov-Shatashvili}
% lookup 0011009
A.~A.~Gerasimov and S.~L.~Shatashvili,
``Stringy Higgs mechanism and the fate of open strings,''
JHEP {\bf 0101}, 019 (2001),
{\tt hep-th/0011009};
% lookup 0103056
G.~Chalmers,
``Open string decoupling and tachyon condensation,''
JHEP {\bf 0106}, 012 (2001)
{\tt hep-th/0103056};
% lookup 0105076
S.~L.~Shatashvili,
``On field theory of open strings, tachyon condensation and closed  strings,''
{\tt hep-th/0105076};
%\cite{Moore:2001fg}
%\bibitem{Moore:2001fg}
G.~Moore and W.~Taylor,
``The singular geometry of the sliver,''
JHEP {\bf 0201}, 004 (2002)
{\tt hep-th/0111069};
%%CITATION = HEP-TH 0111069;%%
D.~Gaiotto, N.~Itzhaki and L.~Rastelli,
``Closed strings as imaginary D-branes,''
{\tt hep-th/0304192}.


%\cite{Maldacena:1997re}
\bibitem{Maldacena:1997re}
J.~M.~Maldacena,
``The large N limit of superconformal field theories and supergravity,''
Adv.\ Theor.\ Math.\ Phys.\  {\bf 2}, 231 (1998)
[Int.\ J.\ Theor.\ Phys.\  {\bf 38}, 1113 (1999)]
{\tt hep-th/9711200}.
%%CITATION = HEP-TH 9711200;%%

%\cite{Aharony:1999ti}
\bibitem{Aharony:1999ti}
O.~Aharony, S.~S.~Gubser, J.~M.~Maldacena, H.~Ooguri and Y.~Oz,
``Large N field theories, string theory and gravity,''
Phys.\ Rept.\  {\bf 323}, 183 (2000)
{\tt hep-th/9905111}.
%%CITATION = HEP-TH 9905111;%%

%\cite{D'Hoker:2002aw}
\bibitem{D'Hoker:2002aw}
E.~D'Hoker and D.~Z.~Freedman,
``Supersymmetric gauge theories and the AdS/CFT correspondence,''
{\tt hep-th/0201253}.
%%CITATION = HEP-TH 0201253;%%

\bibitem{Giddings}
S.\ Giddings, ``The Veneziano amplitude from interacting string field
theory,'' \NP {\bf B278}
242 (1986).

\bibitem{Sloan}
J.\ H.\ Sloan, `` The scattering amplitude for four off-shell tachyons
from functional integrals,'' \NP {\bf B302}
349 (1988).

\bibitem{Samuel-off}
S.\ Samuel, ``Covariant off-shell string amplitudes,'' \NP {\bf B308}
285 (1988).

%\cite{Taylor:2002bq}
\bibitem{Taylor:2002bq}
W.~Taylor,
``Perturbative diagrams in string field theory,''
{\tt hep-th/0207132}.
%%CITATION = HEP-TH 0207132;%%


%\cite{Polchinski:jq}
\bibitem{Polchinski:jq}
J.~Polchinski,
``Factorization Of Bosonic String Amplitudes,''
Nucl.\ Phys.\ B {\bf 307}, 61 (1988).
%%CITATION = NUPHA,B307,61;%%

\bibitem{DiVecchia}
% lookup 9707068
P.~Di Vecchia, M.~Frau, I.~Pesando, S.~Sciuto, A.~Lerda and R.~Russo,
``Classical p-branes from boundary state,''
Nucl.\ Phys.\ B {\bf 507}, 259 (1997)
{\tt hep-th/9707068}.



%\cite{Fischler:ci}
\bibitem{Fischler:ci}
W.~Fischler and L.~Susskind,
``Dilaton Tadpoles, String Condensates And Scale Invariance,''
Phys.\ Lett.\ B {\bf 171}, 383 (1986).
%%CITATION = PHLTA,B171,383;%%

%\cite{Das:dy}
\bibitem{Das:dy}
S.~R.~Das and S.~J.~Rey,
``Dilaton Condensates And Loop Effects In Open And Closed Bosonic Strings,''
Phys.\ Lett.\ B {\bf 186}, 328 (1987).
%%CITATION = PHLTA,B186,328;%%

%\cite{Fischler:1987gz}
\bibitem{Fischler:1987gz}
W.~Fischler, I.~R.~Klebanov and L.~Susskind,
``String Loop Divergences And Effective Lagrangians,''
Nucl.\ Phys.\ B {\bf 306}, 271 (1988).
%%CITATION = NUPHA,B306,271;%%


%\cite{Bars:2002qt}
\bibitem{Bars:2002qt}
I.~Bars, I.~Kishimoto and Y.~Matsuo,
``String amplitudes from Moyal string field theory,''
{\tt hep-th/0211131}.
%%CITATION = HEP-TH 0211131;%%

\bibitem{lpp}
A.\ Leclair, M.\ E.\ Peskin and C.\ R.\ Preitschopf, ``String field theory on
  the conformal plane (I)'' \NP {\bf B317} (1989) 411-463.

\bibitem{Gaberdiel-Zwiebach}
M.\ R.\ Gaberdiel and B.\ Zwiebach, ``Tensor constructions of open string
  theories 1., 2.,'' \NP {\bf B505} (1997) 569, {\tt hep-th/9705038}; \PL {\bf
  B410} (1997) 151, {\tt hep-th/9707051}.

%\cite{Gross:1986ia}
\bibitem{Gross:1986ia}
D.~J.~Gross and A.~Jevicki,
``Operator Formulation Of Interacting String Field Theory,''
Nucl.\ Phys.\ B {\bf 283}, 1 (1987).
%%CITATION = NUPHA,B283,1;%%

%\cite{Gross:1986fk}
\bibitem{Gross:1986fk}
D.~J.~Gross and A.~Jevicki,
``Operator Formulation Of Interacting String Field Theory. 2,''
Nucl.\ Phys.\ B {\bf 287}, 225 (1987).
%%CITATION = NUPHA,B287,225;%%


\bibitem{cst}
E.\ Cremmer, A.\ Schwimmer and C.\ Thorn, ``The vertex function in Witten's
  formulation of string field theory'' \PL {\bf B179} 57 (1986).

\bibitem{Samuel}
S.\ Samuel, ``The physical and ghost vertices in Witten's string field
  theory,'' \PL {\bf B181} 255 (1986).


\bibitem{Ohta}
N.\ Ohta, ``Covariant interacting string field theory in the Fock
space representation,''
\PR {\bf D34}   (1986), 3785;
\PR {\bf D35}   (1987), 2627 (E).

\bibitem{rsz} L.~Rastelli, A.~Sen, and B.~Zwiebach,  ``Classical Solutions in
  String Field Theory Around the Tachyon Vacuum'',
  Adv. Theor. Math. Phys. 5:393-428 (2002) {\tt hep-th/0102112}.

%\cite{Giddings:1986bp}
\bibitem{Giddings:1986bp}
S.~B.~Giddings and E.~J.~Martinec,
``Conformal Geometry And String Field Theory,''
Nucl.\ Phys.\ B {\bf 278}, 91 (1986).
%%CITATION = NUPHA,B278,91;%%

%\cite{Giddings:1986wp}
\bibitem{Giddings:1986wp}
S.~B.~Giddings, E.~J.~Martinec and E.~Witten,
``Modular Invariance In String Field Theory,''
Phys.\ Lett.\ B {\bf 176}, 362 (1986).
%%CITATION = PHLTA,B176,362;%%

%\cite{Zwiebach:1990az}
\bibitem{Zwiebach:1990az}
B.~Zwiebach,
``A Proof That Witten's Open String Theory Gives 
A Single Cover Of Moduli Space,''
Commun.\ Math.\ Phys.\  {\bf 142}, 193 (1991).
%%CITATION = CMPHA,142,193;%%


%\cite{Zemba:rf}
\bibitem{Zemba:rf}
G.~Zemba and B.~Zwiebach,
``Tadpole Graph In Covariant Closed String Field Theory,''
J.\ Math.\ Phys.\  {\bf 30}, 2388 (1989).
%%CITATION = JMAPA,30,2388;%%

%\cite{LeClair:1988sj}
\bibitem{LeClair:1988sj}
A.~LeClair, M.~E.~Peskin and C.~R.~Preitschopf,
``String Field Theory On The Conformal Plane. 2. Generalized Gluing,''
Nucl.\ Phys.\ B {\bf 317}, 464 (1989).
%%CITATION = NUPHA,B317,464;%%

%\cite{Rastelli:2000iu}
\bibitem{Rastelli:2000iu}
L.~Rastelli and B.~Zwiebach,
``Tachyon potentials, star products and universality,''
JHEP {\bf 0109}, 038 (2001)
{\tt hep-th/0006240}.
%%CITATION = HEP-TH 0006240;%%

%\cite{Schnabl:2002gg}
\bibitem{Schnabl:2002gg}
M.~Schnabl,
``Wedge states in string field theory,''
JHEP {\bf 0301}, 004 (2003)
{\tt hep-th/0201095}.
%%CITATION = HEP-TH 0201095;%%

%\cite{Schnabl:2002ff}
\bibitem{Schnabl:2002ff}
M.~Schnabl,
``Anomalous reparametrizations and butterfly states in string field  theory,''
Nucl.\ Phys.\ B {\bf 649}, 101 (2003)
{\tt hep-th/0202139}.
%%CITATION = HEP-TH 0202139;%%

\bibitem{wt2} W.~Taylor,  ``D-brane effective field theory from string
  field theory'',  Nucl. Phys. B585:171-192 (2000) 
{\tt hep-th/0001201}.

\bibitem{kp} V.~A.~Kosteleck\`y and R.~Potting, ``Analytical construction of a 
  nonperturbative vacuum for the open bosonic string'',
  Phys. Rev. D63:046007 (2001) {\tt hep-th/0008252}.

%\cite{Zwiebach:1997fe}
\bibitem{Zwiebach:1997fe}
B.~Zwiebach,
``Oriented open-closed string theory revisited,''
Annals Phys.\  {\bf 267}, 193 (1998)
{\tt hep-th/9705241}.
%%CITATION = HEP-TH 9705241;%%

%\cite{Zwiebach:1990qj}
\bibitem{Zwiebach:1990qj}
B.~Zwiebach,
``Quantum Open String Theory With Manifest Closed String Factorization,''
Phys.\ Lett.\ B {\bf 256}, 22 (1991).
%%CITATION = PHLTA,B256,22;%%


%\cite{Green:pf}
\bibitem{Green:pf}
M.~B.~Green,
``The Influence Of World Sheet Boundaries On Critical Closed String Theory,''
Phys.\ Lett.\ B {\bf 302}, 29 (1993)
{\tt hep-th/9212085}.
%%CITATION = HEP-TH 9212085;%%

\bibitem{ams}
A.~Adams, J.~McGreevy and E.~Silverstein,
``Decapitating tadpoles,''
{\tt hep-th/0209226}.

\bibitem{McGreevy-Silverstein}
J.~McGreevy and E.~Silverstein, {\it to appear}.



%\cite{Berkovits:1995ab}
\bibitem{Berkovits:1995ab}
N.~Berkovits,
``SuperPoincare invariant superstring field theory,''
Nucl.\ Phys.\ B {\bf 450}, 90 (1995)
[Erratum-ibid.\ B {\bf 459}, 439 (1996)]
{\tt hep-th/9503099}.

%\cite{Witten:1986qs}
\bibitem{Witten:1986qs}
E.~Witten,
``Interacting Field Theory Of Open Superstrings,''
Nucl.\ Phys.\ B {\bf 276}, 291 (1986).
%%CITATION = NUPHA,B276,291;%%

%\cite{Preitschopf:1989gp}
\bibitem{Preitschopf:1989gp}
C.~R.~Preitschopf, C.~B.~Thorn and S.~A.~Yost,
``Superstring Field Theory,''
UFIFT-HEP-90-3
%\href{http://www.slac.stanford.edu/spires/find/hep/www?r=ufift-hep-90-3}
%{SPIRES entry}
{\it Invited talk given at Workshop on Superstring and Particle
Theory, Tuscaloosa, AL, Nov 8-11, 1989}

%\cite{Arefeva:1989cm}
\bibitem{Arefeva:1989cm}
I.~Y.~Arefeva, P.~B.~Medvedev and A.~P.~Zubarev,
%``Background Formalism For Superstring Field Theory,''
Phys.\ Lett.\ B {\bf 240}, 356 (1990).
%%CITATION = PHLTA,B240,356;%%


\bibitem{Sen-rolling}
% lookup 0203211
% lookup 0203265
%\bibitem{Sen:2002nu}
A.~Sen,
``Rolling tachyon,''
JHEP {\bf 0204}, 048 (2002)
{\tt hep-th/0203211};
%\bibitem{Sen:2002in}
A.~Sen,
``Tachyon matter,''
JHEP {\bf 0207}, 065 (2002)
{\tt hep-th/0203265}.

%\cite{Leblond:2003db}
\bibitem{Leblond:2003db}
F.~Leblond and A.~W.~Peet,
``SD-brane gravity fields and rolling tachyons,''
{\tt hep-th/0303035}.
%%CITATION = HEP-TH 0303035;%%

\bibitem{Lambert:2003zr}
N.~Lambert, H.~Liu and J.~Maldacena,
``Closed strings from decaying D-branes,''
{\tt hep-th/0303139}.

%\cite{Bogojevic:1987ma}
\bibitem{Bogojevic:1987ma}
A.~R.~Bogojevic,
``BRST Invariance Of The Measure In String Field Theory,''
Phys.\ Lett.\ B {\bf 198}, 479 (1987).
%%CITATION = PHLTA,B198,479;%%

\bibitem{spectroscopy} L.~Rastelli, A.~Sen, and B.~Zwiebach,  ``Star Algebra
  Spectroscopy'', JHEP 0203:029 (2002) {\tt hep-th/0111281}. 

\end{thebibliography}

\end{document}